\documentclass[
aps,
superscriptaddress,
nofootinbib,
longbibliography,
preprint,
amsmath,
amssymb
]{revtex4-2}

\usepackage[utf8]{inputenc}
\usepackage{subcaption}
\usepackage{graphicx}  
\usepackage{dcolumn}   
\usepackage[dvipsnames]{xcolor} 
\usepackage[colorlinks = true,
            linkcolor  = RoyalBlue,
            urlcolor   = RoyalBlue,
            citecolor  = RoyalBlue]{hyperref}
\usepackage{amsmath} 
\usepackage{amssymb} 
\usepackage{bbm}     
\usepackage{lineno} 
\usepackage{xspace} 
\usepackage{csquotes} 
\usepackage{booktabs} 

\usepackage{pifont}
\usepackage[dvipsnames]{xcolor}
\newcommand{\cmark}{\textcolor{PineGreen}{\ding{51}}}%
\newcommand{\xmark}{\ding{55}}%

\newcommand{\pp}           {pp\xspace}
\newcommand{\ppbar}        {\mbox{$\mathrm {p\overline{p}}$}\xspace}

\newcommand{\eA}           {\ensuremath{\mathrm{e^{\pm}}}\mbox{A}\xspace}
\newcommand{\pPb}          {\mbox{p--Pb}\xspace}

\newcommand{\ee}           {\ensuremath{\mathrm{e^+e^-}}\xspace}
\newcommand{\ep}           {\ensuremath{\mathrm{e^{\pm}p}}\xspace}
\newcommand{\q}            {\ensuremath{\mathrm{q}}\xspace}
\newcommand{\qbar}         {\ensuremath{\bar{\mathrm{q}}}\xspace}
\newcommand{\qqbar}        {\ensuremath{\mathrm{q\bar{q}}\xspace}}

\newcommand{\s}            {\ensuremath{\sqrt{s}}\xspace}
\newcommand{\snn}          {\ensuremath{\sqrt{s_{\mathrm{NN}}}}\xspace}
\newcommand{\pt}           {\ensuremath{p_{\rm T}}\xspace}
\newcommand{\meanpt}       {$\langle p_{\mathrm{T}}\rangle$\xspace}

\newcommand{\Npart}        {\ensuremath{N_\mathrm{part}}\xspace}

\newcommand{\RpPb}         {\ensuremath{R_{\rm pPb}}\xspace}

\newcommand{\Raa}          {\ensuremath{R_\mathrm{AA}}\xspace}
\newcommand{\FFc}{\ensuremath{f(\mathrm{c\rightarrow H_{c}})}\xspace}
\newcommand{\zjet}{\ensuremath{z_{\mathrm{||}}^{\mathrm{ch}}}\xspace}
\newcommand{\nineH}        {$\sqrt{s}~=~0.9$~Te\kern-.1emV\xspace}
\newcommand{\seven}        {$\sqrt{s}~=~7$~Te\kern-.1emV\xspace}
\newcommand{\twoH}         {$\sqrt{s}~=~0.2$~Te\kern-.1emV\xspace}
\newcommand{\twosevensix}  {$\sqrt{s}~=~2.76$~Te\kern-.1emV\xspace}
\newcommand{\five}         {$\sqrt{s}~=~5.02$~Te\kern-.1emV\xspace}
\newcommand{\twosevensixnn}{$\sqrt{s_{\mathrm{NN}}}~=~2.76$~Te\kern-.1emV\xspace}
\newcommand{\fivenn}       {$\sqrt{s_{\mathrm{NN}}}~=~5.02$~Te\kern-.1emV\xspace}

\newcommand{\GeVc}         {\ensuremath{\mathrm{GeV}/c}\xspace}

\newcommand{\TeV}          {\ensuremath{\mathrm{TeV}}\xspace}
\newcommand{\GeV}          {\ensuremath{\mathrm{GeV}}\xspace}
\newcommand{\MeV}          {\ensuremath{\mathrm{MeV}}\xspace}
\newcommand{\GeVmass}      {\ensuremath{\mathrm{GeV}/c^2}\xspace}

\newcommand{\mus}         {\ensuremath{\mu\mathrm{s}}\xspace}
\newcommand{\Taa}          {\ensuremath{\langle T_\mathrm{AA}\rangle}\xspace}
\newcommand{\Diffs}        {\ensuremath{\mathcal{D}_{s}}\xspace}
\newcommand{\ellflow}          {\ensuremath{v_\mathrm{2}}\xspace}

\newcommand{\Dzero}        {\ensuremath{\mathrm{D^0}}\xspace}
\newcommand{\Dplus}        {\ensuremath{\mathrm{D^+}}\xspace}
\newcommand{\Dminus}        {\ensuremath{\mathrm{D^-}}\xspace}

\newcommand{\Dstar}        {\ensuremath{\mathrm{D^{*+}}}\xspace}
\newcommand{\Dstarzero}        {\ensuremath{\mathrm{D^{*0}}}\xspace}
\newcommand{\Ds}           {\ensuremath{\mathrm{D_s^+}}\xspace}
 \newcommand{\Bs}           {\ensuremath{\mathrm{B_s^0}}\xspace}
 \newcommand{\Bplus}           {\ensuremath{\mathrm{B^+}}\xspace}
  \newcommand{\Bc}           {\ensuremath{\mathrm{B^+_c}}\xspace}

\newcommand{\Dsstar}{\ensuremath{\mathrm{D_s^{*+}}}\xspace}

\newcommand{\Dsminus}    
{\ensuremath{\mathrm{D_s^-}}\xspace}
\newcommand{\Lc}           {\ensuremath{\Lambda_\mathrm{c}^+}\xspace}
\newcommand{\Lcminus}           {\ensuremath{\Lambda_\mathrm{c}^-}\xspace}
\newcommand{\LcExcitedOne}{\ensuremath{\Lambda_\mathrm{c}(2595)^+}\xspace}
\newcommand{\LcExcitedTwo}{\ensuremath{\Lambda_\mathrm{c}(2625)^+}\xspace}
\newcommand{\SigmacZero}           {\ensuremath{\Sigma_\mathrm{c}(2455)^{0}}\xspace}
\newcommand{\SigmacZeroExcited}           {\ensuremath{\Sigma_\mathrm{c}(2520)^{0}}\xspace}

\newcommand{\Sigmac}           {\ensuremath{\Sigma_\mathrm{c}^{0,+,++}}\xspace}

\newcommand{\XicZero}      {\ensuremath{\Xi_\mathrm{c}^0}\xspace}
\newcommand{\XicPlus}      {\ensuremath{\Xi_\mathrm{c}^+}\xspace}
\newcommand{\XicPlusZero}  {\ensuremath{\Xi_\mathrm{c}^{+,0}}\xspace}
\newcommand{\Omegac}       {\ensuremath{\Omega_\mathrm{c}^0}\xspace}
\newcommand{\Jpsi}         {\ensuremath{\mathrm{J}/\psi}\xspace}
\newcommand{\ccbar}        {\ensuremath{\mathrm{c\overline{c}}}\xspace}

\newcommand{\Lb}        
{\ensuremath{\Lambda_\mathrm{b}^0}\xspace}
\newcommand{\Lbbar}           {\ensuremath{\bar{\Lambda}_\mathrm{b}^0}\xspace}

\newcommand{\lowptbin}{\ensuremath{0<\pt<1}~\GeVc}

\newcommand{\LcD} {\ensuremath{\Lc/\Dzero}\xspace}
\newcommand{\chiThreeEight} {\ensuremath{\chi_{c1}(3872)}\xspace}

\newcommand{\pythia}       {\textsc{pythia}\xspace}

\DeclareOldFontCommand{\bf}{\normalfont\bfseries}{\mathbf}
\DeclareOldFontCommand{\rm}{\normalfont\bfseries}{\mathrm}

\begin{document}

\title{Towards the understanding of heavy quarks hadronization: from leptonic to heavy-ion collisions}

\author{J. Altmann}
\email[]{javira.altmann@monash.edu}
\affiliation{School of Physics and Astronomy, Monash University, Melbourne, Australia} 
\affiliation{Rudolf Peierls Centre for Theoretical Physics, University of Oxford, UK} 

\author{A. Dubla}
\email[]{a.dubla@cern.ch}
\affiliation{GSI Helmholtzzentrum f{\"u}r Schwerionenforschung, 64291 Darmstadt, Germany}

\author{V. Greco}
\email[]{greco@lns.infn.it}
\affiliation{Department of Physics and Astronomy, University of Catania and INFN-LNS, Catania, Italy} 

\author{A. Rossi}
\email[]{a.rossi@cern.ch}
\affiliation{Universit\'a di Padova and INFN - Sezione di Padova, Padova, Italy } 

\author{P. Skands}
\email[]{peter.skands@monash.edu}
\affiliation{School of Physics and Astronomy, Monash University, Melbourne, Australia} 
\affiliation{Rudolf Peierls Centre for Theoretical Physics, University of Oxford, UK} 

\date{\today}

\begin{abstract}
The formation of hadrons is a fundamental process in nature that can be investigated at particle colliders. As several recent findings demonstrate, with \ee collisions as a "vacuum-like" reference at one extreme, and central nucleus--nucleus as a dense, extended-size system characterized by flow and local equilibrium at the opposite extreme, different collision systems offer a lever arm that can be exploited to probe with a range of heavy-flavour hadron species the onset of various hadronization processes. 
In this review, we present an overview of the theoretical and experimental developments. The focus is on open-heavy-flavour measurements. The comparison with model predictions and connections among the results in electron-positron, proton--proton, proton--nucleus, nucleus--nucleus collisions are discussed. After reviewing the current state, we suggest some prospects and future developments.

\end{abstract}

\maketitle
\newpage
\tableofcontents
\newpage


\section{Introduction}

 Hadronization is a fundamental process in nature representing the transition of a system of quarks and gluons to a state in which the degrees of freedom are colour-neutral quark-composite particles, namely hadrons. In the Early Universe, hadronization is supposed to have taken place about $10-20~\mus$ after the Big Bang \cite{Schwarz:2003du,Castorina:2015ava}. A prominent feature of strong interaction is ``colour confinement'', the phenomenon preventing the isolation and direct observation of free colour charges, i.e. of unbound quarks and gluons, in conditions of temperature and energy density typical of ordinary matter. However, particle accelerators make hadronization accessible in laboratory: in high-energy collisions of hadrons or leptons, quarks and gluons represent (for a very short time of the order of few fm$/c$) the system degrees of freedom. The details of the hadronization process influence the subsequent hadroproduction, the multiplicities of  particles produced, the relative abundances of different hadron species, and the hadron kinematics.  

Quantum Chromodynamics (QCD) is the theory of strong interactions within the Standard Model of Particle Physics. In QCD, colour-charge confinement appears in the static colour potential of a quark-antiquark pair as the presence of a term increasing linearly with distance and dominating the QCD potential at large distances. One of the fundamental properties of QCD is that the coupling constant, $\alpha_{\mathrm{s}}$, reduces with increasing squared 4-momentum transfer ($Q^{2}$), i.e. for short-distance interactions. 
Hadronization involves partonic processes characterized by small momentum transfers (i.e. $\approx \Lambda_{QCD} \sim \, R_H^{-1}$) and hence large values of the strong coupling constant, such that perturbative QCD calculations are not conceivable. Therefore, event generators which model parton production and the subsequent hadronization process event-by-event or (semi-)phenomenological models aimed at reproducing single-particle specific observables are used to infer various features of hadron formation and of its dependence on the properties of the collision system. In any kind of high-energy collision, from \ee{} to nucleus--nucleus, modelling hadron production starting from the partonic stage requires a colour neutralization procedure that groups the quarks and gluons 
into colour-neutral objects, i.e. the final hadrons. In event generators, the quarks and gluons involved in this process can arise from the initial parton--parton scatterings, subsequent processes like parton-showers, and string-breaking. Other models assume the presence of a bulk of partons with given kinematic properties, without tracing their history before hadronization.

The hadrons containing a charm or beauty quark (heavy-flavour hadrons), are always formed in the hadronization of a charm or beauty quark produced prior to hadronization, in initial hard-scatterings or in gluon splittings in parton showers. In these processes, contrary to the soft interactions occurring at hadronization time, large $Q^{2}$ ($Q^{2}>(2 m_{\mathrm{c,b}})^{2}$) are possible. The hadronization process determines the relative abundances of the different heavy-flavour hadron species and the hadron kinematics, but, contrary to what occurs for light-flavour hadrons, the hadronization process itself generally does not influence the total heavy-flavour hadron production 
which instead matches the production rate of the parent heavy quarks. The large values of the charm and beauty quark masses ($m_{b,c} \,>> \,\Lambda_{QCD}$) allow to employ perturbative QCD (pQCD) calculations of heavy-quark production down to zero momentum. For these reasons, heavy-flavour hadrons can be regarded as \enquote{calibrated} perturbative probes of the hadronization, which can provide unique information on this fundamental process, complementary to what can be inferred from studies of light-flavour production and of the event properties.

The transition of a heavy quark into a given hadron species requires the modelling of hadronization and of the soft non-perturbative processes involved. The simplest effective description of this transition, adopted in QCD calculations based on a factorization approach like FONLL~\cite{Cacciari:1998it,Cacciari:2012ny} and GM-VFNS~\cite{Kniehl:2004fy,Kniehl:2005mk,Kniehl:2012ti,Helenius:2018uul}, is to use fragmentation functions (FFunc). In the factorization approach, the transverse-momentum (\pt) and rapidity ($y$) differential heavy-flavour hadron production cross sections in proton--proton (pp) are computed as the convolution of three terms: (i) the parton distribution functions (PDFs) of the incoming (anti)protons, (ii) the partonic cross section, calculated as a perturbative series in powers of the strong coupling constant $\alpha_\mathrm{s}$, and (iii) the fragmentation functions, which parameterize the probability that a heavy-quark gives rise to a hadron of a given species carrying a given fraction ($z_{\mathrm{Q}}$) of the parent quark momentum. Thus, FFunc allow to map the quark momentum-differential cross section into the final hadron momentum-differential cross section. The FFunc used in these pQCD calculations have been parameterized for decades from measurements performed in \ee{} or ep collisions~\cite{Braaten:1994bz}. Charm and bottom quark production in \ee{} annihilation offers the cleanest environment for constraining heavy-quark FFunc. However, the usage of FFunc parameterized on \ee{} or ep data in pp collisions requires the assumption that FFunc are \enquote{universal}, independent of the colliding system. Under the universality assumption, once determined in a given collision system, like \ee{}, FFunc can be applied to any reaction. An even more basic assumption, is the universality of fragmentation fractions (FF), which represent the probability that a heavy-quark produces a given heavy-flavour hadron species independently of $z_{\mathrm{Q}}$, i.e. the $z_{\mathrm{Q}}$ integral of the FFunc. The modelling of hadronization in hadronic collisions with universal FFunc and FF relies on the hypothesis that, at large extent, the hadronization of a charm or beauty quark is not sensitive to the hard quark-production process and to the surrounding partonic environment in which hadronization occurs. In \ee{} collisions the FF can be directly measured from the cross sections of the observed hadron species without the need of measuring all existing species, given that the heavy-quark production cross section can be calculated with high precision. Though practical limitations (e.g. kinematic reach of the measurements) may require extrapolation, this in principle allows the important check that the sum of FF of all observed weakly-decaying charm or beauty hadrons equals unity, thus the verification that all weakly decaying states are known.
Though the universality assumptions are not a rigorous prediction of QCD and can be considered reasonable only within certain (quite vaguely defined) kinematic conditions, validating them has been an important goal of experimental measurements of heavy-flavour hadron production since the formulation of QCD and of the parton model. Hadron-to-hadron production cross section ratios are particularly sensitive to FFunc, FF, and in general, to heavy-flavour hadronization mechanisms, because the heavy-quark production cross-section, which depends on PDF and the hard-scattering process, substantially cancel in the ratios. 
The production of specific heavy-flavour hadron species has been measured in different energy regimes and \enquote{hadronic environments}: in \ee{} collisions at LEP and B factories, in ep collisions, in proton-(anti)proton (pp/p$\rm \bar{p}$) collisions, and in heavy-ion collisions (nucleus--nucleus, AA, or proton--nucleus, p--A).
Before the Large Hadron Collider (LHC) era, the universality assumption had been tested mainly by comparing cross-section ratios of different D and B meson species in \ee, \ep, and pp (\ppbar) collisions, while data on baryon-to-meson cross-section ratios in hadronic collisions were not available or precise, and basically limited to $\Lc$ and $\Lb$. The $\Lc/\mathrm{D}$ ratios measured in \ep and \ee collisions were consistent, thus supporting universality. Hadronization models developed in event generators, like the Lund string fragmentation model in \pythia~\cite{Sjostrand:2006za,Skands:2014pea} or the cluster model in HERWIG~\cite{Bahr:2008pv}, capable to reproduce several feature of \ee data and pp data, were not expecting, in their standard implementation, significant differences in the hadron-species ratios from \ee to pp collisions, also supporting the hypothesis that modifications to the hadronization process could be small. On the contrary, the recent heavy-flavour baryon measurements in hadronic collisions at LHC energies~\cite{LHCb:2019fns,LHCb:2023wbo,ALICE:2020wfu,ALICE:2020wla,ALICE:2022ych,sigmac,ALICE:2023sgl,Acharya:2021dsq,xic13tev,ALICE:2022cop,ALICE:2021npz,CMS:2023frs,ALICE:2024ozd,LHCb:2018weo,LHCb:2023cwu,ALICE:2021bib} indicated a significant difference in the fragmentation fractions of charm and beauty quarks into charm and beauty baryons compared to those measured in \ee{} and ep collisions. In the beauty sector, the CDF experiment had found a difference in the $\Lb/B$ ratio in \ppbar collisions at Tevatron with respect to LEP values~\cite{Aaltonen:2008zd}: at that time, however, the result was related to the different jet momenta involved in the two measurements (much higher at LEP) but not traced back to a breakdown of factorization, a modification of the hadronization process, or some other cause. 
The LHC measurements challenge the assumption that heavy-quark hadronization is a universal process across different colliding systems. The breaking of the universality can be related to  effects that make heavy-quark hadronization dependent on the underlying event.

In heavy-ion collisions, hadronization of heavy quarks (as well as that of light partons) can occur in a rather different way. In these collisions a colour-deconfined state of QCD matter is formed, the quark-gluon plasma (QGP)~\cite{ALICE:2022wpn,CMS:2024krd,STAR:2005gfr}. The heavy-quark hadronization process happens in a hot and dense environment with a large abundance of light coloured thermal particles nearby with which the heavy quark can recombine, neutralizing its colour charge. The recombination process, frequently modeled as a 2 → 1 coalescence (or 3 → 1 in the case of baryon formation), is a fundamental hadronization mechanism crucial for reproducing several experimental observations. These include shifts in baryon-to-meson ratios and the collective motion of final-state hadrons, linked to the system's expansion, both in the light and heavy-flavour sectors.
In heavy-ion collisions also other observables are sensitive to the hadronization processes, like the nuclear modification factor and the anisotropic flow coefficients. The nuclear modification factor (\Raa) is defined as the ratio of the yield measured in AA collisions to the pp cross section scaled by the average nuclear overlap function \Taa~\cite{dEnterria:2020dwq}, a quantity proportional to the number of binary nucleon--nucleon collisions, for the considered centrality interval. In the absence of nuclear effects, $\Raa=1$ is expected. The anisotropic flow coefficients are the Fourier coefficients ($v_{\rm n}$) of the distribution of the heavy-flavour hadron azimuthal angle ($\phi$) with respect to the reaction plane defined by the nuclei impact-parameter vector, which is the vector connecting the centres of the colliding nuclei, and the beam direction. The coefficient of the second-order harmonic (\ellflow) is called elliptic flow. It encodes the efficiency by which the hydrodynamic evolution of the medium converts to momentum space the initial spatial azimuthal anisotropy of the fireball (and of the produced partons), which, in non-central collisions, principally arises from the non-zero eccentricity of the nuclei overlap region. The initial asymmetry gives rise to azimuthally anisotropic pressure gradients which drives the hydrodynamic expansion of the system.

In this review, we present an overview of the state-of-the-art of theoretical models and experimental measurements related to the characterization of heavy quark hadronization from \ee{} to heavy-ion collisions. Theoretical models, including recent developments, are discussed in Sect.~\ref{sect:Theory}. The experimental results are presented in Sect.~\ref{sect:exp}. The discrepancies between the measurements and theoretical predictions are discussed in detail. Finally, in Sect.~\ref{sect:future} we suggest new measurements and theoretical ideas that we consider important to achieve further progresses on the understanding of hadronization. 

\section{Models of hadronization in high-energy collisions}
\label{sect:Theory}

This section presents an overview of theoretical models for hadronization in high-energy collisions, with particular emphasis on aspects relevant to heavy-flavour hadron production in hadron-hadron and heavy-ion collisions.
Table~\ref{tablemodels} shows a schematic overview of the available Monte Carlo generators and phenomenological models highlighting their main features. The check marks indicate whether a model has the potential to reproduce the indicated aspects and effects, regardless of how well the predictions match the existing data.

In section~\ref{sec:MCmodels}, we briefly summarize the foundations and basic aspects of the two main dynamical models of  hadronization implemented in Monte Carlo (MC) event generators, the string and cluster models. 

In section~\ref{sec:baryons}, we turn to the charm and beauty baryon sectors specifically, and consider both extensions to the baseline MC models as well as other key approaches like coalescence/recombination and thermal/statistical models.

\begin{table}[t]
\caption{Overview of hadronization models. The check marks indicate whether a model has the potential to reproduce the indicated aspects and effects, regardless of how well the predictions match the existing data.}
\label{tablemodels}
\begingroup
\setlength{\tabcolsep}{2pt} 
\renewcommand{\arraystretch}{0.9} 
\scalebox{0.935}{%
\begin{tabular}{p{3.75cm}ccccccc}
\toprule 
 & \multicolumn{1}{p{2.3cm}}{\setlength{\baselineskip}{3ex}
\centering \textsc{Pythia} } & \multicolumn{1}{p{2cm}}{\setlength{\baselineskip}{3ex}\centering \textsc{Pythia} \textsc{Angantyr}} &  \multicolumn{1}{p{1.6cm}}{\setlength{\baselineskip}{3ex}{\centering \textsc{Herwig} }} & \multicolumn{1}{p{1.5cm}}{\centering\textsc{Epos}}& \multicolumn{1}{p{1.5cm}}{\centering SHM}  & \multicolumn{1}{p{1.5cm}}{\setlength{\baselineskip}{3ex}{\centering Coal\-escence}} &  \multicolumn{1}{p{1.6cm}}{\setlength{\baselineskip}{3ex}{\centering \textsc{Pow\-lang}}} \\
\midrule
$\mathrm{e^+e^-}\to \mathrm{hadrons}$ & \cmark & \cmark & \cmark & \cmark & (\cmark)$^7$ & \xmark & \xmark \\
$\mathrm{pp}$ (soft) & \cmark & \cmark & \cmark & \cmark & \cmark &  \cmark & \cmark \\
$\mathrm{pp}$ (hard) & \cmark & \cmark & \cmark & (\cmark)$^6$ &(\cmark)$^8$ & \cmark & \cmark \\
$\mathrm{pA}$ & (\cmark)$^1$ & \cmark & \xmark & \cmark & (\cmark)$^9$  &(\cmark)$^9$  &(\cmark)$^9$  \\
$\mathrm{AA}$ & (\cmark)$^1$ & \cmark & \xmark & \cmark &\cmark  &\cmark &\cmark \\
\midrule
Fully Exclusive Events & \cmark & \cmark & \cmark & \cmark & \xmark &\xmark &\xmark \\
Particle Yields & \cmark & \cmark & \cmark & \cmark & \cmark &  \cmark & \cmark \\
Particle Spectra & 
  \cmark  & \cmark & \cmark & \cmark & \cmark &  \cmark & \cmark \\
\midrule
Charm baryon Enh. & (\cmark)$^{2,3}$& (\cmark)$^{2,3}$ & (\cmark)$^5$ & \cmark & (\cmark)$^{10}$ &  \cmark  &\cmark \\
Strangeness Enh. & 
  (\cmark)$^3$ & (\cmark)$^3$ & \xmark & \cmark & \cmark    &\cmark &\cmark \\
Collective Flow & 
  (\cmark)$^4$  & (\cmark)$^4$ & \xmark & \cmark & (\cmark)$^{11}$ &\cmark &\cmark \\
\bottomrule
\end{tabular}}
\endgroup
\\[2mm]
\raggedright\small
$^1$: Limited functionality, mainly intended for cosmic-ray cascades.\\\vspace*{-2mm}
$^2$: With the QCD Colour-Reconnection model.\\\vspace*{-2mm}
$^3$: With the Rope-Hadronization or Close-Packing models.\\\vspace*{-2mm}
$^4$: With the String-Shoving model.\\\vspace*{-2mm}
$^5$: With the Baryon Colour-Reconnection model.\\\vspace*{-2mm}
$^6$: In \textsc{Epos}~4.\\\vspace*{-2mm}
$^7$: In some SHM approaches, but not in those discussed within this review for open heavy flavour.\\\vspace*{-2mm}
$^8$: Added as a contribution from independent fragmentation associated to a corona.\\\vspace*{-2mm}
$^9$: It can be done without modifying the approach, but it has not been published\\\vspace*{-2mm}
$^{10}$: Obtained by adding further states predicted by the relativistic quark model.\\\vspace*{-2mm}

$^{11}$:  Added coupling SHM to a blast wave with radial anisotropic flow. \\\vspace*{-2mm}

\end{table}

\subsection{Baseline MC Models}
\label{sec:MCmodels}

 The explicit dynamical models that have been used for several decades to model the hadronization process in \ee{}, $\mathrm{ep}$, and $\mathrm{pp}$ collisions in Monte Carlo event generators, are the string and cluster hadronization models. Below, we outline these as used in the PYTHIA~\cite{Sjostrand:2006za,Bierlich:2022pfr} and HERWIG~\cite{Bellm:2015jjp} generators respectively.

\subsubsection{String model -- PYTHIA}

Hadronization in PYTHIA~\cite{Sjostrand:2006za,Bierlich:2022pfr} is based on the semi-classical Lund string fragmentation model~\cite{Andersson:1983ia}, and in PYTHIA it is mainly implemented to model hadronization from \ee{} up to pp collisions, with the relatively recent Angantyr extension allowing to study also heavy-ion collisions.
In the Lund string model, the ``string'' represents the collapsed colour confinement field formed between a colour charge and anticolour charge in an overall singlet state (otherwise referred to as ``colour-connected'' partons). A key motivation for the modelling of the confinement field as a string is the linear potential --- and therefore constant tension --- observed at non-perturbative scales on the lattice and encapsulated by the QCD Cornell potential~\cite{Eichten:1978tg, Bali:1992ab}. Thus the string is defined by its characteristic constant string tension, $\kappa \sim 1$ GeV/fm, and can be modelled by the dynamics of a relativistic string~\cite{Andersson:2001yu}.

\begin{figure}[t] 
\centering
    \includegraphics[width = 0.95\textwidth]{ 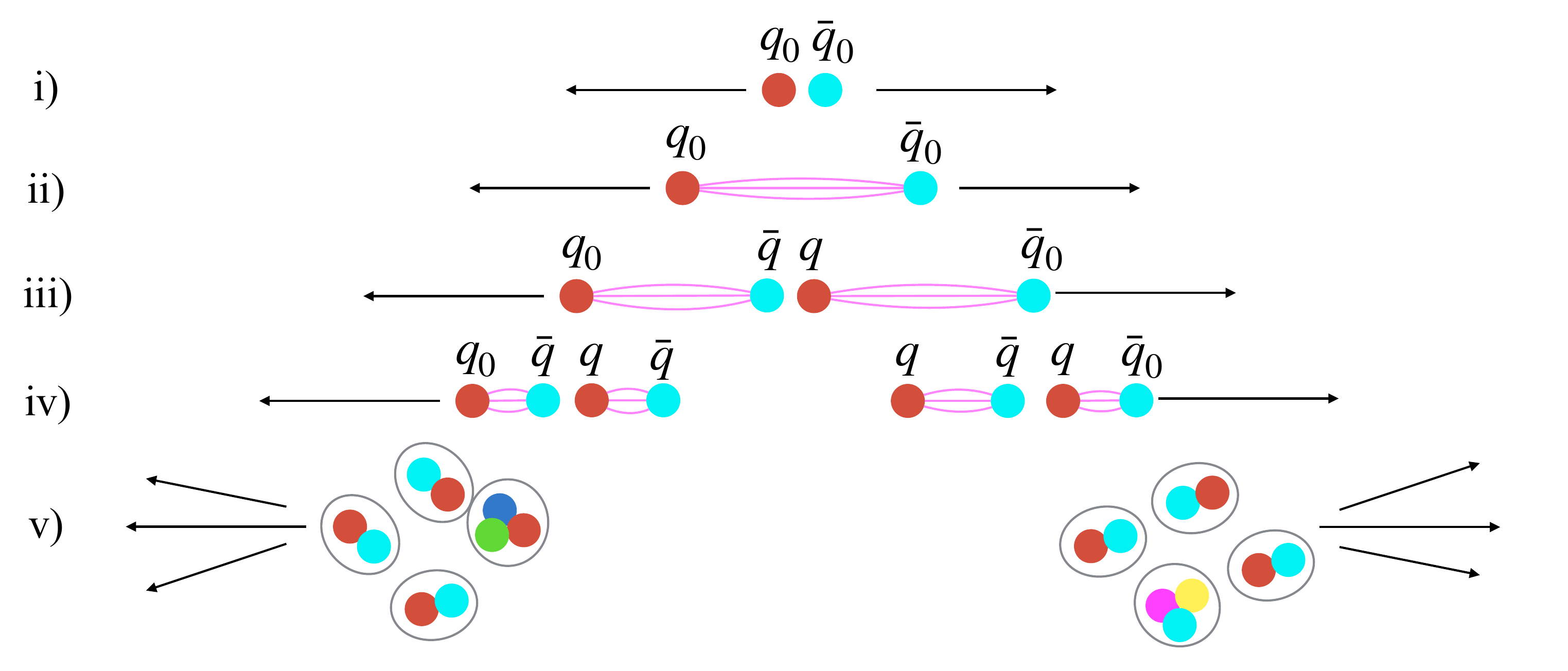}
       \caption{hadronization in the picture of string fragmentation. i) The initial $\q_0$ and $\qbar_0$ begin to move apart. ii) The string begins to form between them. iii) A string break occurs, creating a new $\qqbar$ pair. iv) More string breaks occur. v) Jets of final-state hadrons from the fragmentation process.} 
      \label{stringFrag}
\end{figure}

 In the context of high-energy collisions, colour-connected partons move apart from one another with high energies, stretching the string formed between them. Given sufficient energy, at some point it becomes energetically favourable to break the string rather than keep stretching it. At the site of the string break, a light-flavoured quark-antiquark (or diquark-antidiquark) pair is spontaneously pair produced, resulting in two smaller string systems. Note that in this context, light-flavoured refers to up, down and strange quarks. This string-breaking, or ``fragmentation'' process, occurs until there is no longer sufficient energy to keep breaking the string, resulting in on-shell final-state hadrons. A basic sketch of this fragmentation process can be seen in Fig.~\ref{stringFrag}. 

The string breaks are modelled via a quantum tunnelling process analogous to electron-positron pair production in the presence of a strong electric field described by the Schwinger mechanism~\cite{Schwinger:1951nm}. The modified form of the Schwinger mechanism for QCD is given by 
\begin{equation}
    \exp\left( \frac{-\pi m_q^2}{\kappa} \right)\exp \left(\frac{-\pi p_{\perp q}^2}{\kappa}\right) \equiv \exp\left(\frac{-\pi m_{\perp q}^2}{\kappa}\right)~,
    \label{schwinger}
\end{equation} 
where $m_\perp^2 = m^2 + p_\perp^2$ is called the transverse mass and $\kappa$ is the string tension mentioned above. 

 The Gaussian form of the Schwinger mechanism implies a strong suppression of breakups with high transverse momenta with respect to the string axis, and likewise a very strong suppression of heavier quarks relative to lighter ones. 
In particular, it is impossible to produce charm and heavier quarks this way, since, e.g., $\exp ( - \pi m_\mathrm{c}^2 / \kappa) \lesssim 10^{-11}$. This also reflects that high-$p_\perp$ excitations and heavy quarks should only be produced by perturbative mechanisms (partonic scattering processes and parton showers) and not by non-perturbative ones like string breaks. 

 Once a flavour and a $p_{\perp{q}}$ for the next string break in the iterative process has been selected, the next step is to select a hadron type compatible with the flavour of the current endpoint and that produced in the breakup. Since spin-1 mesons are heavier than spin-0 ones in QCD, the observed ratios between vectors and pseudoscalars are generally smaller than 3 (and equivalently for spin-3/2 baryons compared to spin-1/2 ones), however the baseline string model does not predict what these factors should be\footnote{We note the recent work on making the model more predictive in this regard~\cite{Bierlich:2022vdf}.}; instead, these are determined from data, via tuning\footnote{In the context of heavy quarks, we note that the default PYTHIA Monash tune has a known issue with the $\Dstar/\Dzero$ ratio being too low. This can be improved  by increasing the parameter \texttt{StringFlav:mesonCvector} from 0.88 to 1.5, which we recommend for charm studies.}. 

 To simplify the on-shell constraints of final-state hadrons, string breaks are performed such that on-shell hadrons are fragmented off directly from either string end. This is permissible since the individual string breaks are causally disconnected in the Lund model, meaning breaks can be performed in any order. Given this casual disconnect, a sensible constraint is that it should not matter which end of the string we start from; this limits the possible forms of the ``fragmentation functions'', $f(z)$, we can choose to a family of ``left-right symmetric'' ones~\cite{Andersson:1983ia}, parameterized by the two constants $a$, which may in principle be flavour-dependent, and $b$, which is flavour-universal:
\begin{equation}
    f(z) = N\frac{1}{z}(1-z)^a\exp \left( \frac{-bm_\perp^2}{z} \right)~,
    \label{fragFunc}
\end{equation}
where $N$ is a normalization constant that just ensures unit integral. 
Note that this fragmentation function does not have quite the same interpretation as the ones sometimes used in analytical contexts. The type shown here governs the probability distribution for the longitudinal component of the string-fragmentation process, in each step of the fragmentation. Given the transverse mass of the produced hadron, the function gives the probability the hadron will take a fraction $z$ of the total energy of the string. The $a$ and $b$ parameters must be determined from data, via tuning; see, e.g.,~\cite{Hamacher:1995df,ALEPH:1996oqp,Skands:2014pea,Amoroso:2018qga,Bellm:2019owc,Jueid:2023vrb}.

 Given a massless quark model, a Minkowski space-time diagram for a high energy $q\bar{q}$ pair moving apart from one another would result in the partons moving along a straight line along the light-cone. However given a heavy quark endpoint, this would no longer hold as a good approximation. Consider a scenario with a heavy quark and massless antiquark. Starting in the rest frame of the heavy quark, the equations of motion of strings show that the heavy quark will be dragged by the string field along a hyperbola rather than a straight line, reducing the area available for a string break. The Bowler modification~\cite{Bowler:1981sb} takes this into consideration by introducing an additional factor, $1/z^{r_Q b m_Q^2}$, into the fragmentation function. Here a multiplicative factor of $r_Q$ is introduced to allow for flexibility and is set by the parameters \texttt{StringZ:rFactC} and \texttt{StringZ:rFactB}. Provided uniquely defined quark masses these parameters ought to be unity. There is however some deviation from unity in the default PYTHIA tune. 

The above covers how string breaks occur, however one must also determine which partons are connected via strings in the first place. The default PYTHIA tune (called the Monash 2013 tune)~\cite{Skands:2014pea} forms strings by colour-connecting partons in what is known as the Leading-Colour (LC) limit. This approximation takes the number of QCD colours ($N_C$) to infinity, meaning that each colour in an event is uniquely matched to a single anticolour, resulting in unambiguous string configurations. In this limit, only so-called ``dipole'' string configurations are allowed. These dipole strings are characterized by two endpoints, a colour and an anticolour charge endpoint (typically a quark and an antiquark), and can have any number of intermediate gluons between them which form \enquote{kinks} on the string. Examples of a dipole-type strings can be seen in the left-hand side images of Fig.~\ref{CR}.

Given the LC limit (and thus only dipole strings), diquark endpoints and diquark-antidiquark pair creation are the only mechanisms for baryon production in this model. As string breaks cannot produce heavy-flavour quarks, heavy-flavoured partons are therefore necessarily string endpoints, and thus the only mechanism for heavy-flavour baryon production via fragmentation in the LC limit is diquark-antidiquark pair creation next to a heavy-flavoured endpoint, while diquark production has some fixed probability of occurring (controlled by parameter \texttt{StringFlav:probQQtoQ} in PYTHIA) in the baseline string fragmentation model. 

The LC approximation may be quite good in the environment of \ee{} collisions (more specifically in the absence of multi-parton interactions (MPI)), as corrections to the LC approximation are naively expected to be $1/N^2_C$ suppressed. However, significant deviations from it are expected in pp collisions, particularly for baryon production, where the survival of the strict LC topology should be heavily suppressed. 


\subsubsection{Cluster model -- HERWIG} 

\begin{figure}[ht!]
    \begin{center}
    \begin{subfigure}{0.49\textwidth} 
        \includegraphics[width = \textwidth]{ 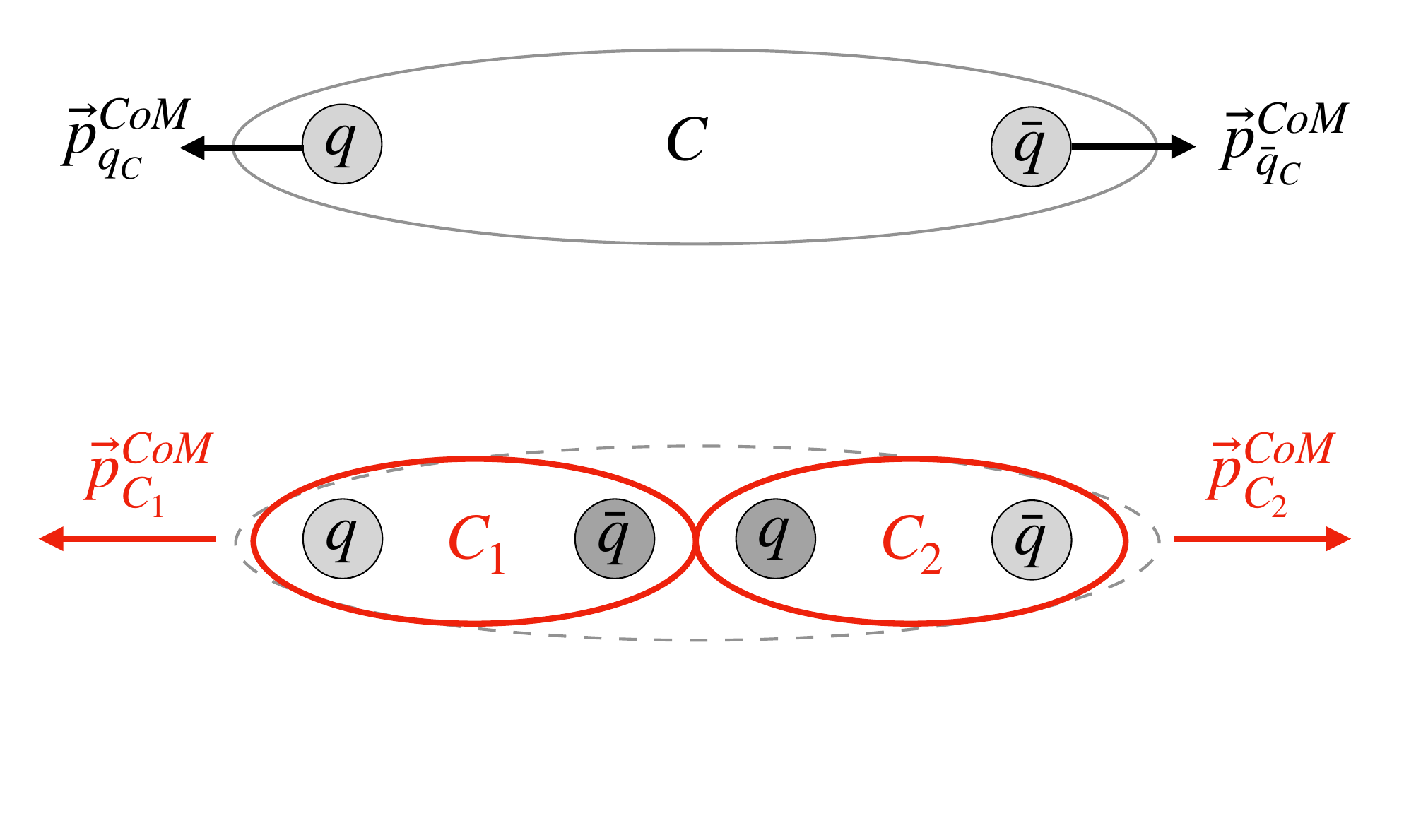}
        \caption{Cluster fission of cluster $C$ into two clusters, $C_1$ and $C_2$.}
        \label{fig:CF}
    \end{subfigure}
    \begin{subfigure}{0.49\textwidth}
        \includegraphics[width = \textwidth]{ 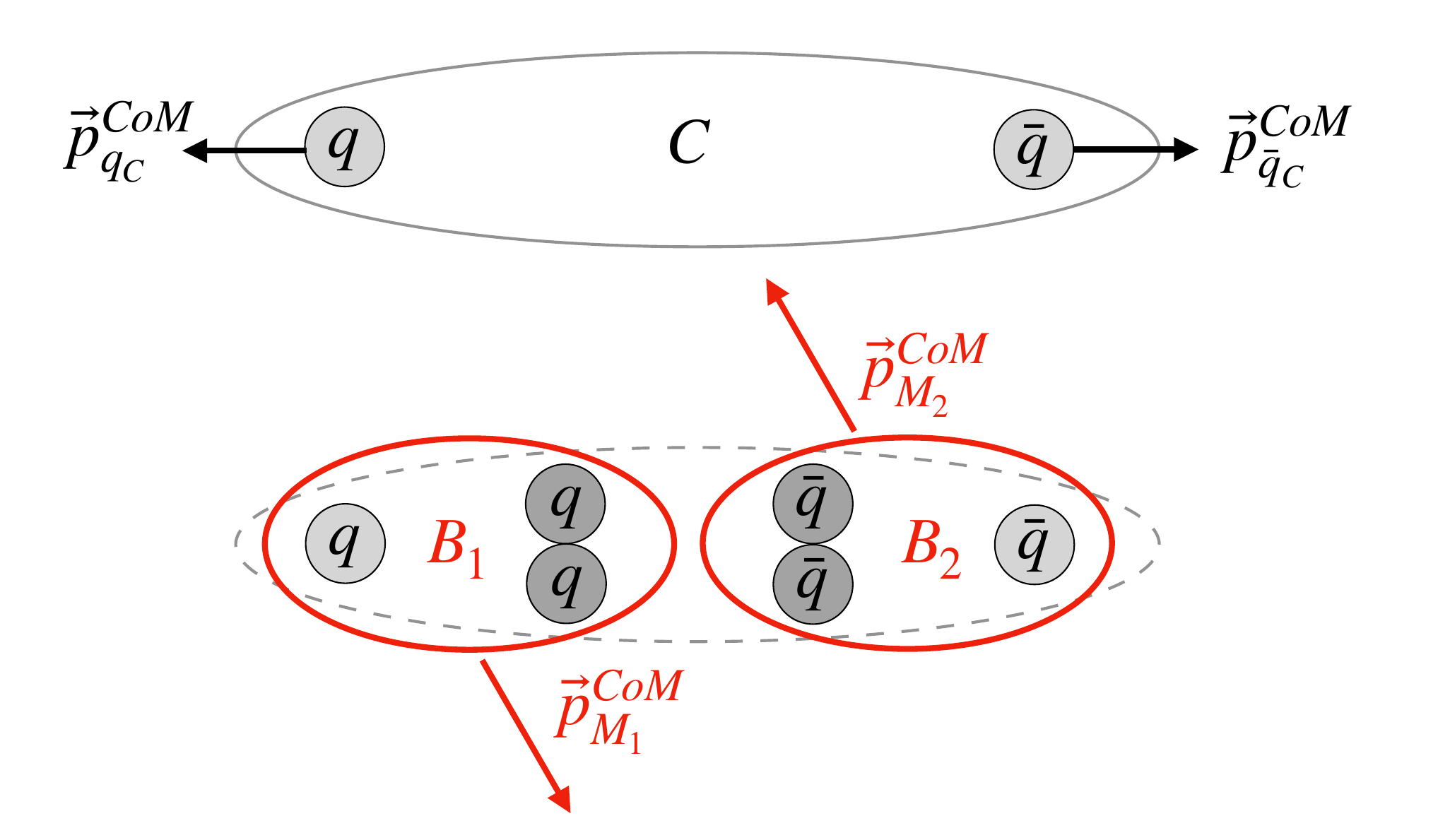}
        \caption{Cluster decay of cluster $C$ into two \newline baryons, $B_1$ and $B_2$.}
        \label{fig:CD}
    \end{subfigure}
    \end{center}
    \caption{Schematics of the cluster fission and decay processes used in cluster hadronization, with top image showing the initial cluster in the centre-of-mass frame, and the bottom image shows the system after the cluster fission/decay process has occurred.}
    \label{fig:clusterCR}
\end{figure}

 Hadronization in HERWIG~\cite{Bellm:2015jjp} uses the cluster model~\cite{Webber:1999ui} which is based on the idea of colour pre-confinement. The cluster model starts by splitting gluons non-perturbatively, $g \rightarrow \qqbar$, after the parton shower. Assuming LC colour-connections in the baseline model, each colour-singlet $\qqbar$ pair forms a cluster. These $\qqbar$ clusters are often referred to as ``mesonic clusters'' (see Sect.~\ref{sec:clusterCR} for other cluster types). Given these clusters, there are two main steps in the cluster hadronization process; cluster fission and cluster decay which are illustrated in fig.~\ref{fig:clusterCR}.

 Cluster fission deals with large mass clusters where clusters with mass, $M$, above a threshold $M_{max}(q_1,\bar{q}_2)$ recursively undergo cluster fission. Working in the centre-of-mass frame of the initial cluster, a light $\qqbar$ pair is created from the vacuum with some flavour probability $P_q$, and with given momenta to align with the initial $\qqbar$ pair. For the example shown in Fig.~\ref{fig:CF}, the initial cluster, $C$, is reduced via cluster fission to two smaller clusters, $C_1$ and $C_2$ with masses $M_1$ and $M_2$ respectively, which are sampled by 

\begin{equation}
	M_{1,2} = m_{1,2} + (M-m_{1,2} -m_q) \mathcal{R}^{1/P}.
\end{equation} 

 Here $m_i$ is the constituent mass of each quark in the initial cluster, and $m_q$ the constituent mass of the new $\qqbar$ pair (special treatment of clusters with beam remnants can be found in Ref.~\cite{Bahr:2008pv}). This distribution of masses is controlled via a random number, $\mathcal{R}$, between 0 and 1, and the parameter $P$ (\texttt{PSplitLight}, \texttt{PSplitCharm} or \texttt{PSplitBottom}), along with satisfying the conditions $M > M_{i}$ and $M_{i} > m_i + m_q$. It is important to note that no diquark formation is permitted in the cluster fission process.

 After the cluster fission process comes cluster decays, a process which maps each cluster onto two hadrons as seen in Fig.~\ref{fig:CD}. This process is similar to cluster fission however it has hadronic mass constraints applied. When a cluster decays, the direction of the decay is typically isotropically chosen in the cluster rest frame. Here both quark-antiquark and diquark-antidiquark formation is allowed, allowing mesonic clusters to decay into either two mesons or a baryon and an antibaryon as per the example shown in fig.~\ref{fig:CD}. For clusters with perturbatively produced partons (thus particularly important for heavy-flavour hadron formation), the decay products retain the direction of the cluster rest frame up to a possible Gaussian smearing through an angle $\theta_{smear}$, such that

\begin{equation} 
	\cos\theta_{smear} = 1 + \texttt{ClSmr} \log\mathcal{R},
\end{equation}

 where the parameter \texttt{ClSmr} can be set independently for clusters containing light, charm and bottom quarks. Similarly to PYTHIA, the LC limit is insufficient to describe baryon production in pp collisions~\cite{Gieseke:2017clv,Bellm:2019wrh}. 

\subsection{Enhancing charm and beauty baryon yields}
\label{sec:baryons}

Several models based on different developments and assumptions, like the inclusion of beyond leading-colour (BLC) effects on string~\cite{Christiansen:2015yqa} or cluster~\cite{Gieseke:2017clv} formation, or the inclusion of hadronization via coalescence~\cite{Oh:2009zj,Song:2018tpv,Minissale:2020bif,Plumari:2017ntm}, or the addition of a set of yet-unobserved higher-mass charm-baryon states in a statistical-hadronization framework~\cite{He:2019tik,He:2019vgs,Andronic:2021erx}, expect an enhancement of heavy-flavour baryon production relative to that of meson states, in hadronic collisions with respect to \ee collisions.  

\subsubsection{Extensions to the default (Monash 2013) PYTHIA tune \label{sec:pythiaExt}}

 One model critical for modelling baryon production in pp collisions using PYTHIA is the so-called QCD-based colour reconnection (CR) model~\cite{Christiansen:2015yqa}. Extending the string model beyond the leading-colour approximation reintroduces colour-space ambiguities due to the $N_C = 3$ finite colour structure of QCD. Modelling partons using SU(3) colour-space means that each colour is not necessarily unique, allowing multiple different partons to be colour-connected and causing ambiguity in where confining potentials arise. To determine between which partons the confining fields form, the configuration that minimizes the string ``length", or the action potential of the system, is chosen. This ``string length" is a Lorentz-invariant measure called the lambda measure~\cite{Andersson:1998tv}, which is a measure of the energy density per unit length of the string. By requiring minimized string lengths between colour-connected partons, strings are thus able to form between partons not produced in the same hard scatterings, allowing topologies to span across different MPI. The so-called QCD-based CR model developed within PYTHIA 8 stochastically reintroduces the SU(3) colour correlations, doing so using SU(3) algebra to describe the probabilities associated with overlaps in colour-space. These overlaps become particularly relevant for hadronic collisions at high energies, where MPI produce densely populated partonic environments.

\begin{figure}[ht!]
    \begin{center}
    \begin{subfigure}{0.8\textwidth} 
        \includegraphics[width = \textwidth]{ 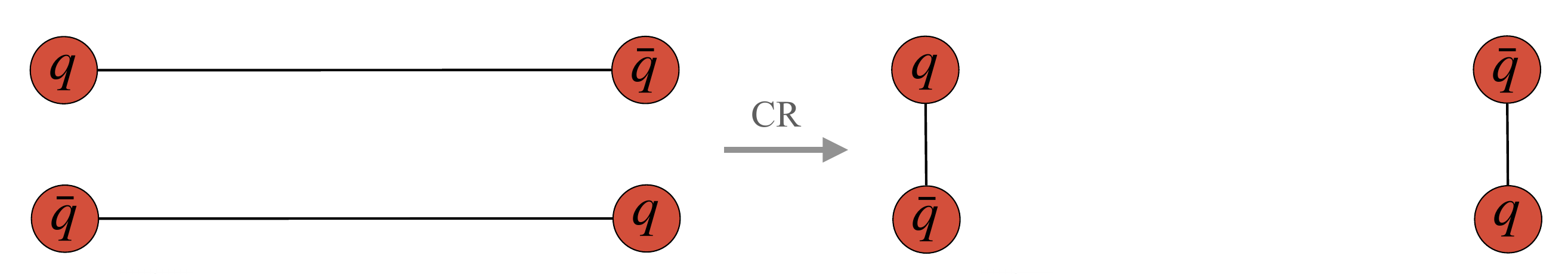}
        \caption{Dipole-type reconnection.}
        \label{fig:dipoleCR}
    \end{subfigure}
    \begin{subfigure}{0.8\textwidth}
        \includegraphics[width = \textwidth]{ 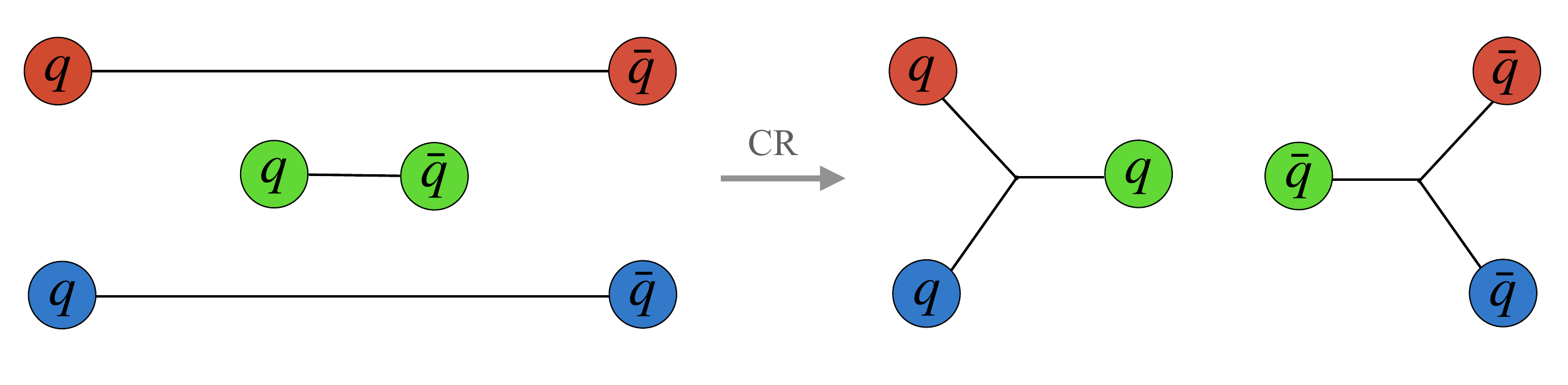}
        \caption{Junction reconnection.}
        \label{fig:juncCR}
    \end{subfigure}
    \begin{subfigure}{0.8\textwidth}
        \includegraphics[width = \textwidth]{ 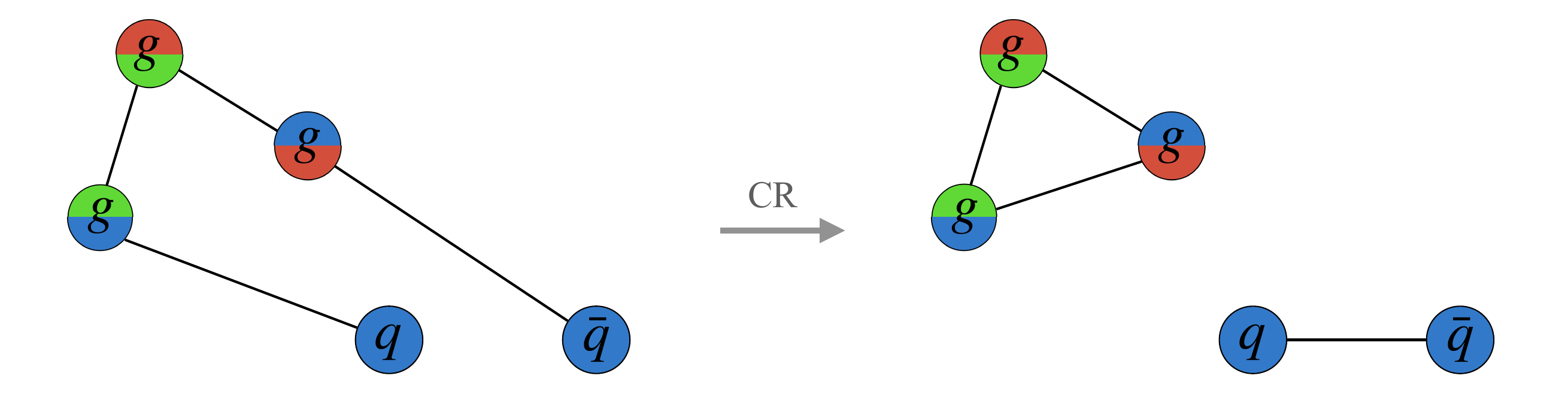}
        \caption{Gluon-loop formation.}
        \label{fig:gluonCR}
    \end{subfigure}
    \end{center}
    \caption{For each reconnection type, the left images show a LC string configurations, and the right-hand side images show possible alternative colour reconnected configurations. Note that colour-octet gluons carry a colour and an anticolour charge, which are represented by the colours on the top and bottom half of the gluons respectively.}
    \label{CR}
\end{figure}

 Given the colour singlet requirement with the inclusion of sub-leading colour effects, there are three general types of string configurations to consider: dipole strings, junction topologies, and gluon loops. Both junctions and gluon loops are explicitly beyond LC configurations. Dipole reconnections and gluon loops make use of the colour-anticolour singlet state, whereas junction topologies occur due to the colour-neutral red, green, and blue colour combination. Fig.~\ref{CR} demonstrates examples of each of these CR types, where the length of the lines representing strings are meant to give an indication of the string length which we want to minimize via CR. The left images show an example LC configuration, and the right show beyond LC strings after allowing CR.

\begin{figure}[ht!]
    \begin{center}
    \includegraphics[width = 0.6\textwidth]{ 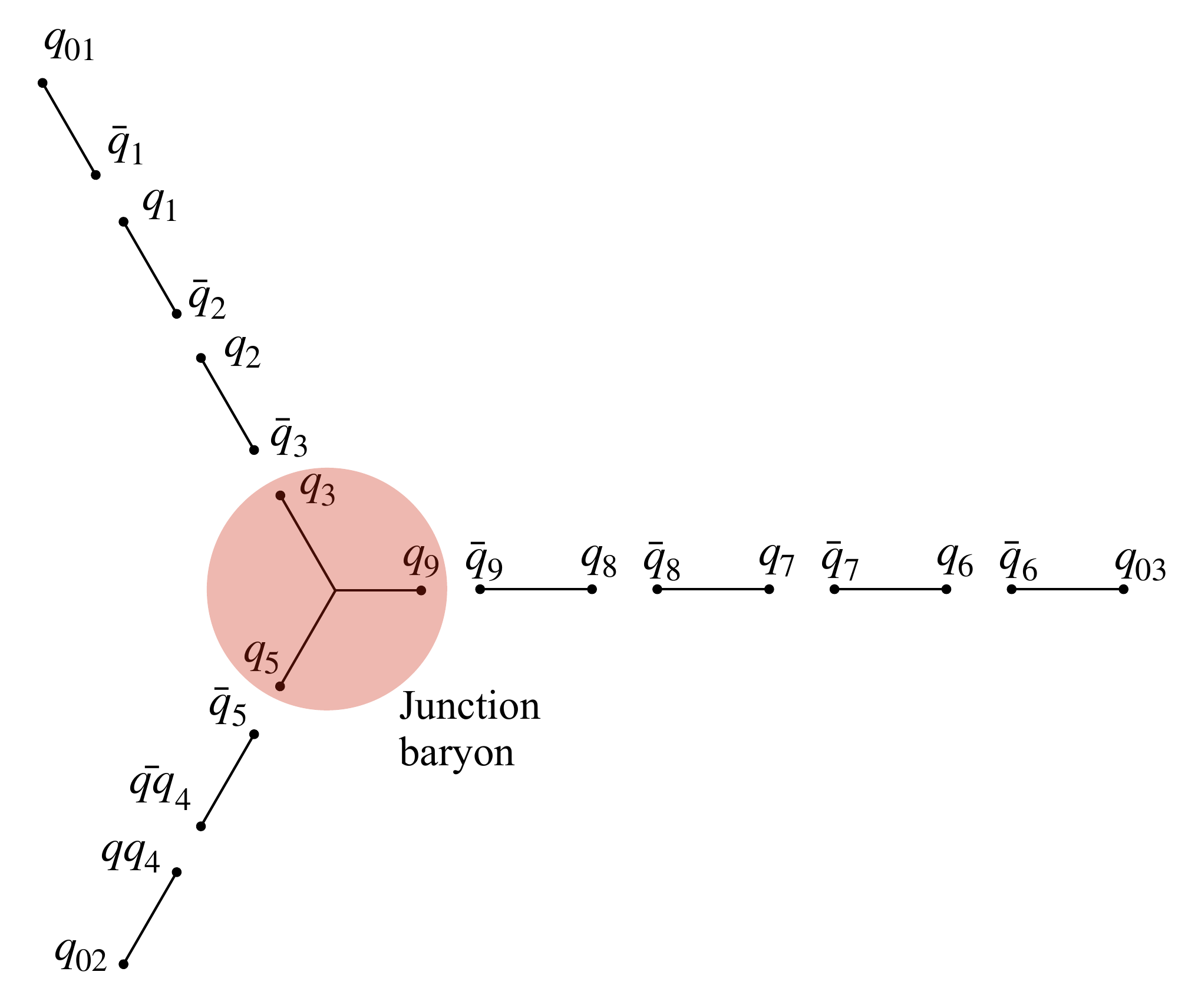}
    \end{center}
    \caption{A sketch of a junction system after fragmentation, with the junction baryon consisting of quarks $q_3 q_5 q_9$.}
    \label{juncFrag}
\end{figure}

 Beyond LC effects are particularly important for heavy-flavour baryon production due to the formation of junction topologies. Junctions provide a new mechanism for baryon production via fragmentation whereby the baryon number of the junction system is preserved. The hadronization of these topologies is depicted in Fig.~\ref{juncFrag}, where each junction ``leg" fragments in a similar manner to dipole string fragmentation. The quarks formed by breaks nearest to the junction itself form a baryon, which in the Fig.~\ref{juncFrag} example has quarks $q_3q_5q_9$ forming the ``junction baryon". This additional baryon production mechanism leads to baryon enhancement as shown in the left panel of Fig.~\ref{Barenh} for the \Lc/\Dzero yield ratio case. As heavy-flavour (charm and beauty) quarks cannot be created via string breaks, heavy-flavour hadrons from fragmentation necessarily come from string breaks nearest to the string endpoints. Therefore heavy-flavour baryons can either be created via diquark creation next to a heavy-flavour endpoint, or from junction baryons. For a junction baryon to be heavy-flavoured, the junction leg containing the heavy-quark endpoint cannot fragment. The effects of this can be observed by examining the \pt dependence of heavy-flavour baryons. 

 In the standard fragmentation framework, diquark-antidiquark pair creation has some fixed probability of occurring with no \pt dependence, making it insufficient to describe the measured increase of heavy-baryon production at low \pt. However, junction baryons are expected to predominantly sit at low \pt due to the way dipole strings reconnect into junctions. Junction reconnections mostly occur between dipoles in different jets, and given the large phase space available, these jets are likely to have large opening angles. As the junction forms between these generally large opening-angled jets, the junction (and therefore the junction baryon) is expected to sit at low \pt, making them particularly interesting for \pt studies of heavy-flavour baryons.

 Junction topologies are also expected to affect heavy-flavour asymmetries. This is due to how junction reconnections map three dipole configurations onto two junctions, as seen in Fig.~\ref{fig:juncCR}, resulting in both junctions and antijunctions forming in equal quantities. This would therefore contribute towards both heavy-flavour baryons and antibaryons, reducing the baryon asymmetry (particularly at low \pt) and in turn also altering the meson asymmetries in the heavy-flavour sector.

 In \ee{} collisions, modifications to the leading-colour picture are small, suppressed by both colour and kinematics factors. However, in dense partonic environments such as pp collisions, MPI increase the number of possible subleading connections, counteracting their naive $1/N^2_C$ suppression. This model represents a significant step in the right direction of probing for the first time the effects of sub-leading colour on hadronization in a systematic way.

 To understand baryons produced via diquark-antidiquark pair creation, it is important to look at how diquarks are treated in PYTHIA. Using standard string fragmentation, these diquarks are given a constituent mass and treated on equal footing as quark-antiquark string breaks.  However there is an alternative approach, called the popcorn mechanism~\cite{Andersson:1984af, Eden:1996xi}, which models diquark creation as a stepwise process using virtual $\qqbar$ fluctuations of colour charges on the string. For example, given a $\qqbar$ string with colours red and antired, then two virtual fluctuations with colours green-antigreen and blue-antiblue can potentially combine to form a real red-antired diquark-antidiquark pair. Furthermore, again using the colour structure of the virtual fluctuations, mesons can therefore be created between such diquark-antidiquark pairs. Thus unlike the standard string fragmentation mechanism where a diquark-antidiquark pair must be created next to one another, the popcorn mechanism allows for the decorrelation of the baryons produced. 

 These virtual fluctuations can also in principle colour reconnect with a diquark endpoint, meaning the constituent quarks forming the diquark will contribute towards two separate hadrons. Given a virtual $\qqbar$ fluctuation with the correct colour structure, one quark from the initial diquark could form a meson with the fluctuation $\qbar$, and the other would thus recombine into a diquark with the fluctuation $\q$. This means that the first hadron produced from a diquark string end may not be a baryon, and therefore a heavy-flavoured diquark does not necessarily guarantee a heavy-flavoured baryon. 

 Another feature of the popcorn model is the suppression of leading-baryon production, and is controlled in PYTHIA by parameters \texttt{StringFlav:suppressLeadingB}, \texttt{StringFlav:\newline heavyLeadingB} and \texttt{StringFlav:lightLeadingB}. Though strictly speaking this should be a suppression of early production time baryons, a simpler version is implemented where the baryon production at the end of the string (otherwise known as a rank 1 baryon) is suppressed. This is a reasonable simplification as the rank 1 baryons are associated with lower average $\Gamma$ values compared to higher ranks (breaks in the centre of the string), where  $\Gamma$ is a variable associated with the proper time of the string break, $\Gamma = \kappa^2\tau^2$. However, $\langle\Gamma\rangle$ of heavy-flavour rank 1 baryons is approximately the same size as the limiting value of $\langle\Gamma\rangle$ for light flavours, meaning that no suppression is expected for leading heavy-flavour baryons relative to string breaks in central regions of a string. As heavy-flavour baryons are expected to only come from fragmentation near string ends, the suppression of endpoint baryon production needs careful handling.

 To complete a comprehensive study on heavy-flavour hadron production in PYTHIA, one must also consider models that alter the light-flavour quark sector, particularly strange quark production rates. Strangeness enhancement has been observed at ALICE~\cite{ALICE:2017jyt, Acharya:2019kyh}, which shows a rise in strange-hadron production with respect to charged multiplicity that neither the default tune nor QCD-based CR in PYTHIA can describe. To describe the multiplicity dependence seen at ALICE, additional collective effects are required on top of the standard fragmentation framework. The two primary models proposed to describe this strangeness enhancement are rope hadronization~\cite{Bierlich:2016faw, Bierlich:2017sxk} and close-packing~\cite{Fischer:2016zzs}. Both models are readily available for use in PYTHIA 8.311, however the close-packing implementation is not catered to strangeness in jets or \ee collisions as of yet. 
 Such models have shown to successfully result in increased strange-hadron production. Both models work by increasing the string tension in more densely packed string environments, which are correlated with high multiplicity events. An increased string tension translates to a greater energy density per unit length of the string, allowing for higher mass particles to be created via string breaks with a greater probability, enhancing both strange quark and diquark production. The distinction between the two models lies in how the altered string tension is calculated. Rope hadronization treats a collection of strings as a coherent structure, where a larger collection of strings results in higher string tensions, and the order of string breaks becomes important. Close-packing proposes a simpler mechanism, whereby the number of surrounding strings determines the rise in string tension and directly changes the string tension. 

 Another avenue for exploration is alternatives to the Schwinger mechanism, with one such model already implemented in PYTHIA being the thermodynamical string breaking model~\cite{Fischer:2016zzs}. This alternative model uses a thermal exponential instead of the Gaussian form provided by the Schwinger mechanism to model pair creation due to string breaks. The ``temperature" dependence, $T$, in the exponent is directly related to the string tension $\kappa$ via $T \propto \sqrt{\kappa}$, and particularly alter the \pt spectra of string breaks. 

\subsubsection{Colour reconnections in HERWIG}
\label{sec:clusterCR}

\begin{figure}[ht!]
    \begin{center}
    \begin{subfigure}{0.45\textwidth} 
        \includegraphics[width = \textwidth]{ 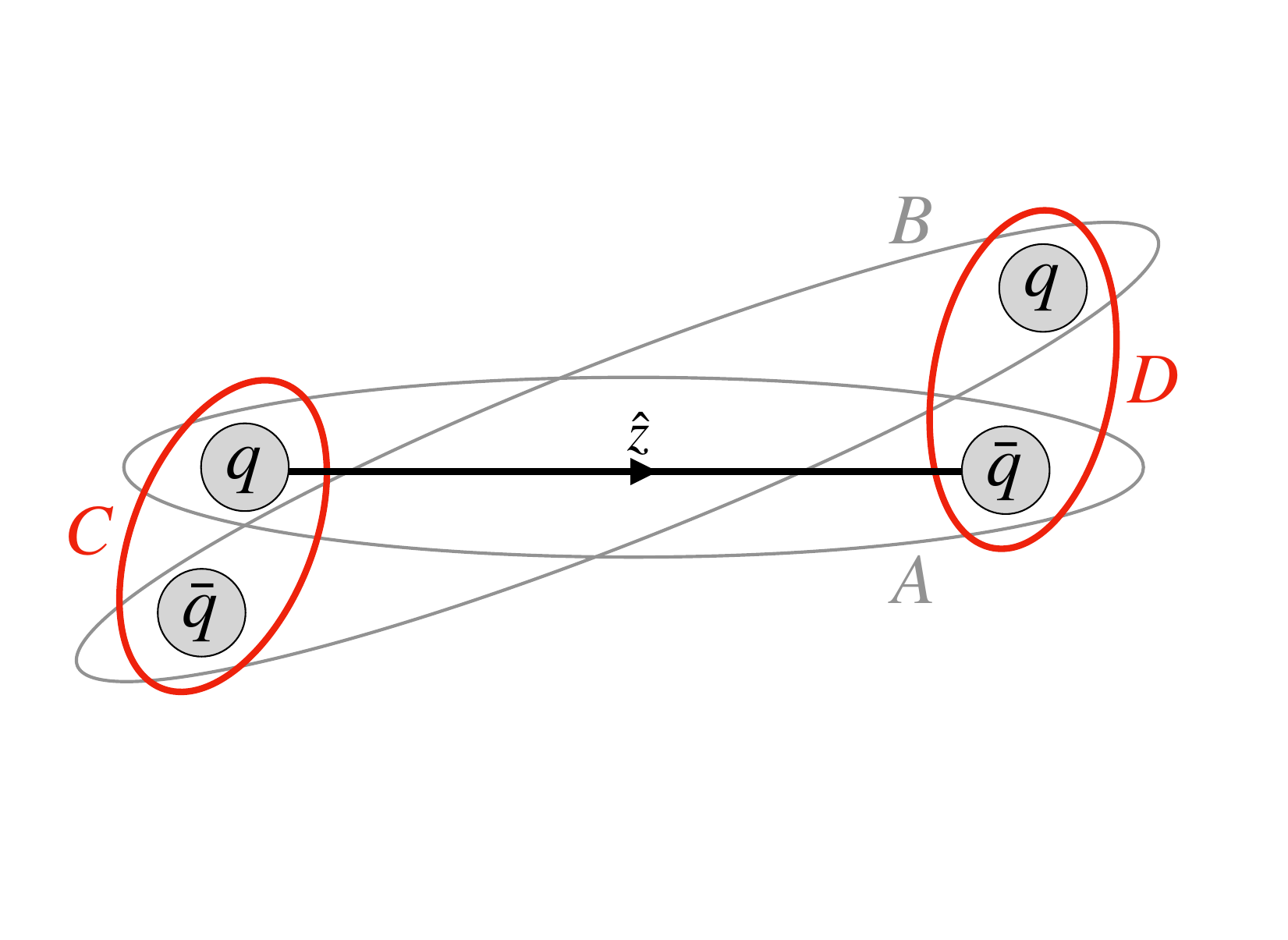}
        \caption{Mesonic reconnection.}
        \label{fig:clusterMeson}
    \end{subfigure}
    \begin{subfigure}{0.45\textwidth}
        \includegraphics[width = \textwidth]{ 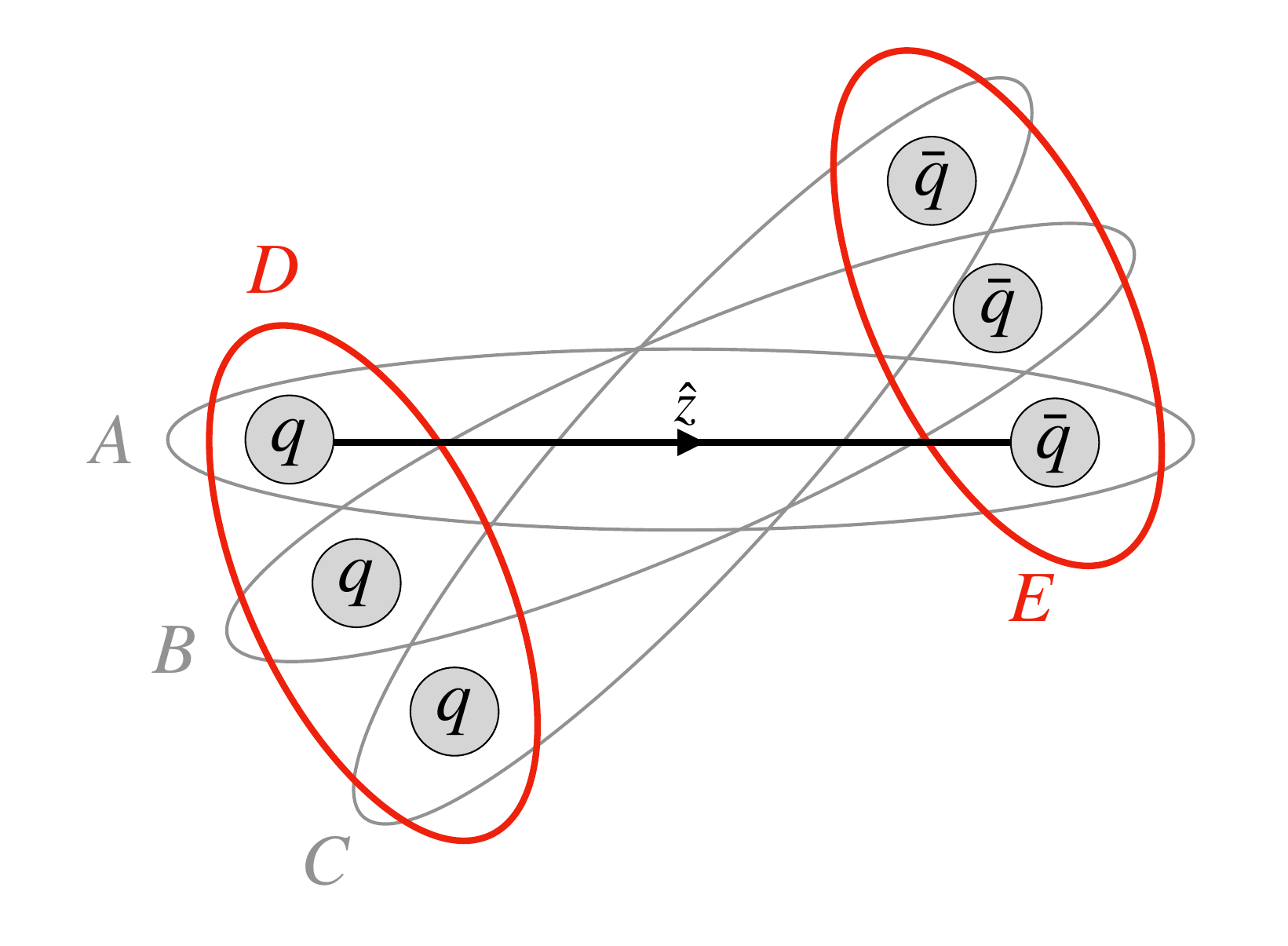}
        \caption{Baryonic reconnection.}
        \label{fig:clusterBaryon}
    \end{subfigure}
    \end{center}
    \caption{The grey clusters represent the initial cluster, and given colour reconnections the red represents the possible alternative cluster formations that are chosen with acceptance probability (a) $P_M$ or (b) $P_B$.}
    \label{clusterCR}
\end{figure}

HERWIG also allows for colour reconnections~\cite{Gieseke:2017clv} in their hadronization model, allowing for both mesonic ($\qqbar$) and baryonic ($\q\q\q$ and $\qbar\qbar\qbar$) cluster reconnections. In order to determine whether to reconnect clusters, the closeness in phase space using a rapidity measure is considered along with a probability to accept the reconnection. The procedure works by first selecting a cluster (call this cluster A) and boosting to its centre-of-mass frame. The centre-of-mass frame momenta then defines a z-axis, with the positive z-direction defined by the direction of the antiquark. In this frame, the rapidity of the remaining clusters are calculated and are used as a measure of closeness between which partons these clusters should form. For a given cluster (call this cluster B), the possible reconnection types available are determined by the rapidities of the $\qqbar$ pair; $y(\q_B) > 0 > y(\qbar_B)$ allows for mesonic reconnections, $y(\qbar_B) > 0 > y(\q_B)$ allows for baryonic reconnections, and if neither conditions are satisfied then no reconnection type is considered. For a mesonic reconnection, the configuration with the maximal $y_{sum} = |y(\q_B)| + |y(\qbar_B)|$ is considered and the reconnection is accepted with a probability of $P_M$. This is demonstrated in Fig.~\ref{fig:clusterMeson} where initial clusters A and B reconnect to clusters C and D. For baryonic reconnections, the two largest $y_{sum}$ values are considered and the reconnection of clusters A, B and C to form clusters D and E (as seen in Fig.~\ref{fig:clusterBaryon}) is accepted with probability $P_B$. This process is then repeated for all other clusters. Similarly to string junctions in PYTHIA, baryonic clusters contribute to an additional baryon production mechanism to the LC hadronization model as the baryon number of the cluster is preserved.

\subsubsection{Statistical hadronization with additional baryon resonances}
\label{shm}

The Statistical Hadronization Model (SHM) assumes that hadrons are produced according to the density of thermal yields $N_h^{th}(T)$. In ultra-relativistic heavy-ion collisions, the SHM successfully reproduces the yields of meson and baryon ground states with one chemical freeze-out temperature of $T_{cf} \simeq 156$ MeV~\cite{Andronic:2017pug}, which represents the temperature of the medium at which the different particle abundances get fixed. The other model parameters are the baryo-chemical potential $\mu_B$ and the volume of the systems, which are found to be $\mu_B$ = 0.7 $\pm$ 3.8 MeV, and
V = 5280 $\pm$ 410 $\rm fm^3$.
Its extension to charm (and beauty) hadrons needs to consider that the charm quarks are produced in the initial stage of the collisions with a multiplicity that is quite larger than the one expected from thermal charm production at $T_{cf}$. To this purpose, a charm-quark fugacity factor, $g_{\rm c}$, is introduced in the model to account for the excess of the initial charm yield wrt to the thermal one. 
A charm balance equation is needed to guarantee the conservation of charm quarks for a given value of $g_c$, while allowing the relative abundances of the different open and hidden charm hadrons to follow the statistical thermal ones, it can be written as:

\begin{equation}
    \frac{N_{\rm c\bar c}}{V}=\frac{1}{2} g_{\rm c}\sum_{h_{oc,1}^i} n^{{\rm th}}_i \,
  + \, \frac{1}{2} g_{\rm c}^2  \sum_{h_{oc,2}^j} n^{{\rm th}}_j,  
  + \, g_{\rm c}^2 \sum_{h_{hc}^k} n^{{\rm th}}_k, \, 
  \label{Eq:SHM}
\end{equation}

where $N_{\rm c\bar c}$ in the total charm yield in the rapidity range considered, $n_{i,j,k}^{{\rm th}}$ are the (grand-canonical) thermal yields
for open charm with one charm quark ($h_{oc,1}$) 
and two charm quarks ($h_{oc,2}$) 
and for hidden charm ($h_{hc}$). 
In principle, Eq.~\ref{Eq:SHM} should include the hadrons with three charm quarks but, in the current implementations, they are discarded under the reasonable assumption that their contribution to the sum is negligible.
Also, from Eq. (\ref{Eq:SHM}) we can see that the fugacity $g_c$ can be seen as proportional to the ratio of total charm yield to the thermal charm one, in the limit where multicharm and hidden charm hadron yields are negligible. 
Fugacity factors in elementary collisions are introduced to account for the non-equilibrium yield of quarks heavier than $u$ and $d$, thus for strangeness as well as for charm. However, due to the dominance of hadrons with one charm quark this does not affect baryon-to-meson yield ratios of single charm hadrons, because both baryons and mesons contain one charm. On the other hand, one can also notice that doubling $N_{\rm c \bar c}$ implies doubling $g_{\rm c}$, with a consequent enhancement by a factor of 2 of single-charm hadrons and a factor of 4 increase of double-charm hadrons. Such a scaling with $N_{\rm c\bar c}$ is expected also in the coalescence model described in the next paragraph.

Such an approach, which correctly predicts the relative abundances of D-meson states, has been challenged by the large $\Lc/\Dzero$ values measured in pp, p--Pb, and Pb--Pb collisions, which will be discussed  in the Sect.~\ref{sect:exp}.
In particular, 
using the resonance states listed in PDG~\cite{Workman:2022ynf} the charm baryon yields are substantially underpredicted by the SHM.  
However, it is argued in Refs.~\cite{He:2019tik,He:2022tod} that the charm and beauty baryon enhancement is to be attributed to the enhanced feed-down from “missing” charm- and beauty-baryon states in the PDG with respect to the predictions of the relativistic quark model (RQM)~\cite{Ebert:2011kk} and lattice QCD calculations~\cite{Padmanath:2014lvr}.
Including the additional baryon resonances makes a decisive difference and enables a significant heavy-flavour baryon enhancement. The additional baryon states almost double the fraction of the ground-state \Lc in the system relative to the PDG scenario when a hadronization temperature of 170 MeV is used.  
The SHM is usually employed to evaluate the total hadron yield. However, in the heavy-flavour sector for pp collisions, it has been coupled to the concept of independent fragmentation functions to compute pertinent \pt spectra by the TAMU group~\cite{He:2019tik}. In this approach, a thermal distribution function is exploited to determine the fragmentation fractions, but when comparing to experimental data the baryons and mesons are non-thermally distributed in \pt but according to the fragmentation functions. The model employs a modern fragmentation function formalism developed for heavy quarks that depend on one parameter $r$ defined as follows for each hadron species. For $\Dzero$ and $\Lc$ the related parameters ($r_{\Dzero}$, $r_{\Lc}$) are tuned to reproduce the measured \pt spectra. For higher meson (M) and baryon (B) resonances, the related parameters are obtained by rescaling those of $\Dzero$ and $\Lc$ according to their mass, $r_{\mathrm M[B]}/r_{\Dzero[\Lc]}=((m_{\mathrm{M[B]}}-m_{\rm c})/m_{\mathrm{M[B]}})/((m_{\Dzero[\Lc]}-m_{\rm c})/m_{\Dzero[\Lc]})$.
The results of this model calculation are reported in the middle panel of Fig.~\ref{Barenh} for the \Lc/\Dzero yield ratio: a significant enhancement at low \pt is observed. 
The uncertainty band in the model is obtained by varying the assumptions on the branching ratios of excited charm-baryon state decays to the ground state.
In heavy-ion collisions, the TAMU model employs a recombination approach that inherits the SHM yields, similarly to the pp-collision case, but evaluates the $\pt$ distribution functions according to a recombination model based on the Boltzmann collision integral (see next section).

More recently, also the standard SHM approach, originally set up to predict the \pt-integrated yields of the different hadrons has been extended, when applied to the charm sector, to supply a first tentative  modelling of the \pt distribution in AA collisions~\cite{Andronic:2021erx,Andronic:2023tui}. A two-component model containing a thermalized distribution with a radial flow plus a contamination from a \enquote{corona shell}, to which a \pt distribution like that of pp collisions is associated, is implemented. The method shows a qualitative agreement with the data mainly in the low \pt region\
($\pt \lesssim 2 \, {\rm GeV}/c$), while already at intermediate \pt there is a significant lack of yield that cannot be compensated only by the corona contribution. This is in line with the general finding of all dynamical approaches to charm in-medium interaction, namely that charm gets close to equilibrium in the low \pt region~\cite{Capellino:2023cxe, Capellino:2022nvf}, but it might retain a long-$p_T$-tail of non-equilibrium.

\subsubsection{Hadronization via coalescence}
\label{sec:theoryCoalescence}

The idea of quark coalescence in heavy-ion collisions dates back to about two decades ago and was applied to account for the first observation of an enhancement in the $p/\pi$ ratio at intermediate $\pt$~\cite{Greco:2003mm,Greco:2003xt,Fries:2003kq,Fries:2003vb}.
Contrary to the fragmentation process, where the final heavy-flavour hadron always carries smaller energy than the parent heavy quark, the idea of the coalescence model comes from the fact that partons in a dense medium (even if not necessarily thermalized) combine their transverse momentum to produce a final-state meson or baryon with higher transverse momentum. Thus, with coalescence the average hadron \pt is pushed to larger values with respect to fragmentation. 


In heavy-ion collisions it has been possible to associate to hadronization by coalescence signatures going beyond the particle-yield ratios, predicting a quark-number scaling of the azimuthal anisotropy of the production that is quantified by the elliptic flow~\cite{Greco:2003mm,Molnar:2003ff,Fries:2008hs}. 
According to this the $v_2(\pt)$ should scale into a universal curve given by $1/n_q * v_2(\pt/n_q)$ with $n_q=2$ for meson and $n_q=3$ for baryons.
A moderate breaking of such scaling is expected mainly at low \pt in realistic coalescence models due to the finite width effect of the hadron wave function, feed-down by hadron resonance decays, and radial flow correlations~\cite{Fries:2008hs,Greco:2004ex}. In addition, it has been shown that off-shell dynamics with the constraint of energy conservation drives the $\pt$ scaling toward an $m_{\rm T}$ scaling~\cite{Ravagli:2007xx} at low \pt. Furthermore, at intermediate \pt the interplay with the fragmentation function induces a further explicit significant breaking of the scaling~\cite{Minissale:2015zwa}.
Data at the Relativistic Heavy Ion Collider (RHIC) and the LHC confirmed that at intermediate $\pt$ (around the peak value of $v_2(\pt)$) the elliptic flow follows two different branches depending on their
valence quark number such that $v_{2M}(\pt/2)/2 \simeq v_{2B}(\pt/3)/3$ for light quark hadrons.

The coalescence approach has been extended to the charm sector to study the production of D mesons~\cite{Greco:2003vf}. There is a quite general consensus that including this process is necessary to correctly reproduce the main observables like $\Raa(\pt)$ and $v_2(\pt)$~\cite{Das:2015ana, Scardina:2017ipo}. It has to be noticed that being the mass of the light and heavy-flavour quarks quite different, one expects that a D meson at a certain \pt is formed from the recombination of a light quark quark of transverse momentum $p_{q}\simeq \frac{m_{\rm q}}{m_{\rm q}+m_{\rm c}} \pt$ with a charm quark with $p_c\simeq \frac{m_{\rm c}}{m_{\rm q}+m_{\rm c}} \pt$~\cite{Lin:2003jy}. Therefore, having usually $m_{\rm c} \simeq 4 m_{\rm q}$, this implies that on average the D meson gets about 0.2 $\pt$ from the light quark and 0.8 $\pt$ from the charm quark. Hence, one envisages that $v_{2, \rm D}(\pt)\simeq v_{2,{\rm q}}(0.2 \,\pt)+ v_{2,\rm c}(0.8 \, \pt)$. This simple argument must be complemented considering the finite width of the wave function and the slope in $\pt$ of quark distribution functions as well as the contribution from the fragmentation, which is non-negligible at intermediate $\pt$.

The coalescence plus independent fragmentation approach predicted already in 2009 that in AA collisions
one could expect a quite large $\Lc$  production and $\Lc/\Dzero \sim O(1)$~\cite{Oh:2009zj}, a value about one order of magnitude larger than that measured in \ee collisions in which the \Lc fragmentation fraction was estimated to be $f(c \rightarrow \Lc) \simeq 0.06$~\cite{Lisovyi:2015uqa}. The first measurements made by the STAR Collaboration and ALICE Collaborations in Au--Au collisions at \snn = 200 GeV and Pb--Pb collisions at \snn = 5.02\TeV, respectively, qualitatively confirmed such a prediction~\cite{STAR:2019ank,ALICE:2018hbc}. Successive measurements by ALICE and CMS allowed to better profile the evolution of $\Lc/\Dzero$ ratio with \pt and centrality~\cite{ALICE:2021bib,CMS:2023frs}.

\textbf{Catania model:} This model has been applied extensively to particle production not only in the light sector but also to charm baryon and meson production in both AA collisions~\cite{Greco:2003vf, Scardina:2017ipo, Plumari:2017ntm} and pp collisions
~\cite{Minissale:2020bif}. The model has been recently expanded to provide first predictions for multi-charm baryons~\cite{Minissale:2023bno} that are expected to be measured in heavy-ion collisions at the LHC with the future ALICE 3 experiment~\cite{ALICE:2022wwr}.
The Catania approach implements hadronization of charm quarks via both coalescence and independent fragmentation and assumes the presence of a flowing thermalized medium of light quarks (u, d, s).
In pp collisions the general expectation was that a QGP would not be created. However, experimental measurements provide strong indications for the formation of an expanding medium in small collision systems~\cite{ALICE:2017jyt, CMS:2015fgy, ALICE:2020wfu, ALICE:2020wla}.
In addition, viscous hydrodynamics and transport calculations supply a reasonable description of experimental measurements~\cite{Weller:2017tsr, Greif:2017bnr, Sun:2019gxg}. Recently, an approximate quark-number scaling of hadron elliptic flows was also shown for small collision systems~\cite{Zhao:2020wcd}. 
This motivated the extension of the Catania model assuming a small fireball of expanding QGP even in pp collisions.

In the model, a blast wave parameterization~\cite{Retiere:2003kf} for light quarks at the hadronization hypersurface with the inclusion of a contribution from mini-jets is considered. For charm quarks, the spectra from FONLL calculations are used in pp collisions, while in AA collisions the spectra are taken after collisional energy loss in the medium evaluated through a Quasi Particle Model (QPM) implemented in a Boltzmann transport dynamics or in a Langevin equation through the corresponding drag and diffusion coefficients ~\cite{Scardina:2017ipo,Sambataro:2022sns,Sambataro:2023tlv}. 
The coalescence approach is based on a phase space description of the system created in AA collisions. The basic formula is written in terms of the Wigner function describing the spatial and momentum distribution of quarks in a hadron. In particular, the momentum spectrum of hadrons formed by coalescence can be written as:

\begin{eqnarray}
\label{eq-coal}
\frac{dN_{H}}{dyd^{2}P_{T}}=g_{H} \int \prod^{N_{q}}_{i=1} \frac{d^{3}p_{i}}{(2\pi)^{3}E_{i}} p_{i} \cdot d\sigma_{i}  \; f_{q_i}(x_{i}, p_{i}) \, {\cal{C}}_H(x_{1}...x_{N_{q}}, p_{1}...p_{N_{q}})\, \delta^{(2)} \left(P_{T}-\sum^{n}_{i=1} p_{T,i} \right),
\end{eqnarray}

with $g_{H}$ indicating the statistical factor
giving the probability that two (or three) random quarks have the right quantum numbers to match the quantum number of the hadron (flavour, spin, isospin) in a colourless combination. For D mesons the statistical factor $g_{\rm D}=1/36$ gives the probability that two random quarks have the right colour, spin, and isospin to match the quantum number of the considered mesons. For $\Lc$ the statistical factor is $g_{\Lc}=1/108$.
$N_{\rm q}$ is the number of quarks that form the hadron, $N_{\rm q}=2$ for mesons and $N_{\rm q}=3$ for baryons.
The $d\sigma_{i}$ denotes an element of a space-like hypersurface, 
while $f_{{\rm q}_i}$ are the quark (anti-quark) phase-space distribution functions for $i^{\mathrm{th}}$ quark (anti-quark) and the coalescence function ${\cal{C}}_H= C^{N_{\rm q}-1} f_{H}(x_{1}...x_{N_{{\rm q}}}, p_{1}...p_{N_{{\rm q}}})$ is directly proportional to the Wigner function which describes the spatial and momentum distribution of quarks in a hadron, $f_H$. The constant $C^{N_{\rm q}-1}$ is an important and specific aspect of the Catania model, not present in the standard coalescence approach. 
It is this constant 
that allows normalizing the coalescence probability for charm quarks such that the total coalescence probability (i.e. summing up all the mesons and baryons channel considered) in the limit $\pt \rightarrow 0$ is $P_{coal}(\pt \rightarrow 0)= 1$. Such a constant $C$ can be determined only by calculating all the charmed mesons and baryons. In the first work
\cite{Plumari:2017ntm} it was evaluated considering only D, \Dstar, \Ds, \Lc, and \Sigmac hadrons. However, more recently the production of $\XicPlusZero$ and $\Omegac$ was observed not to be negligible~\cite{Minissale:2020bif}.
Having fixed such a normalization, the probability of a charm quark to hadronize via coalescence
or fragmentation depends as in the standard approach on its transverse momentum according to Eq.~\ref{eq-coal}. The coalescence probability
is high for low quark momenta and quickly decreases with increasing transverse momentum, meaning that most of charm quarks hadronize via coalescence at low \pt whereas, in the high momentum region, fragmentation is the dominant contribution. 
Imposing the condition $P_{coal}(p \rightarrow 0 )= 1$ is a key aspect underlying the prediction of quite large values of heavy baryons:  $\Lc/\Dzero \sim O(1)$, $\XicZero/\Dzero \sim 0.2$ and $\Omegac/\Dzero \sim 0.1$~\cite{Minissale:2020bif}.
The physical motivation is related to the fact that in the standard coalescence approach there is no dynamics related to confinement.
The renormalization constant $C$ appears linearly for mesons and quadratic for baryons. This is physically motivated at least by two considerations. First, enforcing 
$P_{coal}(\pt \rightarrow 0 )= 1$ implies that 
there is an enhancement of the recombining probability (with respect to the one from a simple phase space overlap of the wave functions) that can be associated to the dynamics of the confining interaction that increases the coalescence probability (here accounted for just by a renormalization of a coupling constant). Such an effect has to be squared for baryons for which two relative distances among the quarks have to be defined.
Second, one has to consider that the density of the medium at which the hadrons are formed is not known, and being the phase transition a cross-over it is likely that there is a density interval over which it occurs.
Therefore the hadrons yields and ratios in a standard coalescence approach 
(i.e. without imposing $P_{coal}(\pt \rightarrow 0 )= 1$) would significantly depend on the specific value of the density chosen. In the Catania approach the probability of forming a hadron scales linearly with the density for mesons and quadratically for baryons. If one chooses a lower density value, implying a reduced standard coalescence probability, the constraint $P_{coal}(\pt \rightarrow 0 )= 1$ increases the value of the
coefficient $C^{N_q-1}$, making the yields and the baryon-over-meson ratios quite stable and nearly independent of the specific value of the density of the background fireball employed to perform the calculation.

Another relevant aspect of the Catania approach is the width of the Wigner function for the various mesons and baryons. These are related to the root mean square charge radius of the hadron taken from the quark model~\cite{Hwang:2001th,Albertus:2003sx}. A Gaussian wave function is assumed for simplicity and it implies that also the associated hadron Wigner function is Gaussian:

\begin{equation}
 f_H(x_{ri},p_{ri})=\prod^{N_{q}-1}_{i=1} 8 \exp{\Big(-\frac{x_{ri}^2}{\sigma_{ri}^2} - p_{ri}^2 \sigma_{ri}^2\Big),}
 \label{wigner}
\end{equation}

with $x_{ri}$ and $p_{ri}$ relative coordinates in space and momentum.
We notice that an important aspect of a proper implementation of the coalescence approach is that, respecting the basic feature of quantum mechanics, the width $\sigma_r$ of the wave function in r-space constrains also the Wigner function width in p-space; for a Gaussian wave function for each relative coordinate we have $\sigma_{ri} \cdot \sigma_{pi} =1$ and 
$\sigma_p$ cannot be used as a further independent parameter.
It can be noticed that the coalescence process, and the related formula, can be factorized as a diquark recombination followed by a diquark-quark coalescence if the diquark binding energy is vanishing.

\begin{figure}[ht!]
    \begin{center}
    \includegraphics[width = 1\textwidth]{ 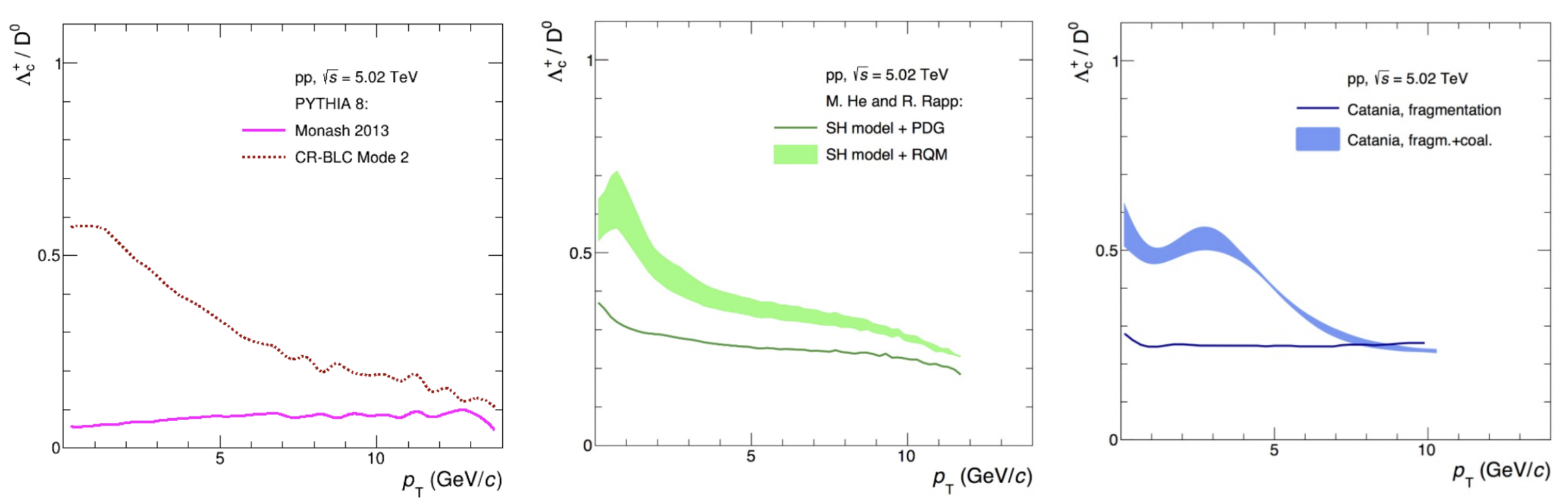}
   \end{center}
    \caption{\LcD ratio obtained by three model calculations in pp collisions at \s~=~5.02~TeV. Left panel: PYTHIA 8  MC generator with (dashed line) and without (solid line) the implementation of junction topologies. Middle panel: SHM with (band) and without (line) the inclusion of the additional charm baryon resonance states predicted by the RQM model. Right panel: Catania model in which charm quark hadronization is implemented via fragmentation only (line) or via fragmentation and coalescence (band).}
    \label{Barenh}
\end{figure}

It is interesting to discuss the dependence of the coalescence results on the specific value of the wave-function width, which is the main parameter of the model.
This can be easily evaluated analytically considering a non-relativistic reduction, which, even if not quantitatively exact, allows having a semi-quantitative estimate of the main aspect involved in the coalescence process.
Assuming a relativistic boost invariant thermal distribution function homogeneous in $r$-space, considering the mid-rapidity region and employing a non-relativistic reduction for the quark energy $E=m+p^2/2m$, discarding the radial flow, and assuming a Gaussian wave function, it can be shown that the coalescence yield for mesons and baryons scales as:

\begin{eqnarray}
N_M \sim \frac{\sigma_r^2}{\sigma_r^2+(2\mu_{12} T)^{-1}} \,\,\,\, \text{and } \,\,\,\,
N_B \sim \frac{\sigma_{r1}^2}{\sigma_{r1}^2+(2\mu_{12} T)^{-1}} \frac{\sigma_{r2}^2}{\sigma_{r2}^2+(2\mu_{(12)3} T)^{-1}},
\label{non-rel-coal} 
\end{eqnarray}

where $\mu_{12}$ is the reduced mass of the two quarks undergoing coalescence and $T$ is the temperature of the medium. The numerator is associated with the probability of forming a meson in $r$-space, which of course increases with the width $\sigma_r$. 
From the denominator of Eq.~\ref{non-rel-coal} we can see that there is another scale that plays a role and is given by $2 \mu T$. This represents the width of the distribution of relative momentum of the coalescing objects. 
The increase of the width $\sigma_r$ leads to an increase of the yield that quantitatively depends on the comparison between the width of the hadron wave function $\sigma_r$ and the inverse of the width of the thermal distribution $1/2 \mu T$. For $\sigma_r^2 << (2 \mu T)^{-1}$ the yield is strongly dependent on the size of the hadron wave function, while for $\sigma_r^2 >> (2 \mu T)^{-1}$ the yield is nearly independent on the hadron wave function.
For the specific case of charm meson and baryons, $\sigma_r^2(\Dzero)= 0.65 \, \rm fm^2$ and $(2 \mu T)^{-1}= 0.5 \rm \, fm^2$, hence an increase of 
$<r^2>$ by a factor of two implies an increase of the yield of about 30\%. Smaller baryons like $\Omega_c$ have the relative widths 
$\sigma_{r1}^2(\Omega_c)= 0.34 \, \rm fm^2$ and $\sigma_{r2}^2(\Omega_c)=0.14 \, \rm fm^2$
and $(2 \mu_{ss}T)^{-1}= 0.6 \rm \, fm^2$ and $(2 \mu_{(ss)c}T)^{-1}= 0.6 \rm \, fm^2$
such that an increase of $<r^2>_{\Omegac}$ by a factor of 2 induces an increase of the yield of about a factor 2.2. Therefore, hadrons that have a large mean square radius and large mass are in general not very sensitive to the parameter $\sigma_r$, if the relation $\sigma_p =  1/\sigma_r$ is employed.
However, on top of this standard coalescence feature, one has to consider the constraint that $P_{coal}(\pt \rightarrow 0 )= 1$ strongly damp such dependence for the D and \Lc hadrons, which dominate the overall charm-hadron yield~\cite{Minissale:2023bno}. This is due to the role played by the C factor described above, while in other coalescence approaches where this factor is not present the effect of modifying the width of the Wigner function has a strong impact on the baryon-to-meson yield ratio, we discuss this point just below in the dedicated paragraph.
The Catania coalescence plus fragmentation model was first applied to AA collisions prediction of \Lc/\Dzero both at RHIC and LHC energy \cite{Plumari:2017ntm}, but soon after it led to a quite good prediction of the \Lc/\Dzero yield ratio in pp collisions, as shown in the right panel of Fig.~\ref{Barenh}. It also provided within the current uncertainties also a reasonable predictions 
for the $\XicZero$ and $\Sigmac$ production, see also Fig.s \ref{ratiotoDo} and \ref{FFfunc} in Sec.III.C.

\textbf{Other coalescence approaches:} In other approaches~\cite{Oh:2009zj, Cao:2019iqs}, applied to AA collisions and based on a formulation formally very similar to Eq.~\ref{eq-coal} and \ref{wigner}, the constraint to have $P_{coal}(\pt \rightarrow 0)= 1$ is also implemented, but it is achieved artificially increasing the width $\sigma_{ri}$ of hadrons by a constant factor, resulting in meson radii about a factor 1.8 larger than the one employed by the Catania model~\cite{Plumari:2017ntm} that relies on the quark model prediction in Ref.~\cite{Hwang:2001th, Albertus:2003sx}. 
The increase of the yield due to an increase of the width $\sigma_{ri}$ is squared for baryons because the relative coordinates of the quarks enter twice for baryons. Therefore, the increase of $\sigma_{ri}$ induces a strong increase of the $\Lc/\mathrm{D}$ ratio. In particular, it has been seen that modifying $\sigma_{ri}$ to satisfy the constraint $P_{coal}(\pt \rightarrow 0)= 1$ results in a $\Lc/\mathrm{D}$ close to the experimental measurements. This is on the one hand similar to the Catania model, but on the other hand the effect of modifying the width of the Wigner function is very different from the Catania model where $P_{coal}(\pt \rightarrow 0)= 1$ is enforced independently on the $\sigma_{ri}$, and the yields of D and $\Lc$ are  marginally affected by a change of the widths by some common factor, because, anyway, due to the C-factor, all low momentum charm hadronize by coalescence, as discussed at the end of the previous paragraph. 
We also notice that increasing $\sigma_{ri}$ as in Refs.~\cite{Oh:2009zj, Cao:2019iqs} also affects the width in momentum space and this should modify the \pt dependence of the spectra of baryons and mesons. A study of such an effect, and a direct comparison to the Catania model, would be instructive but has not yet been performed. 
Ref. \cite{Cao:2019iqs} presents a refinement of the coalescence approach considering the different wave function for $l=1$ states, which leads to a non-Gaussian Wigner function~\cite{Cao:2019iqs}, differently from that of the Catania model that currently has the same Gaussina wave function for all the states. A similar refinement is done also in the hadronization mechanism implemented in the Parton Hadron String Dynamics (PHSD)~\cite{Cassing:2008sv}.

The PHSD model provides a realistic simulation of the entire heavy-ion collision and it also implements a coalescence approach to hadronization of heavy quarks~\cite{Cassing:2008sv}. The quark masses are distributed according to the spectral function of a dynamical quasi-particle model (DQPM)~\cite{Song:2015sfa}, very similar to the QPM used in Catania to describe the heavy-quark dynamics. Then, as in the Catania model, the probability of coalescence is evaluated by a superposition of the quark distribution with the Wigner hadron wave function, but considering both S- and P-wave states. Differently from the Catania model, it does not implement the constraint $P_{coal}(\pt \rightarrow 0)= 1$ and the coalescence process is not instantaneous but occurs during the fireball expansion when the local energy density spans the interval $0.3< \epsilon <0.9 \, \rm GeV/fm^3$. Hadronization for heavy flavour within the PHSD approach has been extensively applied to study charm dynamics in AA collisions; only very recently an extension has been done to pp collision studying the bottomonium production \cite{Song:2023zma}.

Finally, a more simplified approach to hadronization by coalescence is the Quark re-Combination Model (QCM) which, though based on the idea of quark recombination, has several features that are different from the other approaches discussed above. The model is developed in one dimension representing the transverse momentum. The coordinate space is not considered. 
The model obtains the light- and heavy-quark distributions through a fit to experimental data of $\pi$, K, and D mesons~\cite{Li:2017zuj, Song:2018tpv}. Furthermore, hadronization is occurring by coalescence at all momenta, without any contribution from fragmentation. Once the u, d, s, and c quark distribution functions are extracted from the fit, they are used as a source to compute the spectra as a function \pt of all hadron species: p, $\Lambda$, $\phi$, $\Omega$, \Lc etc... 
The role of Wigner hadron function in the standard recombination approach is substituted by a recombination function that is given by a delta function, which imposes quarks with the same velocity to recombine. The ratio between vector and pseudoscalar meson yields is fixed by a parameter $R_{V/P}=3/2$ indicated as thermal weight because it corresponds to the value of $\rm D^*/D \simeq 3\,(M_D/M_{D^*})^{3/2} exp[-(M_D-M_{D^*})/{\it T_H}]\simeq 3/2= {\it R_{V/P}}$, if one assumes a thermal production at a temperature $\rm {\it T_H}=170-175 \, MeV$, as done also in \cite{Rapp:2003wn} in the context D meson production in $\pi N$ collisions, see Sect. III.B. This is the same assumption exploited in the Catania coalescence model even if in this case the temperature, $\rm {\it T_H}=165 \, MeV$ is the one of the hadronizing medium that determines the light and strange quark thermal distribution, while in QCM the concept of temperature does not really play a direct role because the quark distribution functions are deduced from a fit to $\pi$, K, and D mesons experimental data.
A similar ratio parameter is used to fix the ratio between the baryons of the octet ($J^P=(1/2)^+$) and decuplet ($J^P=(3/2)^+$) by the parameter $R_{O/D}=2$.
 
\textbf{Resonance recombination model:} Another main approach to charm hadron formation is the Resonance Recombination Model (RRM) of TAMU group where hadronization is described by scattering of thermal light quarks with heavy quarks into broad D-meson like resonances~\cite{Ravagli:2007xx}. This model is also based on the idea that hadronization occurs by a recombination of quarks in the medium, but the process is modelled like a scattering of two particles into a hadron resonance with a finite width $\Gamma$. The meson phase space distribution is written in terms of the quark and anti-quark distribution as:

\begin{equation}
    \frac{dN_M}{d^3p\, d^3x}= \int \frac{d^3 \vec{p}_1 d^3 \vec{p}_2}{(2\pi^3)}
    f_q(\vec{x},\vec{p}_1) f_{\overline{q}}(\vec{x},\vec{p}_2) 
    \frac{\gamma_M}{\Gamma_M} \sigma_M(s) v_{rel}(\vec{p}_1,\vec{p}_2) \delta^3(\vec p - \vec{p}_1 -\vec{p}_2),
  \label{eq:RRM}  
\end{equation}

where $\sigma(s)$ is the relativistic resonance cross section (Breit-Wigner), $\gamma_M=E_M/M$ and $\Gamma_M$ is
ans assumed width for the formed D meson which is about 0.1 GeV for meson and 0.3 GeV for baryons. 
We can envisage from Eq.~\ref{eq:RRM} and Eq.~\ref{eq-coal} that RRM approach shares with the coalescence approach several features because the main underlying process consists in adding momenta of particles coming from the medium and the relative momenta of quarks determine the formation probability according to the values of $\sigma_M(s)\, v_{rel}$. 
On the other hand, the formation of the hadron is not directly related to the wave function width. 
However, the main features of modifying $R_{\rm AA}(\pt)$ from charm to D meson inducing an enhancement of D-meson $R_{\rm AA}$ at $\pt \gtrsim m_{\rm Q}$ accompanied by an enhancement of the elliptic flow are similar to those resulting from the coalescence approach.
The main difference with respect to coalescence is in the momentum dependence of the probability of hadron formation by recombination. Also in RRM there are two competing hadronization processes, with independent fragmentation applied to the charm quark that does not recombine according to Eq.~\ref{eq:RRM}. However, the recombination in RRM is dominant over fragmentation in a wider $\pt$ range with respect to coalescence~\cite{Zhao:2023nrz}. 
The RRM approach has been more recently applied also to the \Lc production, which is modelled as a two-step process: the formation of a diquark according to Eq.~\ref{eq:RRM} followed by a scattering process between the diquark and the charm quark. An important ingredient introduced in recent years in the modelling of coalescence in the TAMU model are space-momentum correlations between the recombining quarks. Given that recombination is more probable for quarks that are close in both momentum and space, the correlation between the heavy-quark momentum vector and their spatial position in the fireball, induced by the interplay of the heavy-quark interaction in the medium and the medium expansion, influences significantly the heavy-flavour hadron \Raa, \ellflow, and the relative particle-species abundances. 
Furthermore, being the hadronization modelled like a scattering process $2 \rightarrow 1$ going into an off-shell hadron it is possible to conserve the 4-momentum and it has been shown that under the assumption of local thermal equilibrium the D meson spectrum converges to the equilibrium thermal spectrum. 

We note that when going through coalescence objects are formed that have a total energy in the rest frame (an invariant mass) that is distributed within a finite width that is comparable to the object assumed to be formed in RRM. 
Hence, globally, one can say that the energy is conserved looking at the D mesons obtained as to an off-shell distribution of D-meson-like resonances with width of about 0.1-2 GeV. Furthermore, a difference remains w.r.t. RRM, where the exact thermal equilibrium is constrained, while in coalescence one has simply an invariant-mass distribution regulated by the relative quark momentum distribution in the medium and its interplay with the wave function width. This aspect has never been pointed out: a direct comparison of coalescence, RRM, and the diquark-quark in medium hadronization, recently developed by the Turin group within the Powlang model, discussed below, could be interesting.


\textbf{Powlang:} Another approach to in-medium hadronization also based on the idea of quark recombination is the one developed by Powlang~\cite{Alberico:2013bza,Beraudo:2022dpz,Beraudo:2023nlq}, which implements a Langevin evolution of charm quarks produced with a POHWEG+PYTHIA Monte Carlo event generator into a hydrodynamical background. In such an approach a charm quark is recombined with a thermal quark or di-quark on a hadronization hyper-surface identified by the local temperature $T_H$ = 155 MeV. Hence each (anti-)charm quark can randomly form pairs with quarks or di-quarks distributed according to a thermal spectrum corresponding to the
density:

\begin{eqnarray}
n=g_sg_I\frac{T_HM^2}{2\pi^2}\sum_{n=1}^\infty \frac{(\pm1)^{n+1}}{ n}K_2\left(\frac{nM}{ T_H} \right),
\end{eqnarray}

where the mass $M$ of quark and di-quarks are taken in the range of the constituent quark picture: $m_{\rm u/d}=0.33$ GeV, $m_{\rm s}=0.5$ GeV, $m_{(\rm ud)_0}=0.579$ GeV, $m_{(\rm sl)_0}=0.804$ GeV, $m_{(\rm ss)_1}=1.0936$ GeV; while
$g_s$ is the spin degeneracy and $g_I$ the isopin degeneracy.
The recombination occurs if the invariant mass of the pair, $M_{\cal C}$, is larger than the mass of the lightest charmed hadron. Then the cluster decays into a charmed hadron that has the same baryon number and strangeness of the cluster or, for large invariant masses, $M_{\cal C}> M_{max}\simeq 3.8 \,\rm GeV$, it hadronize via string-fragmentation according to PYTHIA 6.4. 
Therefore, differently from the coalescence approach, the probability of hadronizing via recombination is not determined by a hadron wave function with a finite width in momentum space (typically of the order of a few hundred MeV), but rather by the formation-probability of clusters of quite large invariant mass subject to decay. Instead, in a coalescence approach the invariant-mass distribution from the recombination of a light (strange) quark and a heavy quark can be quite narrower due to the constraint of the wave function. Another difference is the assumption of a thermal population of di-quarks that in the modelling acquire also a hydrodynamical radial and elliptic flows.
The assumption in the model of local colour neutralization leads to a strong space-momentum correlation, which provides a substantial
enhancement of the elliptic flow of the final-state charmed hadrons, affecting both their momentum and
their angular distributions. This hadronization scheme has been applied to AA collisions \cite{Beraudo:2022dpz} and very recently to pp collisions \cite{Beraudo:2023nlq} showing a behavior of \Lc/\Dzero similar to the coalescence approach, apart from the tendecy to over estimate the $\Ds$ production especially at low $p_T$.

\subsubsection{EPOS4(HQ)}

The EPOS Monte Carlo event generator~\cite{Pierog:2013ria,Werner:2023jps} is based on the assumption that hadronization in QCD is string-like below a certain critical energy density, while it exhibits thermal (QGP) behaviour above that density. The idea is that the critical density is not reached in \ee collisions nor in low-multiplicity pp collisions. These systems can therefore be described by a conventional Lund string fragmentation model (much like in PYTHIA, though EPOS has its dedicated implementation of string fragmentation). At higher energy densities, which are reached in high-multiplicity pp collisions and in heavy-ion collisions, EPOS invokes a QGP-like medium with non-zero temperature and hydrodynamic properties. 

An elegant interpolation between the two limits is achieved via the \emph{core-corona} picture~\cite{Werner:2007bf,Werner:2023jps}, in which dense (``core'') regions of phase space (above the critical density) are described via the QGP-like thermal modelling of particle production, while more dilute (``corona'') ones (below the critical density $\epsilon = 0.57\, \rm GeV/fm^3$) are described by string fragmentation. Crucially, these two types of regions can coexist \emph{within one and the same collision}, with a smooth transition between low core fractions at high impact parameters and high core fractions at central ones. In terms of observable consequences, this produces an overall modelling that reliably interpolates between vacuum-like string fragmentation in \ee and low-multiplicity pp collisions and a QGP limit in central AA collisions. 

Historically, the scattering modelling in EPOS was based on a zero-momentum limit of cut pomerons and hence did not aim at describing high-$p_\perp$ jet fragmentation (or other perturbative scattering processes). This has changed with EPOS4, which incorporates both factorization for hard processes and saturation for the low-$x$ scatterings.
More specifically related to the purposes of this review, the main aspect of EPOS4HQ is the inclusion of the scattering dynamics of heavy quarks in the ``corona'' medium even in pp collisions and of hadronization via coalescence plus fragmentation~\cite{Zhao:2023ucp,Zhao:2024ecc}. The hadronization mechanism employed is in many aspects very similar to the coalescence plus fragmentation approaches described in Sect.~\ref{sec:theoryCoalescence}. The coalescence for heavy mesons and baryons is evaluated in the Wigner formalism, see Eq.~\ref{eq-coal}. The wave function is a Gaussian and the excited states are included assuming a suppression like in the statistical model as in the Catania model. Heavy quarks not recombining with light partons undergo fragmentation.
Similarly to the SHM-TAMU approach, all the possible excited states are considered, including the unobserved baryon states not listed in the PDG, which are predicted to exist by the quark model and lattice QCD. A more \enquote{technical} difference is the mass of the light quarks which in EPOS are taken to be $m_{\rm u/d}=0.1 \, \rm GeV$, while in the Catania, TAMU, and PHSD models values close to the constituent quark mass of about 0.33 GeV are used, which reduce the violation of energy conservation in the  coalescence process. 
Furthermore, assuming a bulk system of quasi-particles with mass of about 0.1 GeV should lead to a entropy/pressure ratio close to $T_c$ quite larger than lattice QCD data, which are known to be well described by quasi-particle models with masses of about 0.3-0.4 GeV \cite{Plumari:2011mk,Sambataro:2024mkr}. Also in a coalescence approach a small mass with respect to the constituent one tends to favor baryon production especially at intermediate $p_T$ because it enhances more the probability to form baryons with respect to mesons due to the higher density and the larger boost effect of the relative momenta. A more thorough comparison analyzing how the results depend on the quark mass would be quite desirable.

EPOS4HQ has shown to generate for pp collisions ratios of $\Lc/\Dzero$ and $\XicZero/\Dzero$ that are in fair agreement with the experimental data by ALICE and CMS, only if coalescence is included~\cite{Zhao:2023ucp,Zhao:2024ecc}. Moreover, EPOS4HQ being a realistic fully dynamical approach showed that also the elliptic flow $v_2(\pt)$ of $\Dzero$ in pp collisions agrees with the data and origins from the scattering of charm quarks in the medium generated in high-multiplicity events.

\section{Heavy-flavour hadronization - experimental results}
\label{sect:exp}
In the following we explore a selection of experimental measurements on heavy-flavour production sensitive to heavy-quark hadronization. Firstly we discuss results in \ee and \ep collisions. Then the measurements in hadronic collisions are presented using event multiplicity as the order criterion: minimum-bias pp collisions are discussed first and are followed by studies differential in multiplicities, which allow to bridge the pp collisions to both p--A and A--A collisions, as well as to \ee collisions. Measurements in hadronic collisions sensitive to beam remnants and quantum-number conservation, which probe peculiar regions of phase space in which the fragmentation approach was already found (and expected) to not be effective, are discussed separately. 
 
\subsection{Production in ee and ep collisions and spectroscopy}
\label{sect:expEE}

\begin{figure}[tp]
\centering \includegraphics*[width = 0.7\textwidth]{ 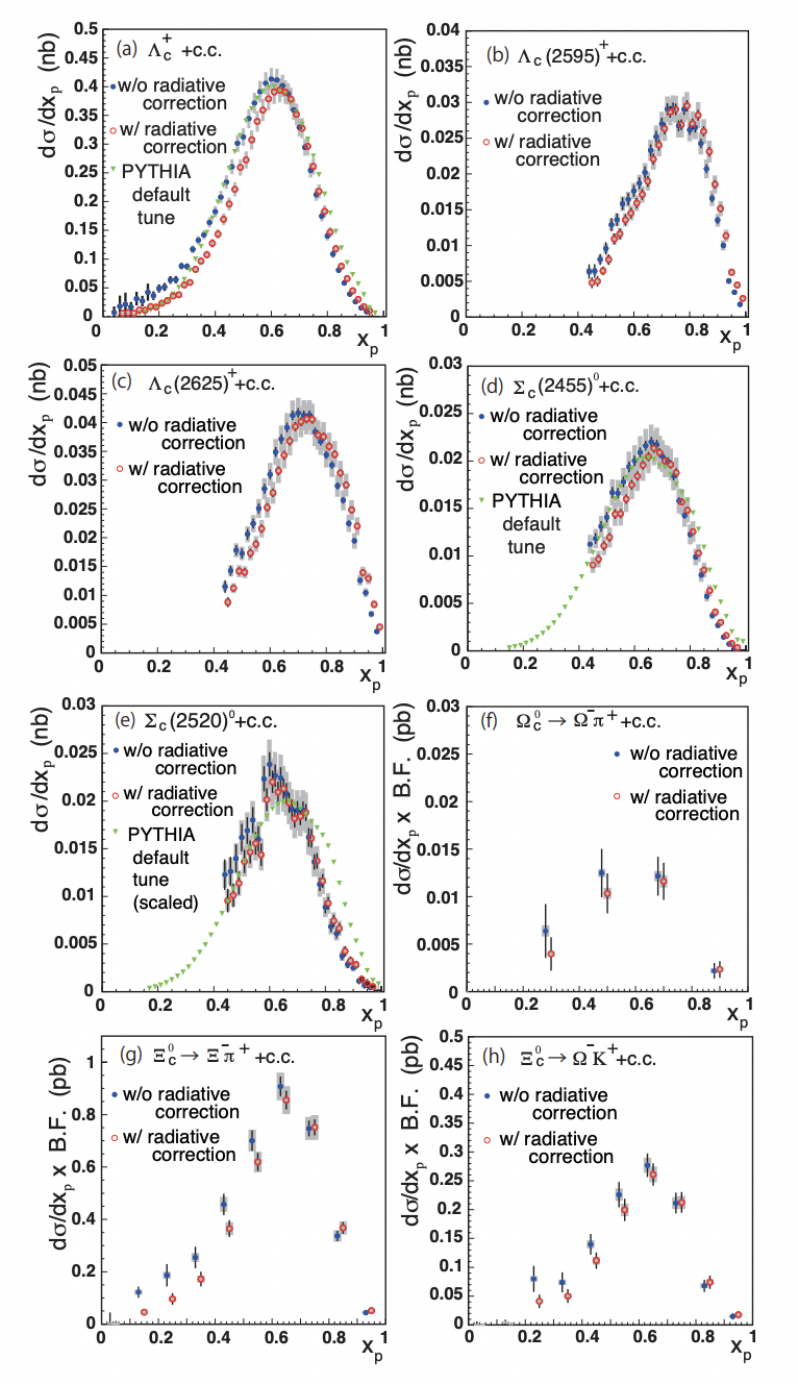}
\caption{Differential inclusive cross sections of charmed baryon production with and without radiative corrections measured by the Belle Collaboration~\cite{Niiyama:2017wpp}. The closed circles are shifted slightly to the left for clarity. Triangle points show PYTHIA
predictions where all radiative processes are turned off.}
    \label{fig:belleCharmBaryonXp}
\end{figure}


Electron--positron collisions present many advantages to study the fragmentation process of heavy quarks. In \ee collisions there is not an hadronic initial state, thus there are no concurrent partonic interactions nor beam remnants, and the centre-of-mass energy of the partonic system is known (up to radiative corrections) and coincides with the $\ee$ collision centre-of-mass energy. The $\rm q\overline{q}$ pair is produced mainly in a colour-singlet topology 
and soft gluon emission can be described as the radiation of a colourless antenna~\cite{Dokshitzer:1991wu}. On the other hand, $\ee$ collisions offer little sensitivity to study heavy-flavour hadron production in events in which heavy-quark production involves gluons in the hard scattering. 
The production cross section and properties of several charm-hadron species were measured at very different collision energies in \ee collisions at both B-factories ($\sqrt{s}\sim 10.5$~\GeVmass) and in $\rm Z^{0}$-boson decays at LEP ($\sqrt{s}\sim 90$~\GeVmass)~\cite{ALEPH:1999syy,OPAL:1996ikk,DELPHI:1999slw,Gladilin:2014tba,Niiyama:2017wpp}.

 A variable typically considered to constrain the heavy-quark fragmentation functions used in theoretical calculations is the beam-momentum fraction $x_{\mathrm{p}}=2p_{\rm{H}}c/\sqrt{s/4 - M^{2}c^{4}}$. The $x_{\mathrm{p}}$-differential cross section of several charm-baryon species measured by Belle~\cite{Niiyama:2017wpp}, namely of $\Lc$, $\LcExcitedOne$, $\LcExcitedTwo$, $\SigmacZero$, $\SigmacZeroExcited$, $\Omegac$ (cross section times branching ratio), $\XicZero$ (cross section times branching ratio), are shown in Fig.~\ref{fig:belleCharmBaryonXp}. The $\Lc$, $\SigmacZero$, and $\SigmacZeroExcited$ are compared to PYTHIA calculations~\cite{Sjostrand:1993yb}. Overall, though differences between data and PYTHIA can be appreciated thanks to the high precision of the data, the $\Lc$ and $\SigmacZero$ $x_{\mathrm{p}}$-differential cross sections are reasonably well reproduced by PYTHIA both in magnitude and in shape. The $\SigmacZeroExcited$ cross section is instead underestimated by about a factor of two by PYTHIA, which also predicts a harder $x_{\mathrm{p}}$ distribution. 

\begin{figure}[ht!]
    \begin{center}
   \includegraphics[width = 0.55\textwidth]{ 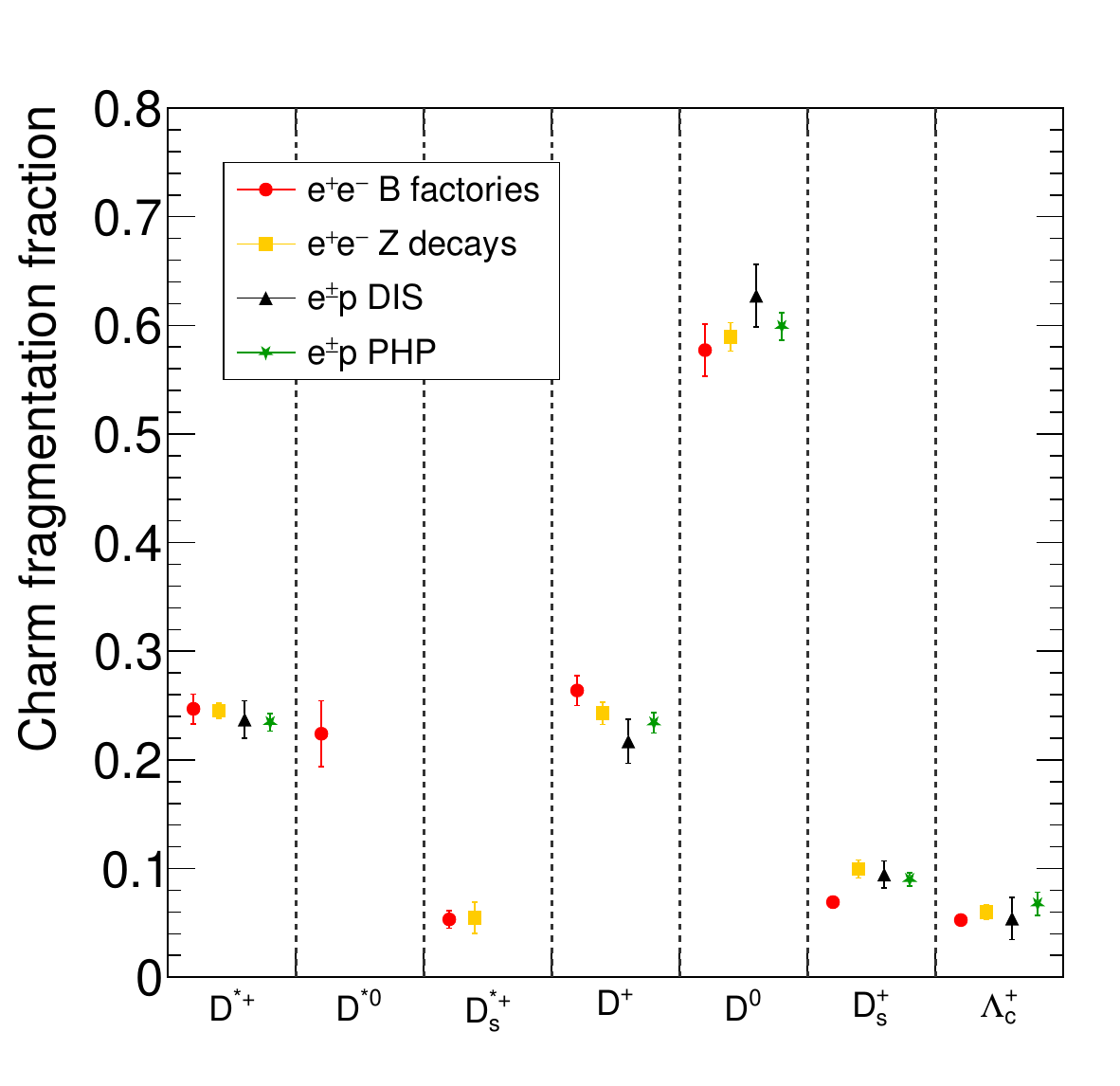}
   \caption{Charm fragmentation fraction to D mesons and \Lc in \ee and \ep  collisions calculated in Ref.~\cite{Lisovyi:2015uqa} by combining measurements from CLEO~\cite{CLEO:1988jcc,CLEO:1990unu}, ARGUS~\cite{ARGUS:1991vjh,ARGUS:1991xej,ARGUS:1988hly}, BABAR~\cite{BaBar:2002ncl,BaBar:2006lxr}, Belle~\cite{Belle:2005mtx}, OPAL~\cite{OPAL:1996ikk,OPAL:1997edj}, ALEPH~\cite{ALEPH:1999syy}, DELPHI~\cite{DELPHI:1999slw,DELPHI:1999qxi}, ZEUS~\cite{ZEUS:2005pvv,ZEUS:2013fws,ZEUS:2007yva,ZEUS:2010cic}, and H1~\cite{H1:2004bwe}. In the calculation, the sum of the fragmentation fractions to all known weakly decaying ground states is constrained to unity and it is assumed that the sum of the fragmentation fractions to $\XicPlusZero$ and $\Omegac$ is a factor of about 0.136 that to $\Lc$ in analogy to strange-quark fragmentation fractions to $\Omega^{-}$, $\Xi^{-}$, and $\Lambda^{0}$.}
   \label{fig:FFatee_ep_Lisoviy}
   \end{center}
\end{figure}

 The precise theoretical knowledge of the total \ccbar cross section allows the direct measurement of the fragmentation fraction of charm quarks to a given charm-hadron species, \FFc, without the need to measure the cross section of the other charm-hadron species. At LEP it was verified (see e.g. ALEPH analysis in Ref.~\cite{ALEPH:1999syy}) that the total charm-production cross section, obtained as the sum of the measured \Dzero, \Dplus, \Ds, and \Lc cross sections, and of an additional contribution for the unmeasured \XicPlusZero and \Omegac cross sections, assumed to be around 2\% of the measured one, is consistent with the Standard Model expectation. In Ref.~\cite{Lisovyi:2015uqa} charm-hadron production measurements in \ee collisions at B factories~\cite{CLEO:1988jcc,CLEO:1990unu,ARGUS:1991vjh,ARGUS:1991xej,ARGUS:1988hly,Belle:2005mtx,BaBar:2002ncl,BaBar:2006lxr} and in $\mathrm{Z^{0}}$-boson decays~\cite{OPAL:1996ikk,OPAL:1997edj,ALEPH:1999syy,DELPHI:1999slw,DELPHI:1999qxi} were analysed together with data from photoproduction (PHP) and deep-inelastic scattering (DIS) in $\mathrm{e^{\pm}p}$ collisions at HERA~\cite{Abramowicz:2013eja,Chekanov:2005mm,ZEUS:2007yva,ZEUS:2010cic,H1:2004bwe}, and with first results from pp collisions at the LHC. As shown in Fig.~\ref{fig:FFatee_ep_Lisoviy}, no significant difference emerged among the charm FF measured in the various collision systems and collision energies. In $\mathrm{e^{\pm}p}$ and hadronic collisions charm-quark pairs are produced in hard-scattering processes from gluons, differently from \ee collisions in which $\mathrm{g\rightarrow c\bar{c}}$ splittings occurs only in a fraction of events. At LEP, the fraction of $\mathrm{Z\rightarrow hadrons}$ events proceeding via $\mathrm{Z\rightarrow \mathrm{q\bar{q}g}, g\rightarrow c\bar{c}}$ is about 3\%, representing less than 20\% of events with charm in Z decays~\cite{OPAL:1999jla,L3:1999nsy,ALEPH:1999syy,ALEPH:2003fll}. The consistency of the FF measured in \ee and $\mathrm{e^{\pm}p}$ collisions was considered as a signal supporting the hypothesis of universality of the fragmentation fractions,
 which was expected to hold also in pp collisions at the LHC. First LHC data on D-meson production supported this hypothesis, which as will be discussed in next sections turned out not to be true.  

\begin{figure}[ht!]
    \begin{center}
   \includegraphics[width = 1\textwidth]{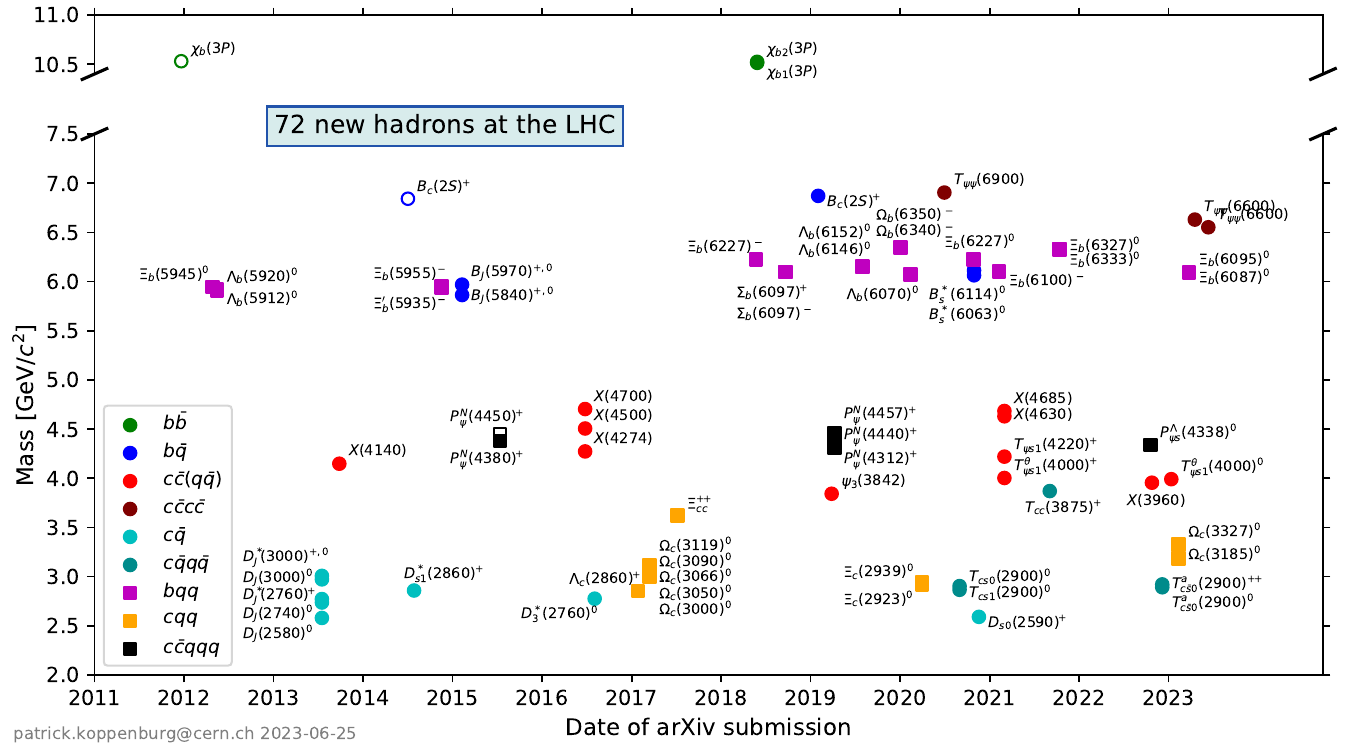}
          \end{center}
   \caption{Hadrons with charm and beauty quarks discovered at the LHC.  Figure from~\cite{LHCb-FIGURE-2021-001-report}.}
   \label{fig:NewHadronsLHC}
\end{figure}

 Spectroscopy measurements provide important pieces of information for understanding heavy-flavour production and hadronization. The measured production yield and \pt-differential spectrum of a given heavy-flavour hadron species, $\mathrm{H_{Q}}$, depends on the number and mass spectrum of existing hadron states energetically accessible, which determines both the direct-production probability of the given state at hadronization and the feed-down contribution. There is a large impact on the heavy-flavour baryon-to-meson ratio expected by models whether one assumes or not the existence of the rich set of excited states expected from the relativistic quark model (RQM)~\cite{Ebert:2011kk} (see Sect.~\ref{shm}), most of which have not yet been observed. Searching for new states in the charm sector has been a major part of the physics program of B factories, whereby exploiting the rather \enquote{clean} environment of \ee collisions allows these states to be looked for both in the direct production from charm-quark hadronization and in the decay of beauty hadrons. A review of relevant spectroscopy measurements at B factories can be found in Ref.~\cite{Kato:2018ijx}. Despite the observation of several states, the number of known states is not comparable to that expected by the RQM. This remains valid also including the significant number of states detected for the first time in hadronic collisions at Tevatron and the LHC~\cite{Klempt:2009pi,LHCb-FIGURE-2021-001-report}. At these colliders, as at LEP, several ground and excited states were discovered and studied also in the open-beauty sector, which is energetically almost precluded to B factories. As reported in Fig.~\ref{fig:NewHadronsLHC}, overall more than 70 new hadrons in the charm and beauty sectors were discovered at the LHC, out of which 19 \enquote{conventional} mesons, 30 conventional baryons, and 23 \enquote{exotic} hadrons, i.e. tetraquark and pentaquark candidates. However only for a few of these states information about the production yield is available. Whether further states exist remains an open question and a major goal of future experimental research at both hadronic and leptonic colliders. In hadronic collisions heavy-flavour baryon production may occur via additional processes than those effective in leptonic collisions. 
This further highlights the importance of measuring the relevant abundances of the various states in both collision systems. 

 The probability that a given state is formed can depend on the hadron internal structure and state quantum numbers, for the determination of which spectroscopy measurements are essential. 
Charm and beauty baryons are also good probes to study diquark effective degrees of freedom. 
Theoretical calculations suggested that heavy-flavour baryons can be interpreted rather simply as a bound state of a diquark and a heavy quark~\cite{Kato:2018ijx,Yoshida:2015tia,Ebert:2005}, a picture that requires experimental confirmation via a systematic comparison of measured and predicted spectra. Reviews on the topic, including discussions extended to exotic states, can be found in Ref.~\cite{Kato:2018ijx,Barabanov:2020jvn,Olsen:2017bmm,Klempt:2009pi,Jaffe:2004ph}.

\begin{figure}[t]
    \begin{center}
   \includegraphics[width = 0.5\textwidth]{ 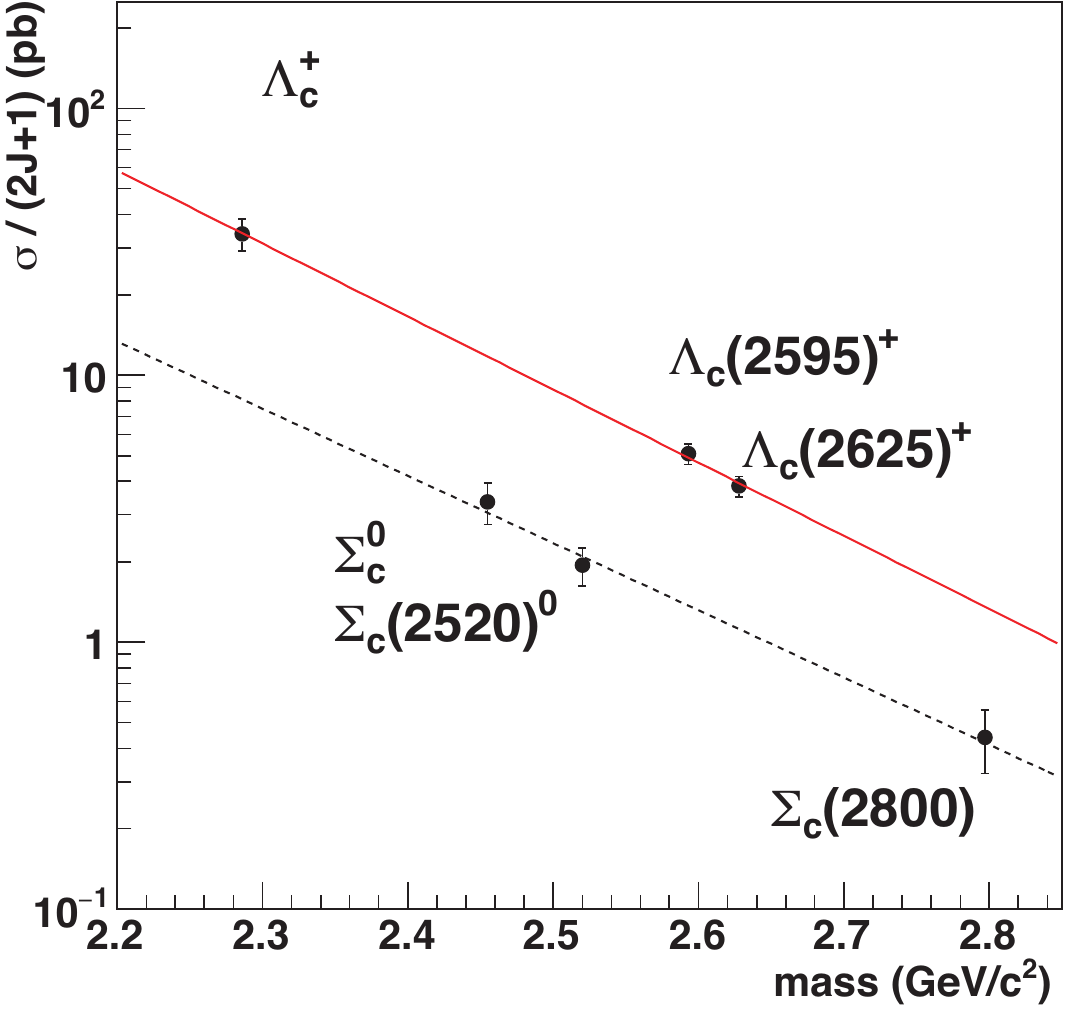}
          \end{center}
   \caption{Charm-baryon direct production cross section scaled by the spin-degeneracy factor as a function of the baryon mass measured in \ee collisions by Belle~\cite{Niiyama:2017wpp}.}
   \label{fig:BelleCharmBaryonsProduction}
\end{figure}

An important example is given by the lower production yields of the isovector $\Sigmac$ states compared to the isoscalar $\Lc$ states in \ee collisions. This can be appreciated looking at Belle result~\cite{Niiyama:2017wpp} in Fig.~\ref{fig:BelleCharmBaryonsProduction}, which shows the direct cross sections of $\Lc$ and $\Sigmac$ states divided by the state spin-degeneracy factor as a function of the particle mass. The $\Lc$ and $\Sigmac$ states, which in terms of valence quarks are all composed of a charm quark and a pair of light u/d quarks, lie on two different exponential lines. In the string-fragmentation picture, charm baryons are formed in \ee collisions via diquark-antidiquark pair creation from string breaks next to a c-quark string endpoint, whereby such diquark-antidiquark pair creation is modelled by a Schwinger tunnelling process. To form a $\Sigmac$ state an isovector diquark is needed, whereas the isoscalar diquark is needed for $\Lc$ states. Isovector diquarks have spin 1 and are heavier than isoscalar spin-0 diquarks, and thus the larger mass suppresses their production probability via the Schwinger mechanism. Though this argument might not take into account the possible role of dynamical diquark degrees of freedom~\cite{Barabanov:2020jvn}, the observed difference of the production yields of hadron states with the same valence quarks is exemplary experimental evidence of the fact that the internal hadron structure can affect the production probability. The impact of this can depend on the properties of the collision system, both because the relative abundances of the hadron \enquote{building blocks} (quarks and diquarks) can differ in different systems and because of the emergence of concurrent formation processes that may overcome limitations and constraints set by the hadron internal structure. Whether diquarks could constitute relevant degrees of freedom in the QGP close to the hadronization phase, possibly affecting hadrochemistry also in the heavy-flavour sector~\cite{Beraudo:2022dpz,He:2019vgs}, is an open question. The structure of \enquote{exotic} hadrons is still unclear. Diquark degrees of freedoms could constitute relevant building blocks also for tetraquarks and pentaquarks production. On the other hand, it is also unknown if nuclear-type forces can bind mesons to other mesons or baryons in \enquote{molecular} states. The study of exotic state properties and production in different collision systems can therefore be an important direction for determining both their nature as well as the relevant degrees of freedom of the systems, as further discussed in Sect.~\ref{sect:future}.




\subsection{Hadronic collisions: beam remnants and quantum-number conservation}

\begin{figure}[ht!]
    \begin{center}
        \includegraphics[width = 0.55\textwidth]{ 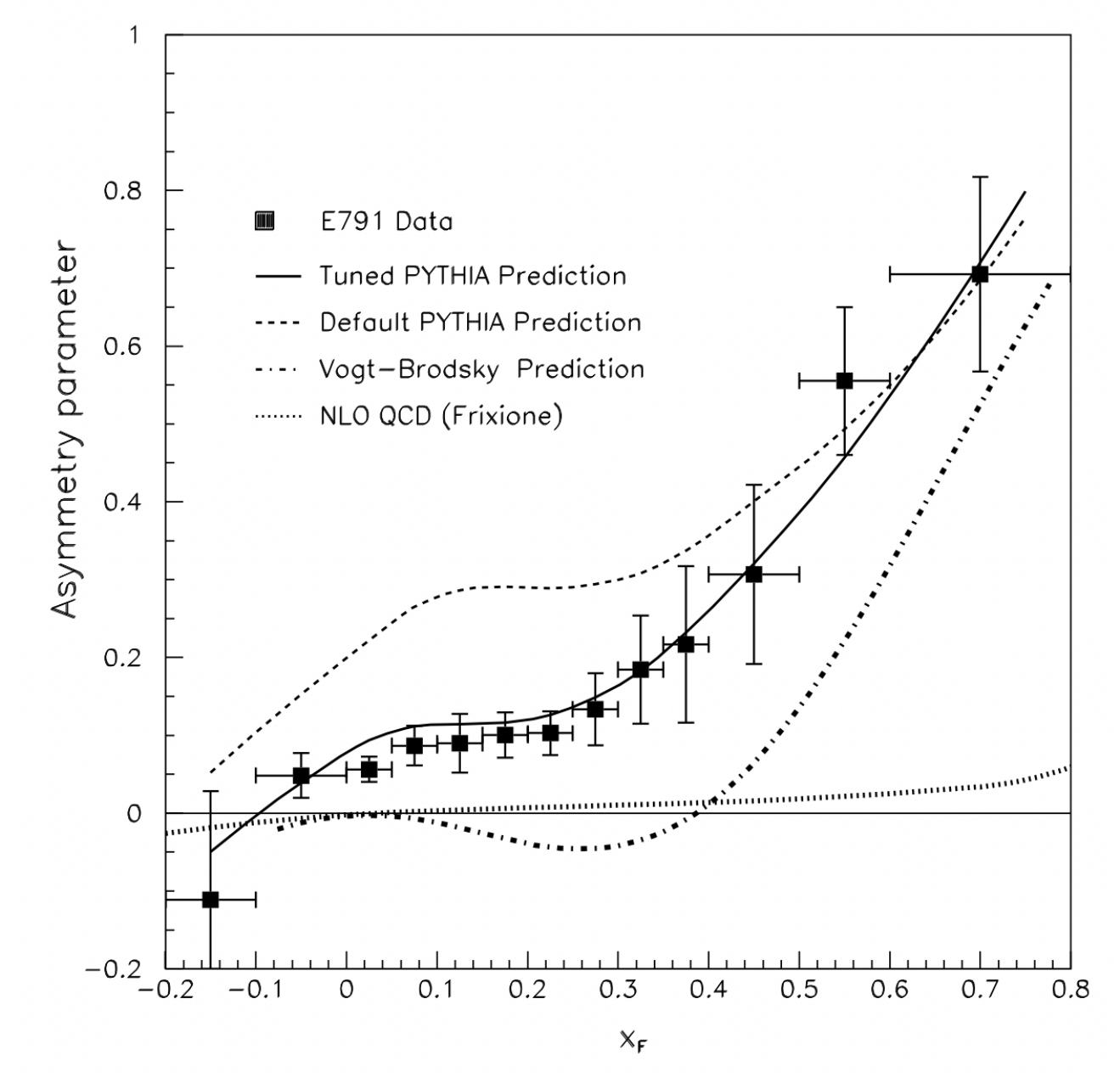}
    \end{center}
    \caption{The $\mathrm{D}^{\pm}$-meson asymmetry $A$ (see text) measured in $\pi^{\mathrm{-}}$--nucleus (C,Pt foils) collisions as a function of $x_{\mathrm{F}}$ compared to model calculations~\cite{E791:1996htn}.}
    \label{asymm}
\end{figure}

Quantum-number conservation affects the relative abundances of the particle species produced at hadronization time leaving a trace that is more evident when phase-space for particle production is limited (e.g. hadronic collisions at low energies, particle decays) or in specific kinematic regions, e.g. close to beam rapidities. However, in high-energy hadronic collisions far from beam rapidities hadronization is not expected to be significantly influenced by the quark content of the beam particles. Of course, quantum-number conservation dictates a departure from the particle ratios observed in a \enquote{vacuum-like} system as \ee, and it spoils the effectiveness of factorization approach using \ee FF in certain kinematic ranges. In many of these cases, the modelling of hadronization involves the recombination of valence quarks with other quarks produced in the event.
Effects that can not be reconciled with a modelling of hadronization via universal fragmentation fractions/functions, were already observed 
at CERN and FNAL in $\pi$--A collisions. An example is the so-called beam-remnant drag~\cite{Norrbin:1999mz}, i.e. the tendency in a hadronic collision to produce at forward rapidity particles that share a valence quark (or diquark) with the remnant of the projectile moving in the same direction. Focusing only on the heavy-flavour sector, asymmetries in the production of \Dminus/\Dplus~\cite{WA82:1993ghz,E769:1993hmj,WA89:1998wdl,E791:1996htn}, \Dsminus/\Ds~\cite{WA89:1998wdl,E791:1997eip}, \Lc/\Lcminus~\cite{WA89:1998wdl,SELEX:2001iqh} and \Lb/\Lbbar~\cite{D0:2015rnb} in various hadronic collisions have been observed. In Fig.~\ref{asymm} the asymmetry $A = \frac{\sigma_{\mathrm{D}^{-}} - \sigma_{\mathrm{D}^{+}}}{\sigma_{\mathrm{D}^{-}} + \sigma_{\mathrm{D}^{+}}}$ measured in $\pi^{\mathrm{-}}$--nucleus collisions (using platinum and diamond foils as target) by the E791 Collaboration is shown as a function of the Feynman $x_{\mathrm{F}}$ variable~\cite{E791:1996htn}. The positive asymmetry values emerging at large $x_{\mathrm{F}}$ indicate that the production of hadrons sharing valence quarks with beam hadron (the pion) is favoured. 

The measurement is compared to predictions based on two PYTHIA 5.7 tunes~\cite{Sjostrand:1993yb}, to the prediction for charm quarks by NLO QCD~\cite{Mangano:1995di, Frixione:1997ma}, and a calculation involving intrinsic charm by Vogt and Brodsky~\cite{Vogt:1994zf}. The NLO calculation shows an asymmetry ratio almost flat in $x_{\mathrm{F}}$ and close to unity. The model from Vogt and Brodsky, in this case specifically calculated for the E791 beam momentum, presents a shape of the asymmetry $A$ similar to the data one, however, the prediction is too low at all $x_{\mathrm{F}}$. 
The PYTHIA predictions are shown with two different tunes. The dashed line shows the prediction from the PYTHIA generator, which is significantly higher than the experimental data for $-0.2 < x_{\mathrm{F}} < 0.4$. This is because PYTHIA predicts a higher overall production ratio of $\rm D^-$ to $\rm D^+$ than is seen in the data. A tuned PYTHIA prediction that is consistent with the experimental data is shown as the solid line. The tuning included an increased mass of the c quark from $m_\mathrm{c}=1.35$~GeV/$c^2$ to $m_\mathrm{c}=1.7$~GeV/$c^2$ and an increased average primordial $k^2_t$ of the partons from (0.44 GeV/$c$)$^2$ to (1.0 GeV/$c$)$^2$. This decreases the likelihood that the remnant $d$ quark can combine with the $\overline{c}$ quark with a small enough invariant mass to collapse to a $\rm D^-$ meson. Anyhow, in order to reproduce the data, models must assume that the heavy quark hadronizes via recombination with some other parton already present in the system, in this case belonging to the beam remnant~\cite{Norrbin:1998bw,Norrbin:2000zc,Braaten:2002yt,Rapp:2003wn}.


\subsection{Proton--proton collisions}
\label{sec:pp}

\begin{figure}[htp]
    \begin{center}
    \includegraphics[width = 0.75\textwidth]{ 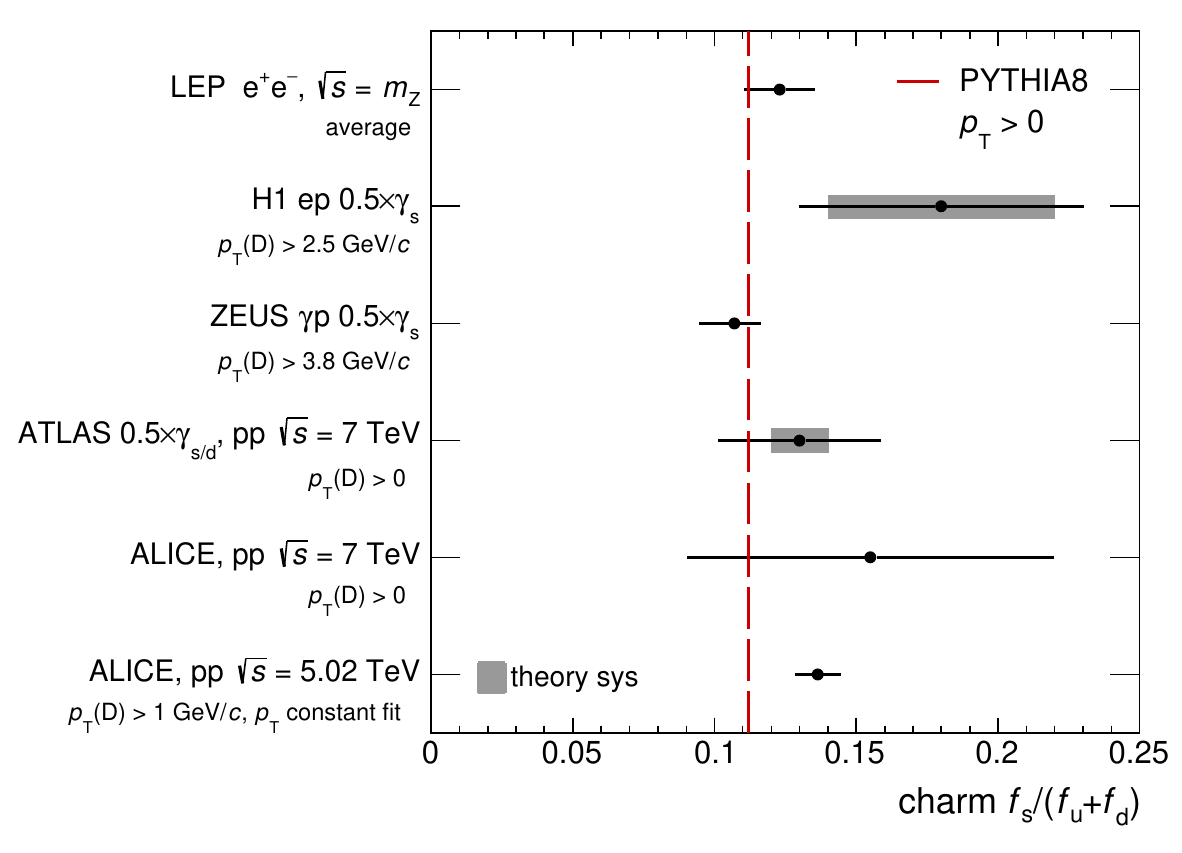}
    \includegraphics[width = 0.75\textwidth]{ 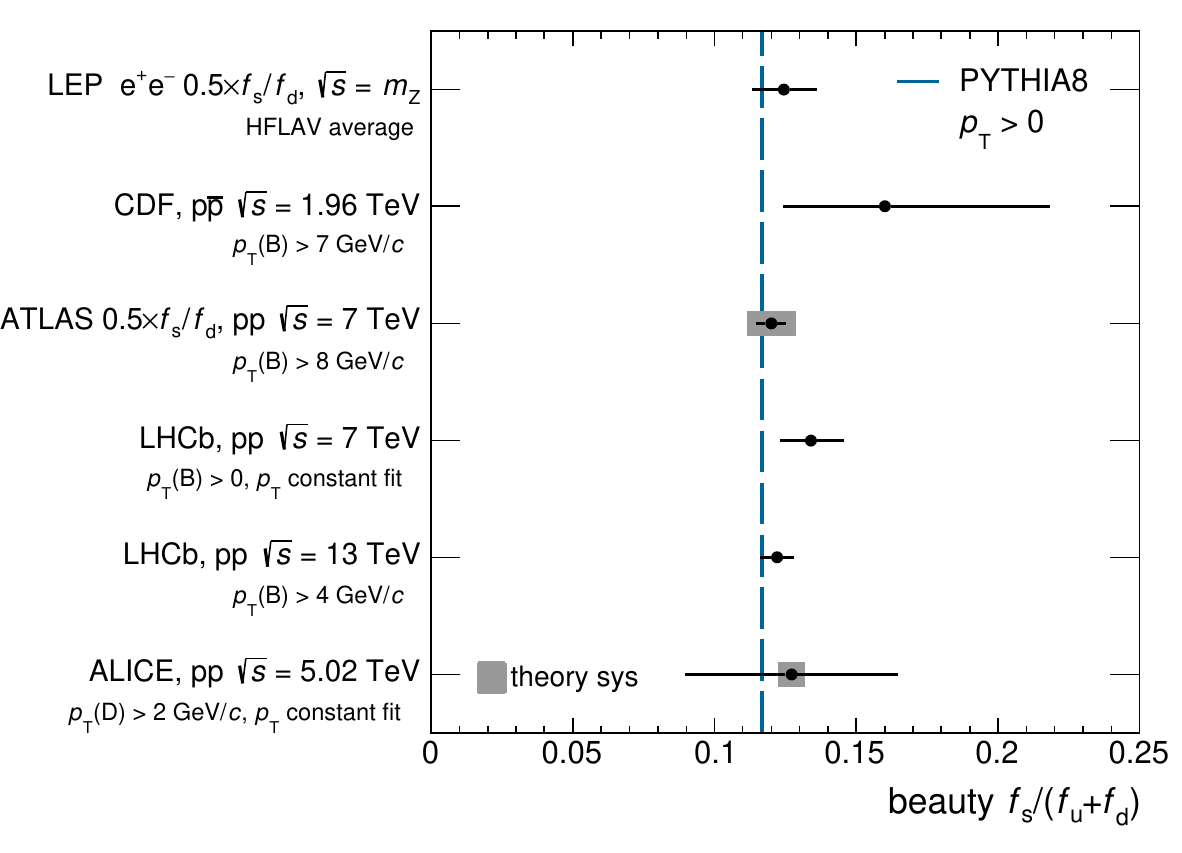}
    \end{center}
    \caption{Charm-quark (top panel) and beauty-quark (bottom panel) fragmentation-fraction ratio $f_\mathrm{s}/(f_\mathrm{u}+ f_\mathrm{d})$ measured by several collaborations in \pp, \ppbar (beauty case only), and \ep (charm case only) collisions, compared to the averages of LEP measurements~\cite{Gladilin:2014tba, HFLAV:2019otj}. The values obtained from PYTHIA 8 simulations of pp collisions with Monash tune~\cite{Skands:2014pea}.}
    \label{fs}
\end{figure}

 As several measurements show~\cite{Andronic:2015wma,Cacciari:2005rk,Acosta:2003ax,CDF:2004jtw,CDF:2006ipg,LHCb:2015swx,ALICE:2021mgk,ALICE:2023sgl,CMS:2017uoy,CMS:2010nis,Sirunyan:2017xss}, \pt-differential cross sections of open charm- and beauty-meson production in pp collisions are successfully described by quantum chromodynamics (QCD) calculations based on the factorization of the soft (non-perturbative) and hard (perturbative) processes and making use of FFs parameterized to reproduce measurements performed in \ee or ep collisions. Implicitly, it is assumed in these calculations that the fragmentation and hadronization of charm and beauty quarks are universal processes independent of the collision system, or at least that they are approximately the same in \ee and pp collisions.
 The charm-meson \pt-integrated cross sections measured by several Collaborations in different colliding systems were used to compute the ratios of production yields among the different D-meson species (\Dzero, \Dplus, \Ds)~\cite{ALICE:2019nxm,ALICE:2021mgk,LHCb:2016ikn,ATLAS:2015igt,CMS:2017qjw}.
The prompt \Ds/(\Dzero+\Dplus) ratio represents the FF of charm quarks to charm-strange mesons $f_\mathrm{s}$ divided by the FF to non-strange charm mesons $f_\mathrm{u} + f_\mathrm{d},$ given that all \Dstar and \Dstarzero mesons decay to \Dzero and \Dplus mesons, and \Dsstar mesons decay to \Ds mesons. The charm-quark fragmentation-fraction ratios $f_\mathrm{s}/(f_\mathrm{u}+ f_\mathrm{d})$ from the ALICE~\cite{ALICE:2021mgk, ALICE:2012gkr}, H1~\cite{H1:2004bwe}, ZEUS~\cite{Abramowicz:2013eja}, and ATLAS~\cite{Aad:2015zix} Collaborations are shown in the upper panel of Fig.~\ref{fs}. 
All the values are compatible within uncertainties and with the average of measurements at LEP~\cite{Gladilin:2014tba}. The experimental
data is also compared to the value obtained from a PYTHIA 8 simulation of pp collisions using the Monash tune~\cite{Skands:2014pea}, which is found to be compatible with the data within the uncertainties.
The fragmentation fractions of beauty quarks to beauty-strange mesons divided by the one to non-strange beauty mesons computed by the ALICE~\cite{ALICE:2021mgk}, CDF~\cite{Aaltonen:2008zd}, LHCb~\cite{LHCb:2011leg,LHCb:2019fns}, and ATLAS~\cite{ATLAS:2015esn} Collaborations are also reported in the lower panel of Fig.~\ref{fs}. All the $f_\mathrm{s}/(f_\mathrm{u}+f_\mathrm{d})$ values measured in pp and $\rm{p\bar{p}}$ collisions are found to be compatible with the LEP average, computed by the HFLAV Collaboration~\cite{HFLAV:2019otj}, and with the value obtained from PYTHIA 8 simulation of pp collisions with the Monash tune. 
The LHCb and CMS Collaborations observed a mild \pt dependence ($\sim$ 10\%) of $f_\mathrm{s}/(f_\mathrm{u}$ ratio in the low/intermetdiate \pt region followed by a flat high-\pt trend, while the ratio $f_\mathrm{d}/(f_\mathrm{u}$ is found to be consistent with unity and independent of \pt and rapidity~\cite{LHCb:2019lsv,CMS:2022wkk}. In general, the production cross sections and cross-section ratios of charm and beauty mesons do not show evidences of violation of the assumption of universality of the fragmentation and 
hadronization processes. However, recent measurements of charm and beauty baryon-to-meson production ratios and fragmentation fractions in pp collisions at LHC energies showed significant deviations from the values measured at \ee{} and ep colliders, demonstrating that the assumption of universality of the hadronization process across collision systems has to be reconsidered.

\begin{figure}[ht!]
    \begin{center}
    \includegraphics[width = 0.45\textwidth]{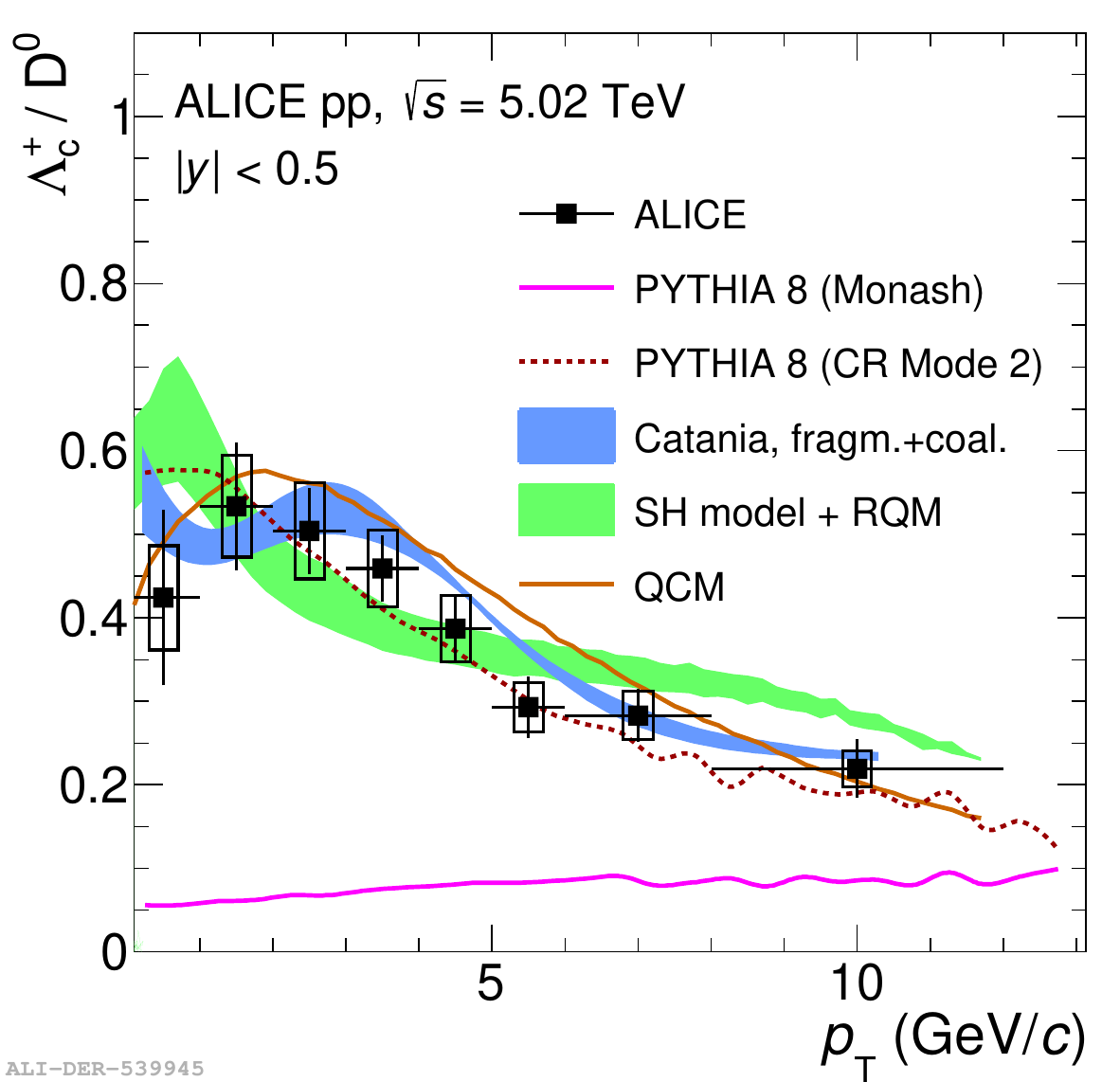}
    \includegraphics[width = 0.51\textwidth]{ 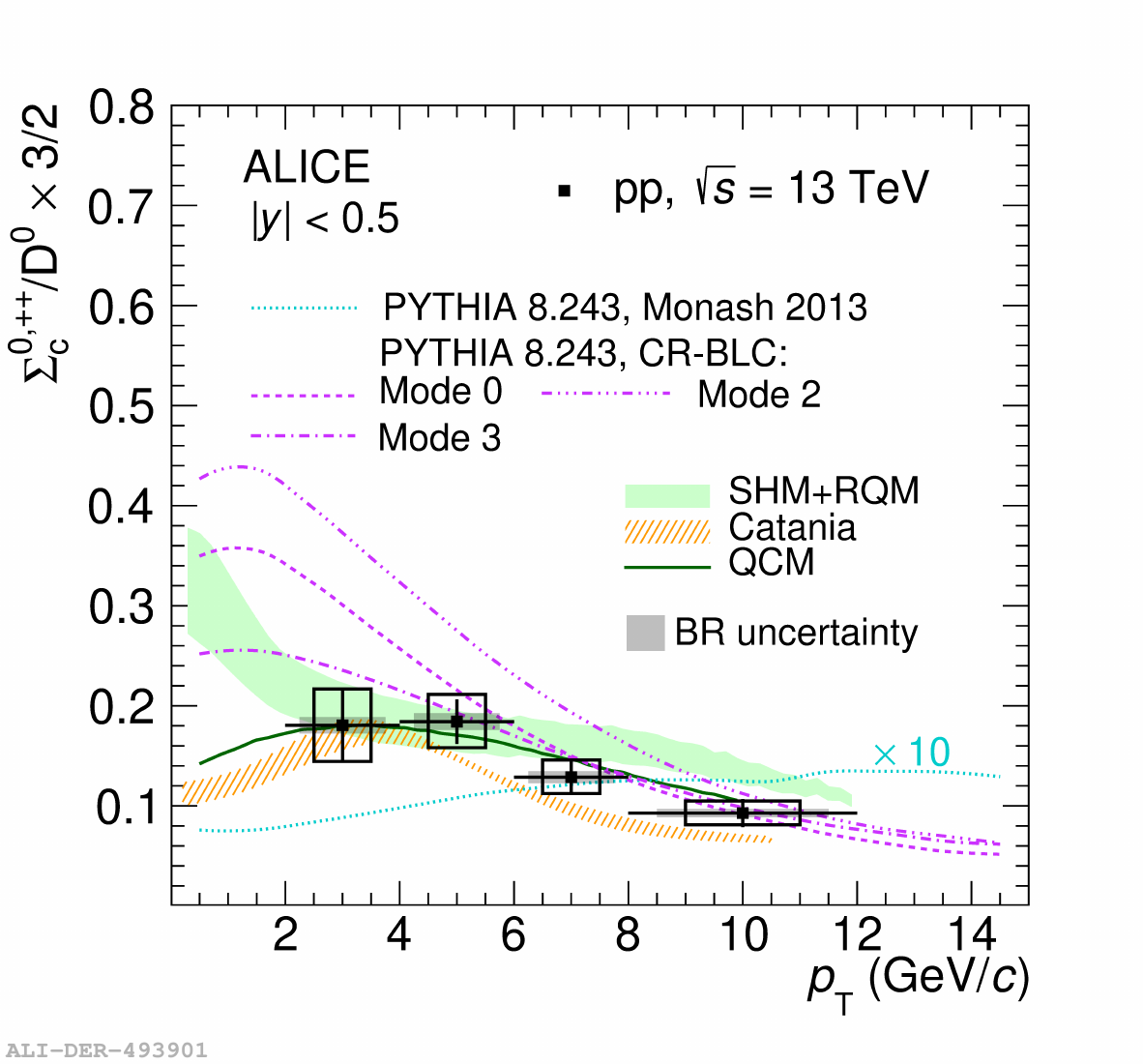}
    \includegraphics[width = 0.49\textwidth]{ 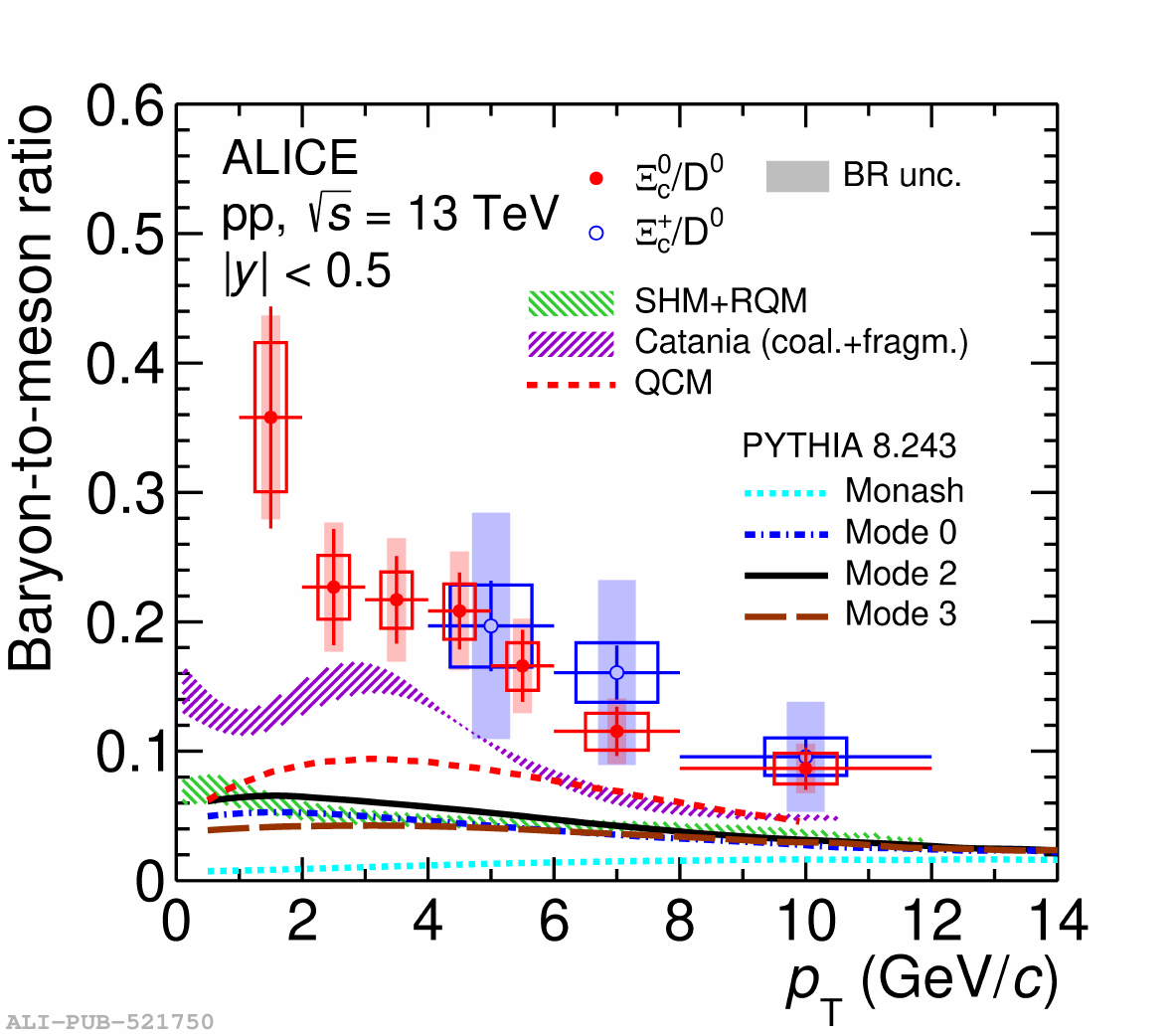}
    \includegraphics[width = 0.45\textwidth]{ 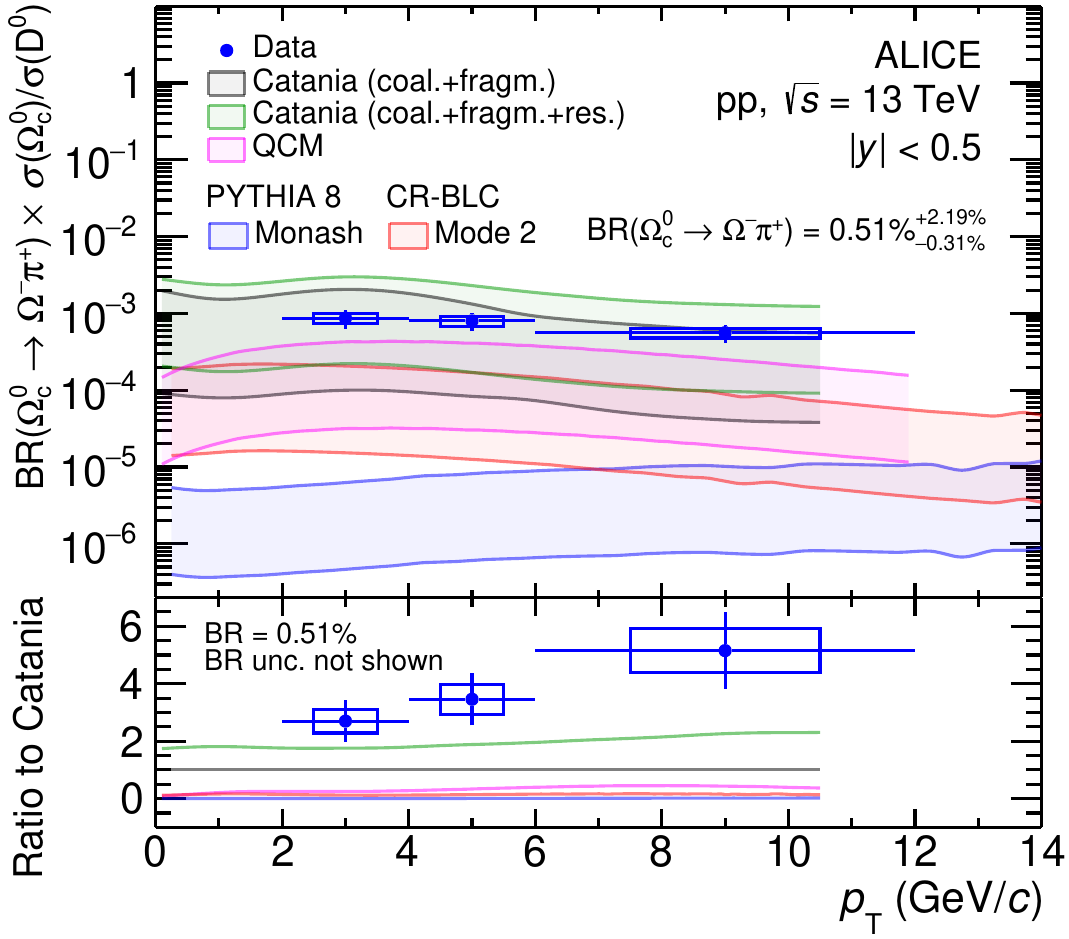}
    \end{center}
    \caption{ \LcD (top left), \Sigmac/\Dzero (top right) \XicPlusZero/\Dzero (bottom left), and \Omegac/\Dzero (bottom right) yield ratios measured by ALICE~\cite{ALICE:2022ych,sigmac,ALICE:2022cop} as a function of \pt in pp collisions in comparison with model calculations~\cite{He:2019tik,Christiansen:2015yqa,Skands:2014pea,Minissale:2020bif,Li:2021nhq}.}
    \label{BMR}
\end{figure}

The \LcD baryon-to-meson yield ratio as a function of \pt at midrapidity in pp collisions at \s = 5.02 TeV~\cite{ALICE:2022ych} is shown in the top-left panel of Fig.~\ref{BMR}. 
The ratio is measured to be 0.4-0.5 at low \pt, and
decreases to around 0.2 at high \pt. The strong \pt-dependence of the \LcD ratio contrasts with the ratios of strange and non-strange D mesons, which do not show a significant \pt dependence. This indicates that, interpreting the ratio within a pure fragmentation scheme, 
the fragmentation functions of baryons and mesons must differ significantly. The data are compared with model calculations in which, as described in Sect.~\ref{sect:Theory}, different hadronization processes are implemented. The Monash tune of PYTHIA 8~\cite{Skands:2014pea} predicts an integrated value of about 0.065 for the \LcD ratio, with a mild \pt dependence. This model calculation significantly underestimates the data at low \pt by a factor of about 5–10, while at high \pt the
discrepancy is reduced to a factor of about 2. The expectations of models including processes that enhance baryon production (see Sect.~\ref{sec:baryons}), such as PYTHIA 8 with QCD-based CR including junction string topologies~\cite{Christiansen:2015yqa}, SHM+RQM~\cite{He:2019tik}, QCM~\cite{Song:2018tpv}, and Catania~\cite{Minissale:2020bif} are also shown. All of these models qualitatively reproduce the data. The data hints at a decrease of the \LcD yield ratio in the interval \lowptbin, though a more precise measurement is needed to reach a firm conclusion about the actual trend and the possible discrepancy with model predictions. The \LcD yield ratio at \s=5~\TeV also by the CMS Collaboration in $|y|<1$. A first measurement in the 5 $<$ \pt $<$ 20 GeV$/c$~\cite{Sirunyan:2019fnc} was recently followed by a second one extended in \pt down to 3 and up to 30~\GeVc~\cite{CMS:2023frs}. In the \pt region covered by both experiments, the results are consistent with each other~\cite{ALICE:2020wla,Rossi:2023xqa}. The \LcD ratio was measured also at \s=7 and 13~\TeV by ALICE: within uncertainty, no significant energy dependence was observed~\cite{sigmac,ALICE:2023sgl}. The more recent CMS results at \s=5~\TeV and ALICE results at 13~\TeV, which extends up to $\pt=24$~\GeV/c, indicate that the \LcD ratio decreases below 0.2 at high \pt, further approaching the values measured in \ee collisions.

\begin{figure}[t]
    \begin{center}
    \includegraphics[width = 0.49\textwidth]{ 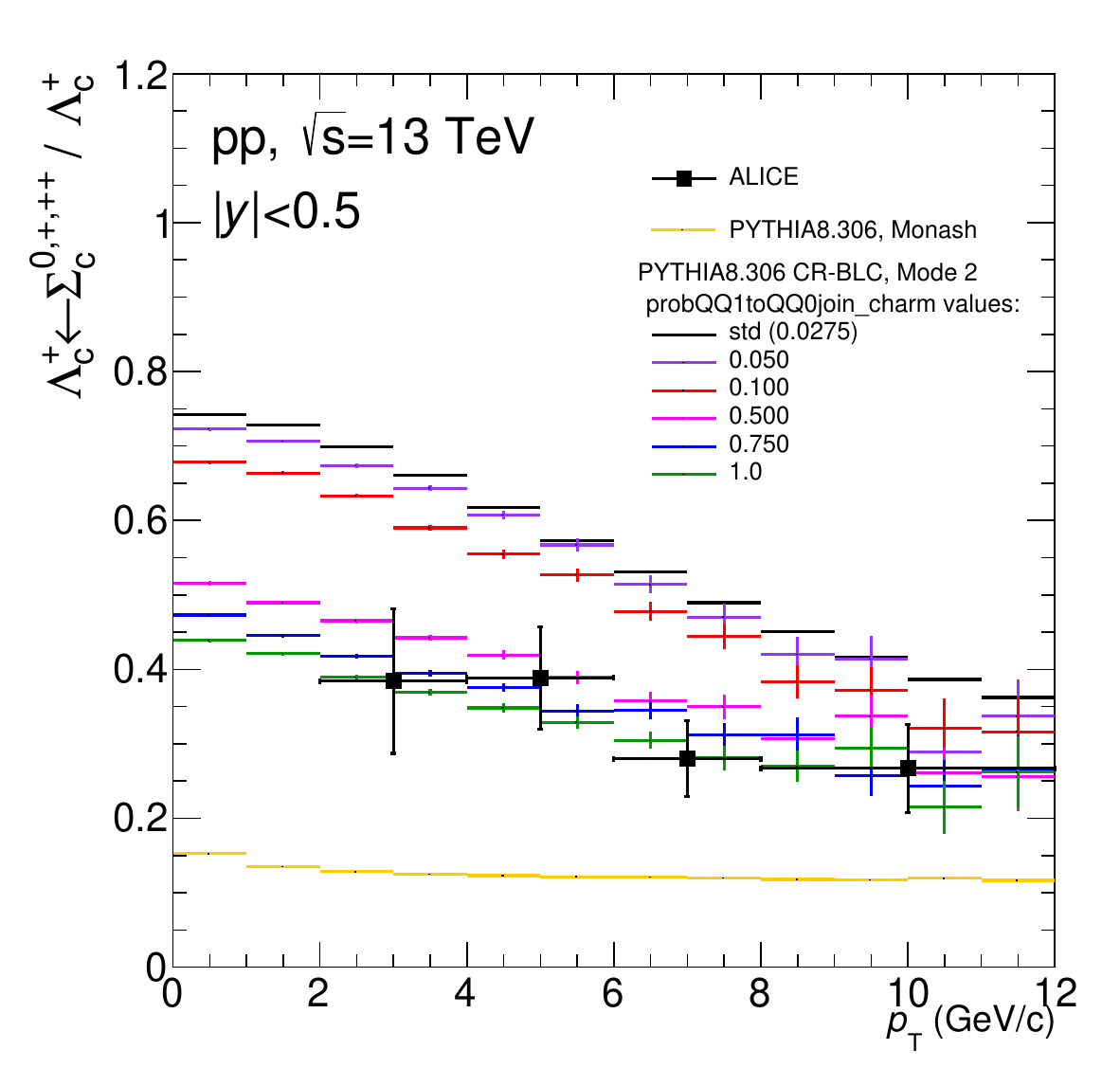}
    \includegraphics[width=0.49\textwidth]{ 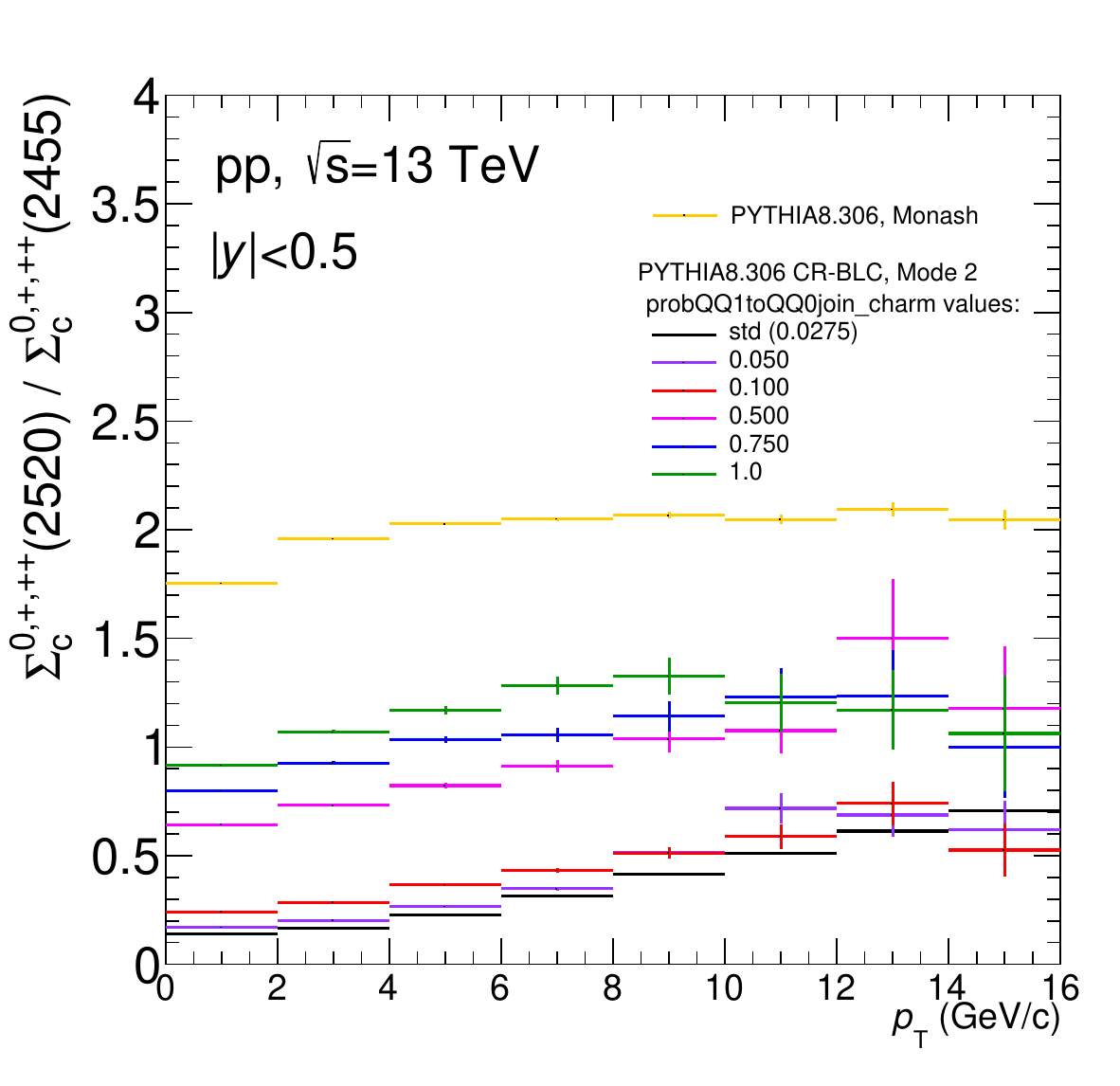}
    \end{center}
    \caption{Left: comparison of the fraction of $\Lc$ from $\Sigmac(2455)$ decay measured as a function of \pt with expectations from PYTHIA 8. The vertical bars on the data points represent the total uncertainty, quadratic sum of statistical and systematic uncertainties. For PYTHIA~8 with CR-BLC, the results obtained when the value of the charm {\sffamily probQQ1toQQ0join} parameter, which regulates the probability that junction diquarks containing charm have spin=1 or spin=0, is increased for charm-light diquarks from the default value of 0.0275 to 1, are shown. Right: expectation for the $\Sigma_{\mathrm{c}}^{0,+,++}(2520)/\Sigmac(2455)$ ratio for the same cases.}
    \label{fig:ScPYTHIA8tuning}
\end{figure}

 Another important measurement related to the enhancement of the \Lc baryon performed by the ALICE Collaboration, is the measurement of the \Sigmac baryon at midrapidity in pp collisions at \s = 13~\TeV~\cite{sigmac}.
As discussed in Sect.~\ref{sect:expEE},  
the production of \Sigmac states is suppressed with respect to \Lc-states~\cite{Niiyama:2017wpp} in \ee{} collisions.
As shown in the top-right panel of Fig.~\ref{BMR}, in pp collisions the \Sigmac/\Dzero ratio is close to 0.2 for 2 $<$ \pt $<$ 6 GeV$/c$, and decreases with \pt down to about 0.1 for 8 $<$ \pt $<$ 12 GeV$/c$, though the uncertainties do not allow firm conclusions about the \pt dependence. From Belle measurements (Table 4 in Ref.~\cite{Niiyama:2017wpp}), the \Sigmac/\Lc ratio in \ee{} collisions at \s = 10.52 GeV can be evaluated to be around 0.17 and, thus, the \Sigmac/\Dzero ratio can be estimated to be around 0.02. Therefore, a remarkable difference is present between pp and \ee{} collision systems. In Ref.~\cite{sigmac} the $\Sigmac$ feed-down fraction for $\Lc$ with $2<\pt<$~\GeVc was measured to be $\Lc\leftarrow\Sigmac/\Lc = \mathrm{0.38\pm 0.06~(stat) \pm 0.06~(syst)}$,
a significantly larger value than the \Sigmac/\Lc ratio measured by Belle and the $\sim$0.13 expectation from PYTHIA 8 Monash tune simulations. This indicates a larger increase of the \Sigmac/\Dzero ratio than of the
direct-\LcD ratio from \ee{} to pp collisions. The larger feed-down from \Sigmac partially explains the
difference between the \LcD ratios in pp and \ee{} collisions. It suggests that the suppression of \Sigmac states observed in \ee collisions, which in the Lund string-fragmentation picture is ascribed to the lower production yield of spin-1 than spin-0 diquarks in string breaking, is absent or reduced in pp collisions. In the models that provide a fair description of the \LcD ratio in pp collisions this suppression mechanism is absent or heavily reduced, and a sizeable contribution to \Lc production from strong decays of \Sigmac states is expected. As shown in Fig.~\ref{BMR}, the QCD-based CR-BLC (for which the three variations defined in Ref.~\cite{Christiansen:2015yqa} are considered), SHM+RQM, and Catania models describe, within uncertainties, both the \LcD and \Sigmac/\Dzero ratios. 
In the SHM, only the hadron mass and spin degeneracy factor determine the hadron yields. In all these models the production of \Sigmac(2455) is expected to increase by significant factors in pp with respect to \ee collisions and to become comparable or even larger than that of direct \Lc. 
In Fig.~\ref{fig:ScPYTHIA8tuning} it is illustrated how the current measurements can provide useful constraints for a better tuning of the PYTHIA~8 QCD-based CR-BLC parameters. Specifically, in junction systems with slow heavy quarks, charm-light diquarks are introduced in an intermediate step during the fragmentation process (see \cite{Altmann:2024icx} for details). These should probably not be thought of as a physical representation of the internal structure of the baryon being formed (see the discussion in section~\ref{sect:expEE}), but mainly as a technical bookkeeping device. 
Nevertheless these diquarks formally need to be assigned a spin quantum number, 0 or 1. 
When the charm-light diquark is combined with a quark to form a baryon, the spin assigned to the charm-light diquark is imprinted on the rate of spin-3/2 baryons since spin=3/2 can only be produced (in an $l=0$ state) if the charm-light diquark has spin=1, not otherwise. 

In PYTHIA~8, this probability is regulated by the parameter {\sffamily probQQ1toQQ0join}, which dictates the suppression of forming a junction diquark with spin-1 compared to spin-0, acting on top of spin-state counting that dictates a 3-to-1 enhancement. The parameter {\sffamily probQQ1toQQ0join} allows for different suppression factors given the heaviest flavour involved in the junction diquark formation is up/down, strange, charm or bottom. A value close to zero implies a large suppression of spin-1 junction diquarks, while {\sffamily probQQ1toQQ0join}=1 implies no suppression. 
Given the argument above that the charm-light diquark should probably not be attributed significant physical meaning, it would be reasonable to expect this parameter not to differ greatly from unity. 

Excited $\Lc$ and $\Sigma_{\mathrm {c}}$ states with spin=3/2 decay preferentially to $\Lc$ rather than to $\Sigmac$(2455)~\cite{Workman:2022ynf}. 
Therefore, increasing the probability of producing spin=1 charm-light diquarks changes the feed-down contribution to $\Sigmac$ and $\Lc$, effectively resulting in a reduction of the $\Sigmac/\Lc$ ratio. 
Thus one can significantly improve the description of the measured feed-down contribution from $\Sigmac$ to $\Lc$ production in PYTHIA by increasing the probability of spin-1 charm-light diquark formation in junctions. 

Though the data uncertainties do not allow a fine tuning of the charm {\sffamily probQQ1toQQ0join} parameter, values around 0.5 or higher seem preferable as seen in Fig.~\ref{fig:ScPYTHIA8tuning}.
Note this is in contrast to the parameter {\sffamily probQQ1toQQ0} which determines the spin-1 to spin-0 probability for light diquark-antidiquark production from standard string breaks, which has a default LEP-tuned value of 0.0275. 
Such a large suppression of spin-1 light diquarks can be justified by mass difference arguments and the smaller binding energy of spin-1 diquarks. Nonetheless there remains large ambiguity in {\sffamily probQQ1toQQ0join} in the light sector, and {\sffamily probQQ1toQQ0join} values for all flavours were not well constrained for the default PYTHIA values nor the QCD-based CR tunes. 
Future measurements of the production of excited states, in particular of $\Lambda_{\mathrm{c}}^{+}(2595)$,  $\Lambda_{\mathrm{c}}^{+}(2625)$ and $\Sigma_{\mathrm{c}}^{0.+.++}(2520)$ will be useful to constrain this parameter, as illustrated in Fig.~\ref{fig:ScPYTHIA8tuning} below.

 The ALICE Collaboration extended the measurement of charm-baryon production to baryons with strange quarks among the constituent quarks, measuring the \XicPlusZero cross section in pp collisions at \s = 5.02 TeV~\cite{Acharya:2021dsq} and \s = 13 TeV~\cite{xic13tev} and the \Omegac~\cite{ALICE:2022cop} cross section times branching ratio in pp collisions at \s = 13 TeV.
The bottom left panel of Fig.~\ref{BMR} shows the \XicPlusZero/\Dzero ratios measured as a function of \pt at midrapidity in pp collisions at \s = 13 TeV. The values observed are generally lower than those of the \LcD ratio, but the \pt dependence is similar. 
The PYTHIA~8 Monash tune significantly underestimates the data by a factor of 23--43 in the low-\pt region and by a factor of about 5 in the highest \pt interval.
All three CR modes give a similar magnitude and \pt-dependence of \XicPlusZero/\Dzero and predict a larger ratio with respect to the Monash tune. However, differently from what is observed for the \Lc and the \Sigmac baryons, they underestimate the \XicPlusZero/\Dzero ratio by a factor 4--9 for \pt$<$4 GeV$/c$. At higher \pt the measured \XicPlusZero/\Dzero ratio is closer to PYTHIA 8 calculations. 
As the CR-BLC models underpredict the \XicPlusZero/\Dzero ratio despite well describing \Lc/\Dzero, the source of this discrepancy is primarily assumed to be the strange content of the \XicPlusZero baryon, where we suspect strangeness enhancement effects particularly at low \pt. There have been some models such as Rope hadronization~\cite{Bierlich:2017sxk} which aim to implement a strangeness enhancement mechanism, however the rope model thus far has not been able to describe both the \XicPlusZero/\Dzero and \Lc/\Dzero ratios simultaneously (or equivalently stated, it cannot describe the \XicPlusZero/\Lc ratio). 
The SHM+RQM model~\cite{He:2019tik}, which captures the magnitude and \pt dependence of the \LcD and \Sigmac/\Dzero yield ratios, also underestimates the \XicPlusZero/\Dzero ratio. 
The measured ratios are also compared to the QCM~\cite{Song:2018tpv} and the Catania~\cite{Minissale:2020bif,Plumari:2017ntm} models that include hadronization via coalescence and  reproduce well the \LcD and \Sigmac/\Dzero ratios. 
They both tend to underestimate the data though, considering the uncertainties, the Catania model remains close to the measured values over the whole \pt interval.
Another interesting observation is given by the fact that the \XicZero/\Sigmac ratio shown in Ref~\cite{xic13tev} is described well by the default PYTHIA~8 Monash tune~\cite{Skands:2014pea}, which significantly underestimates both \XicPlusZero/\Dzero and \Sigmac/\Dzero ratios. Though this could be accidental (the predicted value remains to be explained along with its dependence on the two almost identical particle masses, and the assumed feed-down contributions), it may also signal the removal of a similar suppression factor affecting both \XicZero and \Sigmac production in \ee collisions. 
In particular, in most calculations~\cite{Barabanov:2020jvn}, a similar mass is obtained or assumed for S=1 (ud,uu,dd) diquarks as contained in \Sigmac and S=0 (ds) diquarks as in \XicZero. 
As discussed for the $\Lc\leftarrow\Sigmac/\Lc$ ratio, this could suggest that light diquarks are more easily produced in pp than \ee collisions via other processes than the Schwinger mechanism (e.g.\ colour reconnections or coalescence), or that in pp collisions charm-baryon formation does not proceed via the combination of a charm quark with a light diquark. As mentioned in Sect.~\ref{sec:theoryCoalescence}, the coalescence mechanism can be factorized in terms of a formation of diquark recombining with a further quark. In the limit of negligible binding energy this would correspond to the coalescence model already discussed. However, this is different from the assumption of the existence of a thermal yield of diquarks with masses that are not simply the sum of the two quark masses. In this respect
it is also argued in Ref.~\cite{Yun:2023kym} that a further enhancement of \XicPlusZero/\Dzero ratio 
could be due to the large attraction of the $us$ and $ds$ diquark channel with respect to the $ud$ one which is suggested by a relativistic quark model with a potential having a large screened mass as extracted from recent LQCD data.
Introducing such an additional coalescence component in the Catania model could strongly improve its agreement with experimental data \cite{Yun:2023kym}, but within the current experimental data uncertainties no conclusion can be drawn.
Clarifying the role of light and heavy diquarks will be important for understanding hadronization in pp as well as in heavy-ion collisions, where diquarks could also be effective degrees of freedom of the system close to the phase transition~\cite{Beraudo:2022dpz}.

 The \Omegac baryon is composed of a charm quark and two strange quarks. As it has just been discussed, most of the models can describe the \Lc data better than \XicPlusZero data. This signals a possible difficulty with charm-strange baryons, hence the measurement of $\Omegac$ production represent a crucial step to constrain models and understand whether strange quarks, or strange diquarks, play a peculiar role in charm-baryon formation in pp collisions. 
Moreover, following what discussed above and in Sect.\ref{sect:expEE} concerning the \XicZero/\Sigmac and $\Lc/\Sigmac$ ratios, the comparison of the relative abundances of \XicPlusZero and \Omegac in pp and \ee collisions can provide further information about the role of diquarks because it can be sensitive to the different mass values of $S=1$ (ss) and $S=0$ (sd, su) diquarks~\cite{Ebert:2011kk}. The \Omegac/\Dzero ratio was recently measured by the ALICE Collaboration~\cite{ALICE:2022cop} and is shown in the bottom right panel of Fig~\ref{BMR}, although without an absolute normalization because of the lack of absolute measurements of the branching ratios of \Omegac decays. The uncertainties do not allow conclusions about its possible dependence on \pt. The data are compared with model expectations that were obtained by scaling the \Omegac/\Dzero ratio predicted by the models by an estimate of the BR of the $\rm \Omega_c^0\rightarrow\Omega^-\pi^+$ decay channel, $\rm BR(\Omega_c^0\rightarrow\Omega^-\pi^+)=0.51\% ^{+2.19\%} _{-0.31\%}$~\cite{Hsiao:2020gtc}. The uncertainty band of the models represents the BR uncertainty, which is estimated considering other BR estimations reported in Refs.~\cite{Hsiao:2020gtc, Gutsche:2018utw, Cheng:1996cs, Hu:2020nkg, Solovieva:2008fw, Wang:2022zja}. 
Only for the Catania model the specific uncertainty of the model itself are also included in the uncertainty band~\cite{Minissale:2020bif}. In the bottom panels, the ratios of the various models and of the data to the Catania prediction are shown. The expectations of the models differ significantly, even by orders of magnitude, demonstrating the sensitivity of the measured ratios to the implementation of the charm hadronization process in the models. 
The Monash~\cite{Skands:2014pea} and CR-BLC~\cite{Christiansen:2015yqa} tunes of PYTHIA~8, as well as the QCM~\cite{Song:2018tpv} model underestimate the data significantly. The Monash tune expects a $\rm BR(\Omegac \rightarrow \Omega^-\pi^+)\times \Omegac/\Dzero$ ratio increasing with $\pt$ from about $4\times10^{-7}$ to about $1\times10^{-5}$. The CR-BLC model enhances the ratio by a factor of 12 to 34 with respect to the Monash tune. The prediction of the QCM is larger than that of the CR-BLC model, but it is lower than the data by more than 1.8$\sigma$. The Catania model~\cite{Minissale:2020bif} is again consistent with the data. 
The \Omegac/\Dzero ratio decreases with $\pt$ in the measured $\pt$ range in the CR-BLC, QCM, and Catania models, oppositely to what is expected by Monash.
The \XicPlusZero and \Omegac measurements provide important constraints to models of charm quark hadronization in pp collisions, being in particular sensitive to the description of charm-strange baryon production in the colour reconnection approach, and to the possible contribution of coalescence to charm quark hadronization in pp collisions.
It is interesting to note that in the Catania model, the contribution from fragmentation is subdominant for \XicPlusZero and \Omegac, 
and the main contribution comes from a pure coalescence mechanism, hence these baryon-to-meson ratios are very sensitive to the hadronization process and to the modelling of coalescence. 
Using the \XicZero~\cite{xic13tev} and \Lc~\cite{sigmac} measurements, the ratios $\rm BR(\Omega_c^0\rightarrow\Omega^-\pi^+)\times{\it \sigma}(\Omegac)/{\it \sigma}(\Lc)$ and $\rm BR(\Omegac\rightarrow\Omega^-\pi^+)\times{\it \sigma}(\Omegac)/{\it \sigma}(\XicZero)$ of the cross sections integrated in the $\Omegac$ measured \pt interval were calculated and are compared with the values measured in $\rm e^+e^-$ collisions at $\sqrt{s}=10.52$~GeV by Belle, obtained from the cross sections reported in Table~1 of Ref.~\cite{Niiyama:2017wpp}.
Though the limited $\pt$ and rapidity ranges of the ALICE measurement do not allow for a direct comparison of the pp and $\ee$ data, the ratios measured by ALICE as reported in Tab.~1 of Ref.~\cite{ALICE:2022cop} are larger by a factor of $8.7\pm2.2{\rm (stat.)}\pm0.9{\rm (syst.)}$ and $4.7\pm1.3{\rm (stat.)}\pm0.5{\rm (syst.)}$ for the $\rm BR(\Omega_c^0\rightarrow\Omega^-\pi^+)\times{\it \sigma}(\Omegac)/{\it \sigma}(\Lc)$ and $\rm BR(\Omegac\rightarrow\Omega^-\pi^+)\times{\it \sigma}(\Omegac)/{\it \sigma}(\XicZero)$, respectively. 
This difference represents further evidence that the hadronization process differs in pp and $\ee$ collisions and is sensitive to the density of quarks, colour charges, and system size.

\begin{figure}[ht!]
    \begin{center}
    \includegraphics[width = 0.47\textwidth]{ 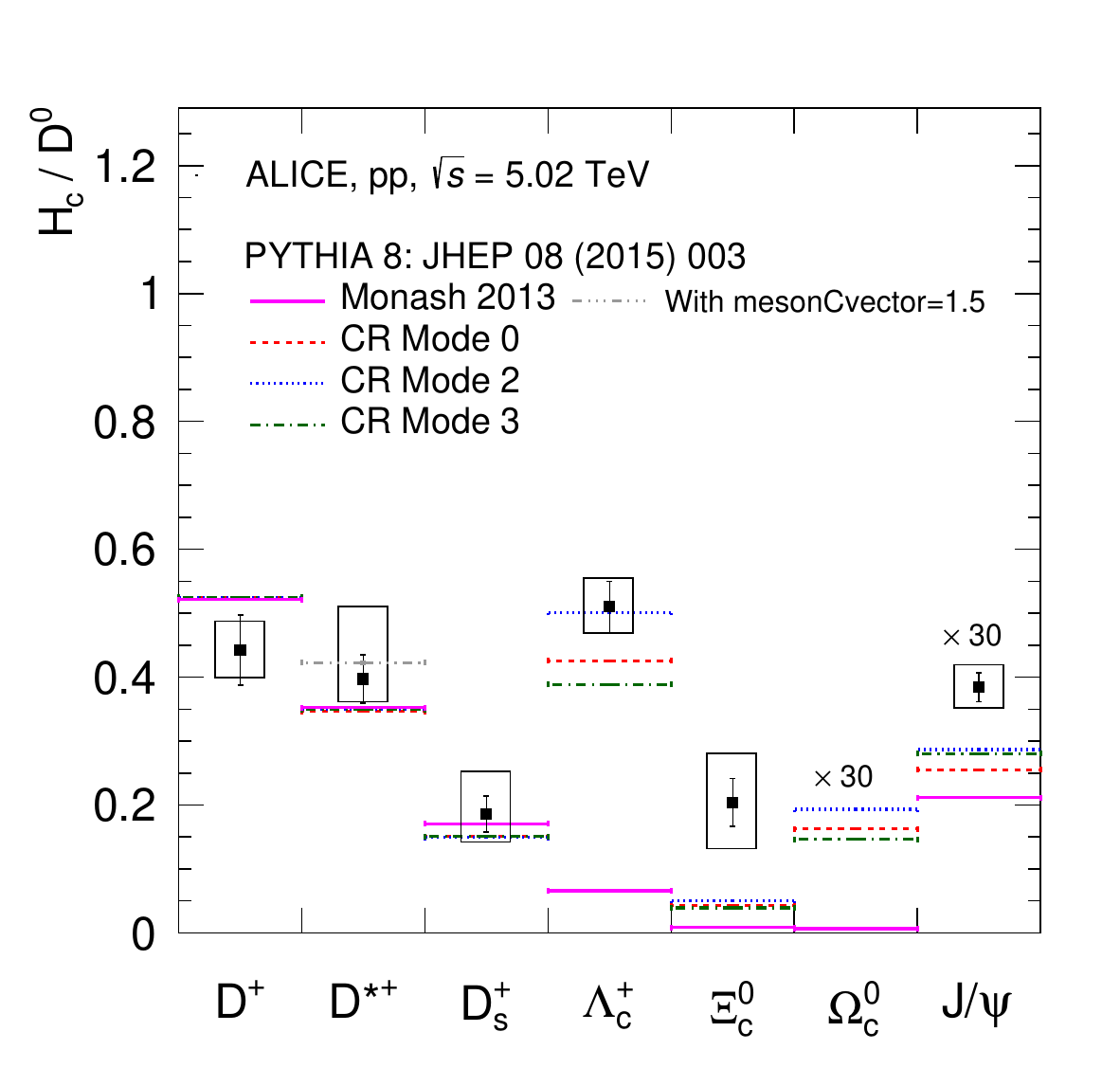}
    \includegraphics[width = 0.49\textwidth]{ 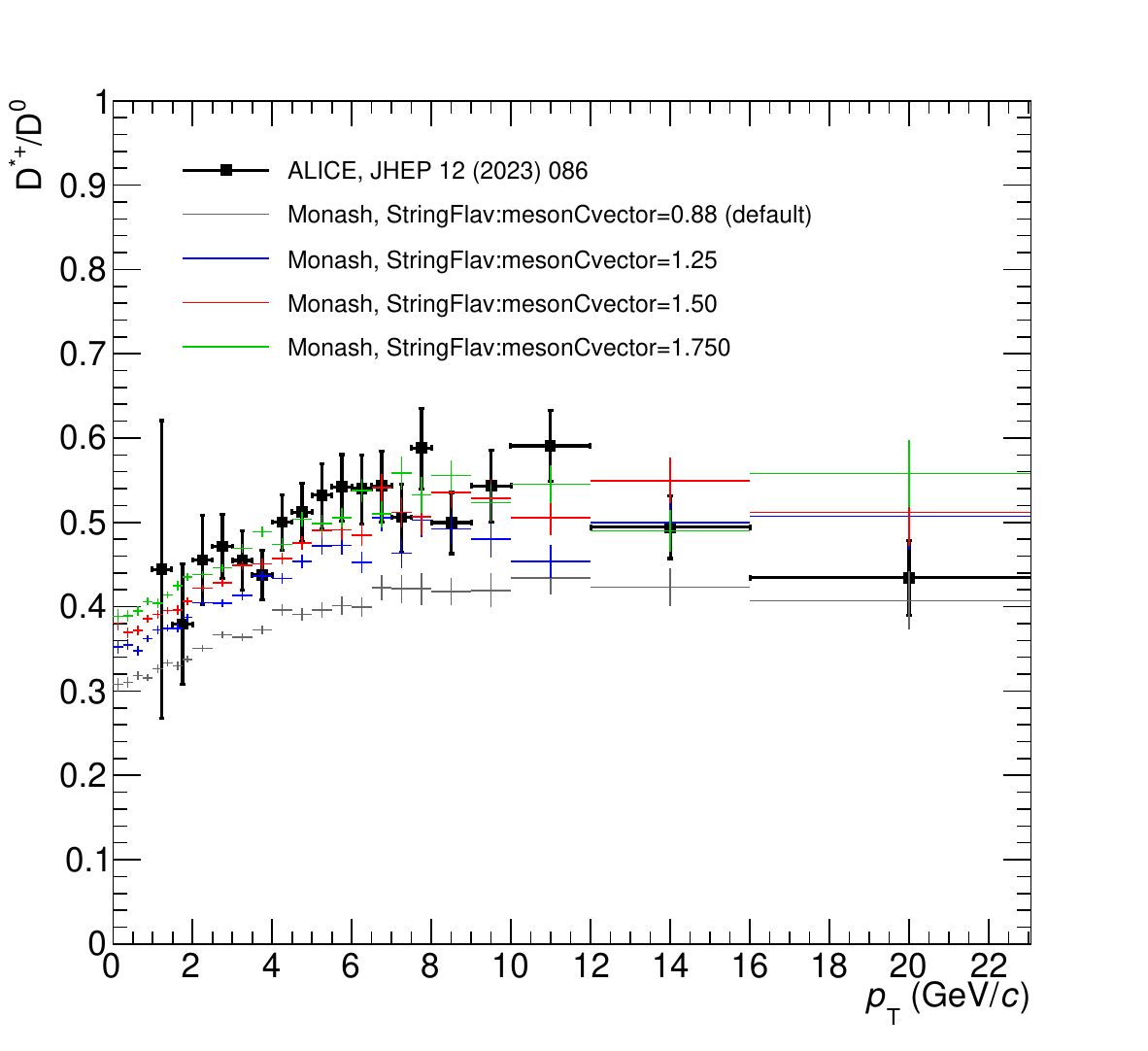}
    \end{center}
    \caption{\pt integrated production cross sections of the various charm meson and baryon species per unit of rapidity at midrapidity normalized to that of the \Dzero meson measured by ALICE in pp collisions at $\s = 5.02~\TeV$~\cite{ALICE:2021dhb}. The measurements are compared with different tunes of PYTHIA 8 calculations. The grey marker shows the values of \Dstar/\Dzero ratio obtained with Monash 2013 PYTHIA8 tune after setting the parameter {\sffamily StringFlav:mesonCvector} to 1.5, which, as shown in the right panel, allows to better describe the \pt-differential \Dstar/\Dzero ratio measured by ALICE in pp collisions at $\s=13$~\TeV~\cite{ALICE:2023sgl} (see text for details).    
    }
    \label{ratiotoDo}
\end{figure}

The low-\pt charm baryon enhancement is reflected in the \pt-integrated yields and their ratios to the \Dzero meson. 
The left panel of Fig.~\ref{ratiotoDo} shows for pp collisions at $\s = 5.02~\TeV$ the \pt-integrated production cross sections at midrapidity per unit of rapidity for various open- and hidden-charm mesons (\Dplus{}, \Ds{}, \Dstar{}, and \Jpsi{})~\cite{Acharya:2019mgn,ALICE:2021mgk,Acharya:2019lkw} and baryons (\Lc{} and \XicZero{})~\cite{ALICE:2020wfu,Acharya:2021dsq}, taking the average of particles and antiparticles, and normalizing to the \Dzero{} meson. The experimental data is compared with results of pp simulations done with PYTHIA~8 using both the Monash tune~\cite{Skands:2014pea} and the tunes implementing CR-BLC~\cite{Christiansen:2015yqa}. For the open charm meson ratios, the PYTHIA~8 predictions obtained with the different tunes are fairly similar and describe the data within uncertainty. The value of the $\Dstar/\Dzero$ ratio obtained with Monash2013 setting $StringFlav:mesonCvector=1.5$ is also reported. As anticipated in Sect.~\ref{sec:MCmodels}, values of this parameter around 1.5 or larger are preferable to better describe the data, as shown in the right panel of the same figure for the \pt-differential \Dstar/\Dzero ratio. However, this is not an artefact of the modelling of $\Dstar$ \pp collision environments. A discrepancy is existent also when considering \ee collisions data from LEP and is a consequence of the parameters \texttt{StringFlav:mesonCvector} and correspondingly \texttt{StringZ:rFactC} being mistuned. The default value of 0.88 for \texttt{StringFlav:mesonCvector} should instead be closer to $\sim$1.5. Contrary to the meson case, and as expected from what discussed about the low-\pt trend of the baryon-to-meson ratios shown in Fig.~\ref{BMR}, significant differences between the model expectations and data are observed when considering the baryon-to-\Dzero ratios. The Monash 2013 tune is observed to underestimate the $\Lc/\Dzero$ and $\XicZero/\Dzero$ ratios by nearly $8\,\sigma$ and $2.3\,\sigma$, respectively. Comparatively, the CR tunes provide larger baryon-to-meson ratios with the Mode~2 tune well reproducing the $\Lc/\Dzero$ ratio. Nonetheless, they still underestimate the $\XicZero/\Dzero$ ratio by about $2\, \sigma$, which again is unsurprising when looking at the \pt distributions in Fig.~\ref{BMR}.
For the $\Jpsi/\Dzero$ ratio the CR tunes provide a better description than the Monash tune, however, all PYTHIA~8 tunes underestimate the measurement. One should note that in the simulations, as in the experimental measurement, the \Jpsi{} cross section consists of the prompt and beauty feed-down contributions. 
From these results, one can conclude that the high baryon-to-meson ratios (\LcD, \Sigmac/\Dzero, \XicPlusZero/\Dzero, and \Omegac/\Dzero), with magnitudes much larger than those of FF ratio measured in \ee{} collisions, clearly indicate that the hadronisation process is not universal and that modelling it solely with FF and FFunc determined from \ee data fails already in pp collisions.
The right panel of Fig.~\ref{ratiotoDo} shows the same experimental data compared in addition to the SHM+RQM and Catania models. For the SHM+RQM for a hadronisation temperature $T_{\rm h} =$ 170 MeV. Similarly to what is observed for the \pt differential ratios, the SHM+RQM model describes the \Lc measurements but underestimates the \XicZero one. On the other hand, the Catania model overpredicts the production of the \Lc baryon but captures the production of the \XicZero baryon.
For the Catania model the $\Dstar/\Dzero$ value shown is the one for the mesons coming from coalescence, because the fragmentation functions employed are inclusive and give only the total contribution to D meson including the $\Dstar$ feed-down. The significantly large $\Lc/\Dzero$ is due to the combination of both a lower $\Dzero$ and a larger $\Lc$ yield with respect to experimental data. However, it has to be noticed that the value in the plot is the original prediction employing the width corresponding to the mean square radii of hadrons from the quark model, hence without any free parameter; a moderate modification of the width parameter would allow to get quite close to the data.

\begin{figure}[ht!]
    \begin{center}
    \includegraphics[width = 0.45\textwidth]{ 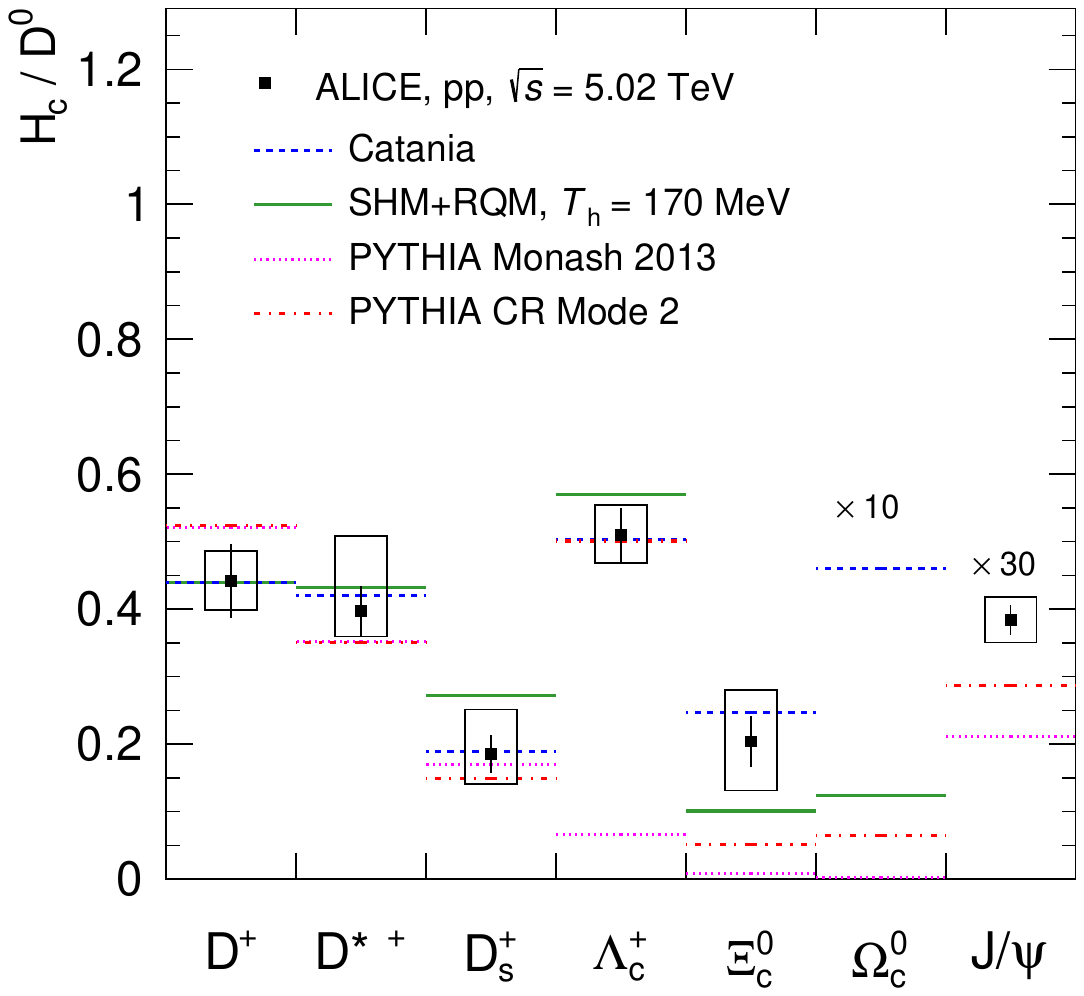}
    \includegraphics[width = 0.52\textwidth]{ 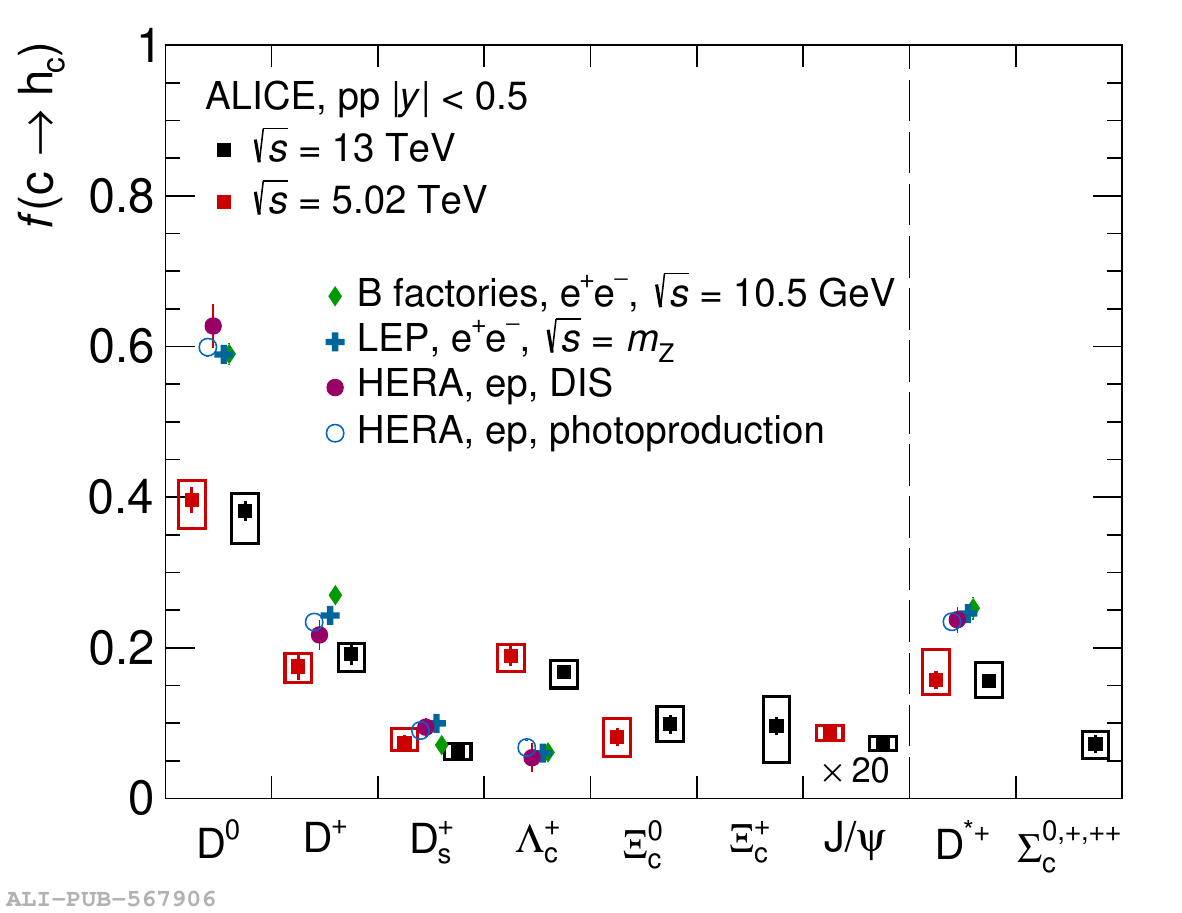}
    \end{center}
    \caption{Left: same as in left panel of Fig.~\ref{ratiotoDo} but with additional comparison to the SHM+RQM~\cite{He:2019tik} and Catania~\cite{Minissale:2020bif} models. Right: charm-quark fragmentation fractions into charm hadrons measured in pp collisions at $\s = 5.02~\TeV$~\cite{ALICE:2021dhb} and at $\s = 13~\TeV$~\cite{ALICE:2023sgl} in comparison with experimental measurements performed in \ee{} collisions at LEP and at B factories, and in ep collisions at HERA~\cite{Lisovyi:2015uqa}. 
    }
    \label{FFfunc}
\end{figure}

The charm FF, f($\rm c \rightarrow H_c$), measured by ALICE in pp collisions at \s = 5.02 TeV~\cite{ALICE:2021dhb} and at $\s = 13~\TeV$~\cite{ALICE:2023sgl}, are displayed in Fig.~\ref{FFfunc} together with values derived from experimental measurements performed in \ee{} collisions at LEP and B factories and in ep collisions~\cite{Lisovyi:2015uqa}. 
An increase by about a factor of 3.3 of the charm FF to the \Lc{} baryon from \ee{} and ep to pp collisions, with a concomitant decrease by about a factor 1.4--1.2 for the FF to D mesons is observed. The significance of the difference is about $5\,\sigma$ for \Lc{} baryons and $6\,\sigma$ for \Dzero mesons. 
In previous measurements in \ee{} and ep collisions, no value for the charm FF to \XicPlusZero{} was obtained. In the analysis performed in Ref.~\cite{Lisovyi:2015uqa} to estimate charm FF, the \XicZero and \XicPlus yields were assumed to be equal and estimated according to the assumption $f({\rm c} \rightarrow \XicPlus{})/f({\rm c} \rightarrow \Lc{})$ = $f({\rm s} \rightarrow \Xi^{-})/f({\rm s} \rightarrow \Lambda^{0}) \sim$  0.066~\cite{Lisovyi:2015uqa}. The fraction $f({\rm c} \rightarrow \XicZero{})$ was measured for the first time in pp collisions by ALICE, which for the $\XicZero/\Lc$ FF ratio obtained the value $f({\rm c} \rightarrow \XicZero{})/f({\rm c} \rightarrow \Lc{})$ = $0.39 \pm 0.07(\mathrm{stat})^{+0.08}_{-0.07}(\mathrm{syst})$~\cite{Acharya:2021dsq}. The probability of a c quark to hadronize in a \XicZero{} baryon is measured to be comparable to the probability to hadronize in a \Ds{} meson, suggesting
that the sum of the ${\rm c}\rightarrow \XicZero$ and ${\rm c} \rightarrow \XicPlus$  fragmentation fractions could be larger than the ${\rm c}\rightarrow \Ds$ one.
Recently, the ALICE Collaboration measured the charm-quark fragmentation fractions in pp collisions at \s = 13 TeV~\cite{ALICE:2023sgl}. In
addition to the charm-quark fragmentation fractions into \Dzero, \Dplus, \Ds, \Dstar, \Lc, and \XicZero hadrons, also the results for the fragmentation into J$/\Psi$ mesons and \XicPlus and \Sigmac baryons were reported. The results are compared with those in pp collisions at \s = 5.02 TeV and no significant energy dependence is observed within the  uncertainties. Overall, charm quarks hadronize into baryons approximately 35-40\% of the time. These results confirm that the baryon enhancement at the LHC with respect to \ee{} collisions is caused by different hadronization mechanisms at play in the parton-rich environment produced in pp collisions. 
The difference between the fragmentation fractions measured at midrapidity in pp collisions at the LHC and those measured in \ee collisions provide strong evidence that the parton-to-hadron transition is not a universal process that can be modelled only with universal (colliding-system independent) fragmentation functions. 

\begin{figure}[ht!]
    \begin{center}
        \includegraphics[width = 0.46\textwidth]{ 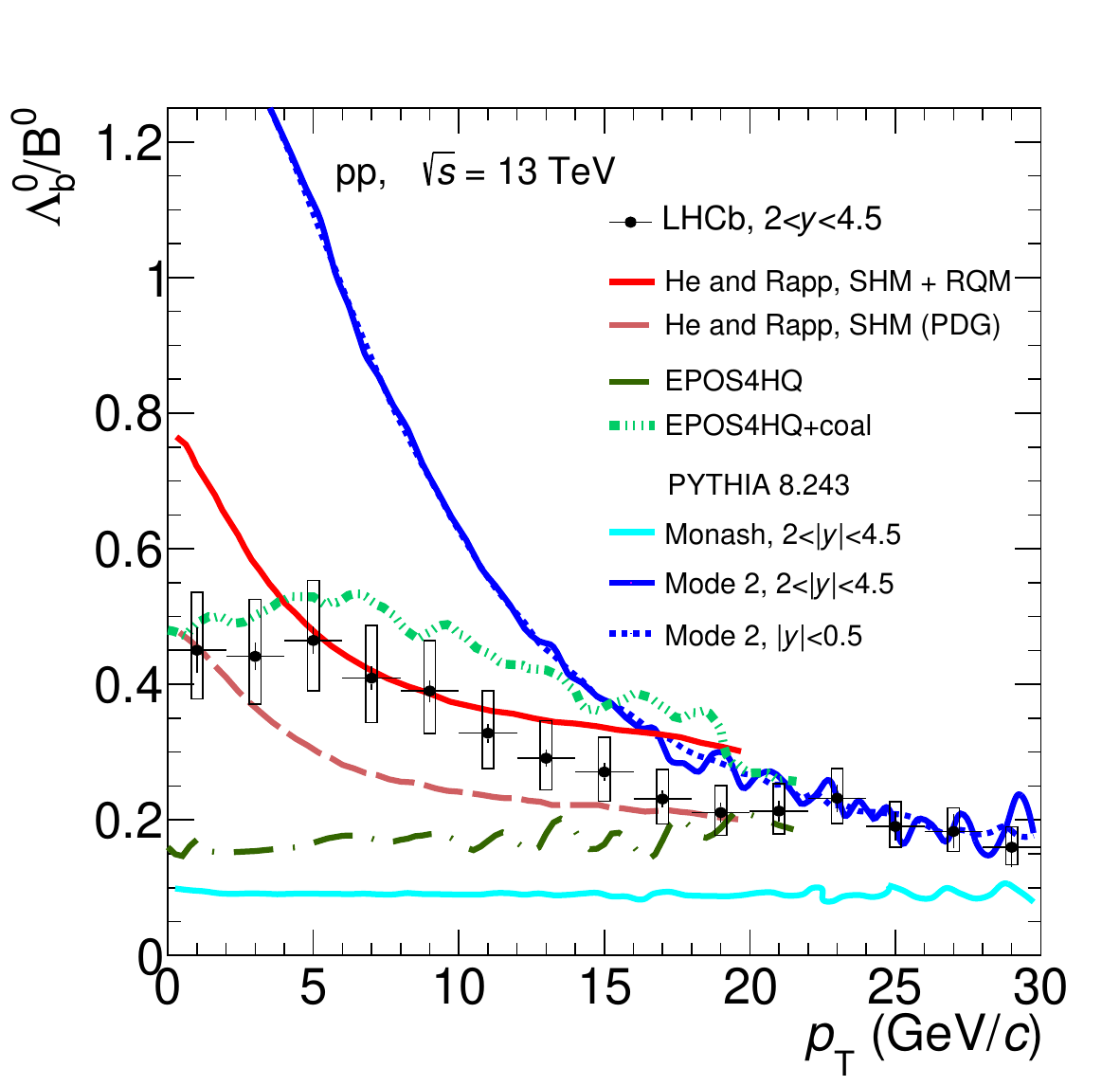}
    \includegraphics[width = 0.5\textwidth]{ 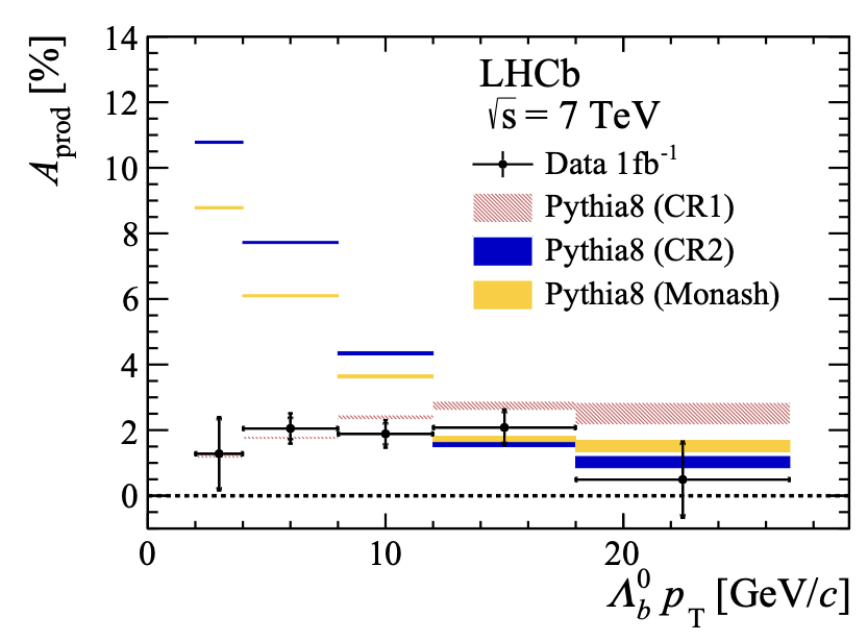}
    \end{center}
    \caption{Left panel: ratios $\mathrm{\Lb}/\mathrm{B^{0}}$ as a function of \pt measured at forward rapidity in pp collisions at $\sqrt{s}$ = 13 TeV by the LHCb Collaboration~\cite{LHCb:2023wbo}. The data are compared to model calculations ~\cite{He:2022tod,Christiansen:2015yqa,Skands:2014pea,Zhao:2023ucp} (see text for details). Right panel: comparison of the $\Lb$ production asymmetry measured by the LHCb Collaboration at \s = 7 TeV~\cite{LHCb:2021xyh} with PYTHIA 8 predictions~\cite{Christiansen:2015yqa,Skands:2014pea}.}
    \label{asymmetry}
\end{figure}

The LHCb and CMS Collaborations, and earlier the CDF Collaboration, made similar observations in the beauty sector when computing the $\Lb$ to $\rm B^{0,-}$ yield ratios~\cite{LHCb:2019fns, LHCb:2011leg, LHCb:2015qvk, LHCb:2014ofc,CMS:2022wkk,LHCb:2019lsv,LHCb:2023wbo,CMS:2012wje,Aaltonen:2008zd} in pp (\ppbar for CDF) collisions at various centre-of-mass energies.
As shown in the left panel of Fig.~\ref{asymmetry}, the ratio is observed to strongly depend on \pt and approaches LEP value at high \pt, which can be estimated to be $\Lb/\mathrm{B^{0}}=0.22\pm0.03$~\cite{HFLAV:2019otj}. At low \pt it is significantly higher than in \ee collisions, reaching approximately twice the value measured at LEP for both the CDF and LHCb experiments. It does not show variations with pseudorapidity, within the LHCb acceptance of pseudorapidity 2 $< \eta <$ 5. 
The ALICE Collaboration measured the production cross sections of \Dzero and \Lc hadrons originating from beauty-hadron decays at midrapidity in pp collisions at a centre-of-mass energy $\sqrt{s}$ = 13 TeV~\cite{ALICE:2023wbx}. A trend consistent with the one of $\Lb/\mathrm{B^{0,-}}$ measured by LHCb at forward rapidity is observed.
In the left panel of Fig.~\ref{asymmetry} the $\rm \Lb/B^0$ measured by the LHCb Collaboration~\cite{LHCb:2023wbo} is compared with PYTHIA 8 calculations using different tunes. The cyan line represents the Monash tune computed for the rapidity window 2--4.5, while the blue solid line shows one of the QCD-based CR-BLC tunes in the same rapidity range. This comparison confirms that the experimental measurement is larger than what is expected by models based on fragmentation tuned on \ee collisions. On the other hand, given the inclusion of junction string topologies and hence junction baryons in the CR-BLC PYTHIA~8 models, though the trend of producing an enhancement of \Lb at low \pt and reducing to LEP values at high \pt is predicted, the magnitude of the enhancement at low \pt is too large and overpredicts the experimental data. Such a discrepancy remains an active question in junction string modelling.
An additional point that is important to highlight is the comparison between the expectations of PYTHIA 8 with CR-BLC in different rapidity ranges shown in the same figure. Simulations of PYTHIA 8 with CR-BLC for $|y| < 0.5$ (dashed blue line) are compared to the forward rapidity one (2 $< |y| <$ 4.5, solid blue line). The results indicate that the model does not expect any dependence on rapidity. 
Additionally, the expectations of a SHM approach with and without the set of b-hadron states predicted by the RQM to exist beyond the currently measured spectrum listed by the particle data group are also shown~\cite{He:2022tod}. The substantial gap between the LHCb data and the model calculation obtained with the PDG scenario (dashed dark-red line) is largely overcome by the feeddown of the ``missing'' baryons included in the RQM calculation (solid red line), leading to a better description of the experimental data at intermediate \pt, while overpredicting the measurements at low and high \pt.
It would also be interesting to see model calculations including coalescence, like the Catania model, to verify whether different baryon enhancements, both in the charm and beauty sector, might be expected in different rapidity windows.
Future measurements of other beauty baryons ($\rm{\Xi_b}$, $\rm{\Omega_b}$) are necessary to investigate a possible enhancement in pp collisions for strange-beauty baryons. 

 The LHCb Collaboration also published differential measurements of the asymmetry between $\Lb$ and $\overline{\Lambda}_b^0$ baryon production rates in pp collisions at centre-of-mass energies of $\sqrt{s}$ = 7 TeV and $\sqrt{s}$ = 8 TeV~\cite{LHCb:2021xyh}. The production asymmetry is defined according to the quantity $A_{\rm prod}$, which represent the relative difference between the production
rates of $\Lb$ and $\overline{\Lambda}_b^0$.
The production asymmetry is measured both
in intervals of rapidity in the range 2.15 $< y < $ 4.10 and transverse momentum in 2 $< \pt <$ 27 GeV$/c$. The results are incompatible with symmetric production with a significance of 5.8$\sigma$ assuming no CP violation in the decay. The asymmetry is increasing with rapidity and there is evidence for a trend with a significance of 4$\sigma$. The comparison between data and various PYTHIA 8 models is shown in the right panel of Fig.~\ref{asymmetry}. 
Among the several tunes of PYTHIA, only that with CR-BLC approximation (red boxes) shows a good agreement with the measurement. The Monash tune (yellow boxes) and the gluon move CR (blue boxes) predict asymmetries that are too large, exhibiting the strongest deviation at low transverse momentum and for the widest rapidity range investigated~\cite{LHCb:2021xyh}. In PYTHIA, a key source of low \pt $\Lb$ production in pp collisions at large rapidity is the combination of a b quark with the beam remnants. Obviously, the same does not apply to the $\overline{\Lambda}_b^0$ as there are not antiquarks in proton beam remnants. The  CR-BLC adds large amounts of low-\pt junctions, which form  $\overline{\Lambda}_b^0$ and $\Lb$ baryons in equal amounts, diluting the asymmetry.

In this context, it is also worth noting that the LHCb Collaboration has measured the \Ds production asymmetry in pp collisions. The \Ds meson does not contain any of the proton's valence quarks, which means that the production asymmetry is in this case not influenced by the valence quarks of the colliding protons. PYTHIA simulations have shown a strong dependence on
both \pt and $y$, whereas the experimental measurements do not~\cite{LHCb:2018elu}.

\subsection{Measurements as a function of multiplicity}
\label{sect:expppvsMult}

The observation of substantially different charm baryon-to-meson ratios in pp collisions compared to \ee collisions evidences differences in the hadronization process in the two systems. Though at high-\pt the ratios seem to approach system-universal values, suggesting the presence of an energy scale delimiting the modification of the hadronization process, the absence of clear differences between the $\Lc/\Dzero$ ratios measured at LEP and B-factory energies and the similarity of the trends observed in pp collisions in the charm and beauty sectors makes it extremely difficult to reconcile pp and \ee data ascribing the difference to a kinematic effect related to the energy of the initial-parton or of the jet, without implicating different properties of the systems. One is instinctively tempted to relate the observed differences to the typically larger number of quarks and gluons produced in hadronic collisions. The parton density is strictly connected to the \enquote{hadronic activity}, which is encoded in the measurable hadron multiplicity. Inevitably, the study of the evolution of the heavy-flavour baryon-to-meson ratios with particle multiplicity was expected to provide important information to better understand the modification of the hadronization process and to model it. As it will be discussed later in the section, it is important to stress that it is not just the number of hadrons produced that differentiates pp and \ee collisions, but rather the way in which hadrons can be formed. It is interesting to compare the multiplicities of charged particles measured in different collision systems and at different energies. A compilation of average charged particle multiplicities measured in hadronic and leptonic collisions is reported in Fig.~19.6 of Ref.~\cite{Workman:2022ynf}. A precise comparison should take into account and correct for the possible inhomogeneities in the criteria adopted to define the set of particles that are counted, like possible different upper limits on particles lifetimes, kinematic cuts not corrected for, event classes (e.g. inelastic or non-single diffractive for hadronic collisions). Implementing these corrections is far beyond the goal of the qualitative comparison that we want to make here. Values of average multiplicity around 28, with quite large variances~\cite{DELPHI:2000ahn}, are reached at top LEP energies. This number is close to the multiplicities of charged particles with $|eta|<2.5$ in pp collisions at LHC energies (2.36 - 13~\TeV) and to the total one in $\ppbar$ collisions at $\sqrt{s}\approx 540$~\GeV measured by UA5 experiment~\cite{UA5:1985hzd}. In pp collisions at LHC, measurements of the total average multiplicities are not available: without aiming at a precise estimate, extrapolations to infinite pseudorapidity range can be done by using either UA5 data~\cite{UA5:1988gup} or \pythia to estimate the ratio of the multiplicity value in a given finite $\eta$ interval ($|eta|<0.5$ or $|eta|<2.5$) to the multiplicity obtained without $\eta$ cuts. In this way, total average multiplicities in pp collisions ranging from about 60 to about 85 from $\sqrt{s}=2.36$~\TeV to $\sqrt{s}=13$~\TeV can be estimated. Therefore, the total average multiplicities at LEP are just a factor 2-3 smaller than those at the LHC. A similar factor applies between the average multiplicities at LEP and at B factories where, as discussed in Sect.~\ref{sect:expEE}, consistent values of $\Lc/\Dzero$ ratio are found. Given that at LEP a considerable fraction of particles is part of high-energy jets (most charged particles are within $\pm 3$ units in rapidity calculated w.r.t. the sphericity axis~\cite{ALEPH:2003obs}), "local" multiplicities in finite $\eta$ intervals are actually comparable to those of LHC. These considerations should make it clear that it is unlikely that the differences among the relative charm-hadron abundances observed in \ee and pp collisions can be related to a simple increase of parton and charged-particle multiplicity without considering the different ways by which the parton and hadrons are formed. The connection of charged-particle multiplicity with MPI in hadronic collisions is a fundamental motivation for studying how charm-quark hadronisation evolves with multiplicity in these collisions.  As discussed in Sect.~\ref{sec:pythiaExt}, the concurrence of several initial scatterings, i.e. of MPI, allow for more and topologically different colour connections, which can impact the hadronization process.
Also the hadronization via coalescence, which requires multiple quark wavefunctions to overlap in position and momentum space, 
is expected to progressively emerge as hadron-formation process with increasing number of quarks produced in the collision. 
Moreover, given the enhancement of multi-strange hadrons observed in high-multiplicity pp collisions ~\cite{ALICE:2016fzo}, one naturally wonders whether the production of heavy-flavour hadrons containing a strange quark~\cite{ALICE:2016fzo} also increases as particle multiplicity increases and whether, if present, this increase should be related to heavy-flavour baryon production.

\begin{figure}[ht!]
    \begin{center}
    \includegraphics[width = 0.8\textwidth]{ 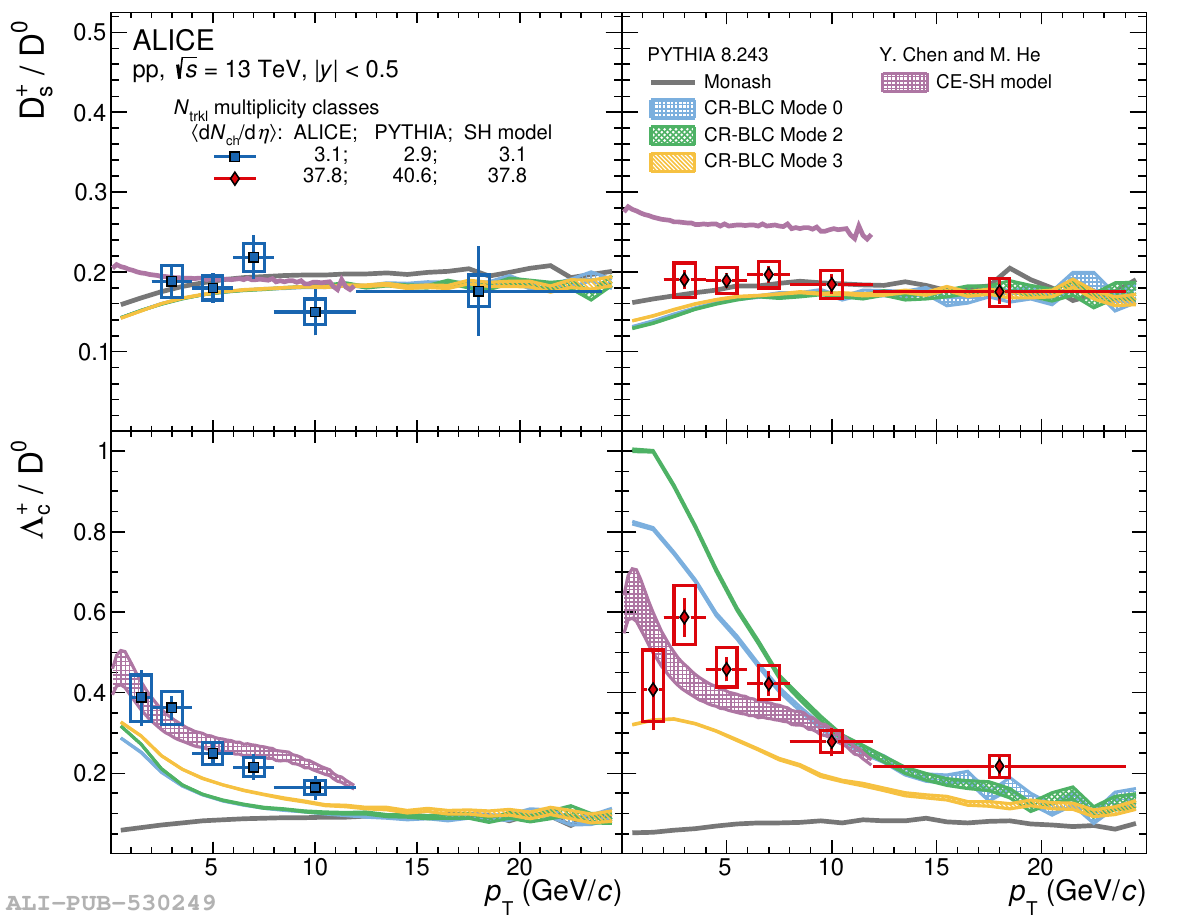}
    \end{center}
    \caption{The $\Ds/\Dzero$ (top) and $\Lc/\Dzero$ (bottom) ratios measured at midrapidity by ALICE in pp collisions at $\sqrt{s}=13$ TeV in the lowest (left) and highest (right) midrapidity multiplicity classes inspected~\cite{ALICE:2021npz}. The measurements are compared to PYTHIA 8 predictions with the Monash~\cite{Skands:2014pea} and the CR-BLC~\cite{Christiansen:2015yqa} tunes, and the CE-SH model~\cite{Chen:2020drg}, calculated in similar multiplicity classes. Figure from Ref.~\cite{ALICE:2021npz}.}
    \label{fig:vsmultalice}
\end{figure}

\begin{figure}[ht!]
    \begin{center}
    \includegraphics[width = 1\textwidth]{ 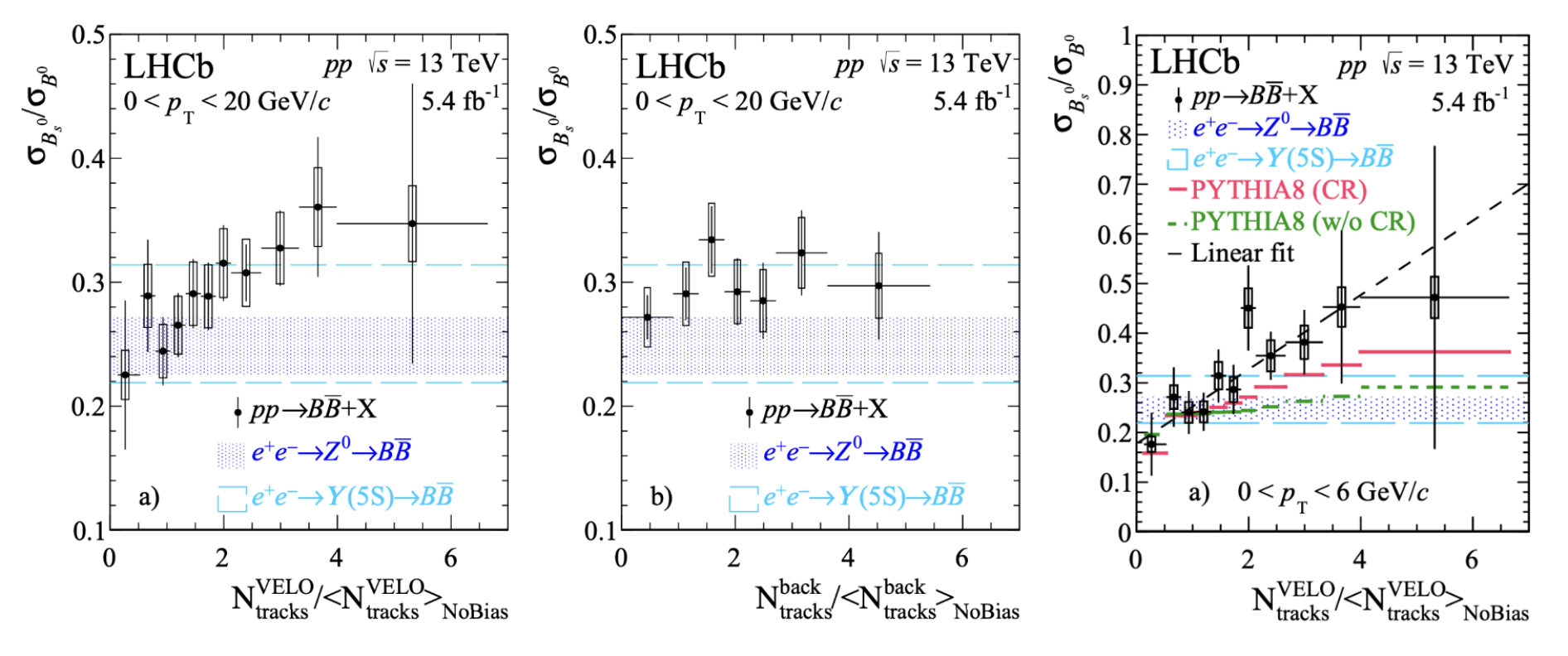}
    \end{center}
    \caption{The left and middle panel display the ratio of cross sections $\sigma_{\rm{B^0_s}}/\sigma_{\rm{B^0}}$ versus the normalized multiplicity of all VELO tracks and backward VELO tracks, respectively. The right panel shows the comparison with PYTHIA 8 calculations for the $0<\pt<6$~\GeVc interval. The horizontal bands show the values measured in \ee collisions. Figure from Ref.~\cite{LHCb:2022syj}.}
    \label{bosb0}
\end{figure}

\begin{figure}[ht!]
    \begin{center}
    \includegraphics[width = 0.57\textwidth]{ 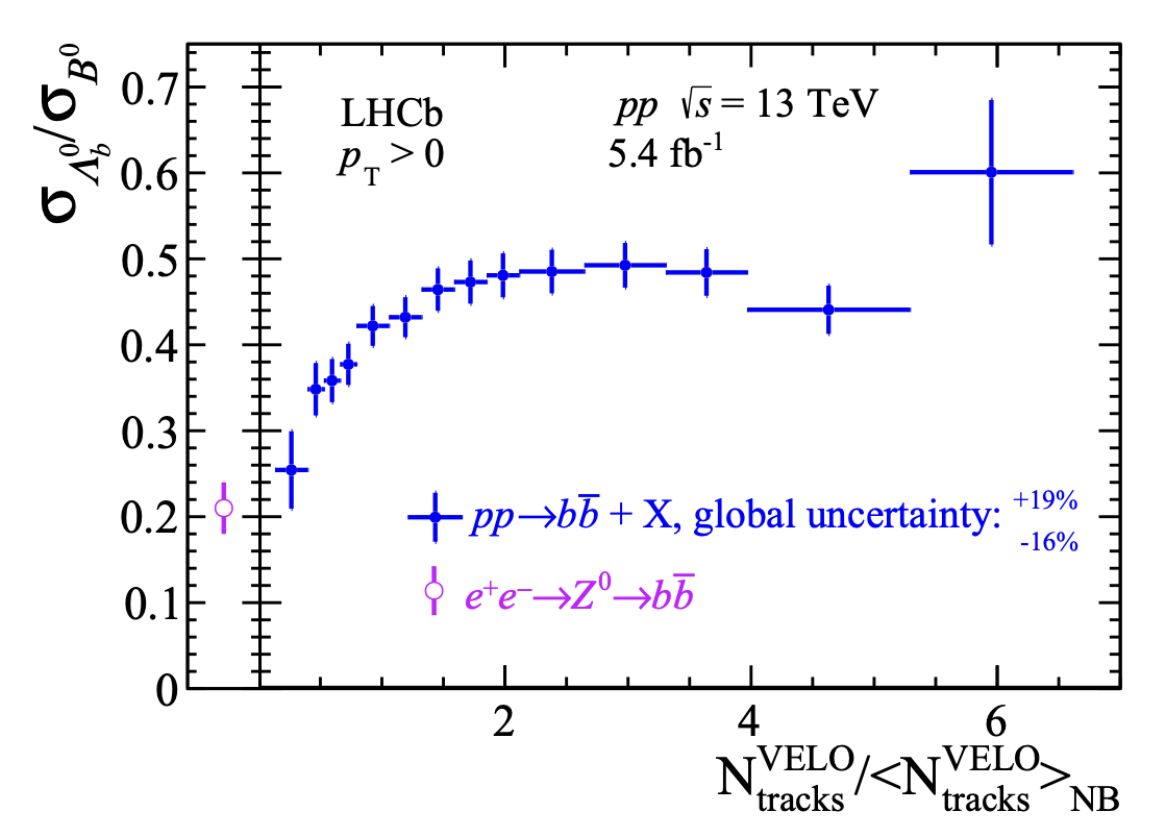}
    \end{center}
    \caption{Ratio of cross sections $\sigma_{\Lb}/\sigma_{\rm B^0}$ measured by LHCb in $2<|y|<4.5$ in intervals of the multiplicity of all VELO tracks, which is reported on the $x$ axis after scaling it by the average multiplicity. The purple point indicates the value measured in \ee{} collisions at LEP. Figure from~\cite{LHCb:2023wbo}.}
    \label{Lbb0}
\end{figure}

The evolution of the $\Ds/\Dzero$, $\rm{B^0_s}/\rm{B^0}$, $\Lc/\Dzero$, and $\Lb/{\mathrm B^{0}}$ cross-section ratios with charged-particle multiplicity was studied in pp collisions at $\sqrt{s}=13$ TeV. Figure~\ref{fig:vsmultalice} shows ALICE measurements of $\Ds/\Dzero$ and $\Lc/\Dzero$ ratios at low and high multiplicities~\cite{ALICE:2021npz}, while the $\rm{B^0_s}/\rm{B^0}$ and $\Lb/{\mathrm B^{0}}$ trends measured by LHCb~\cite{LHCb:2022syj,LHCb:2023wbo} are shown in Fig.~\ref{bosb0} and Fig.~\ref{Lbb0}, respectively. As discussed in what follows, these results provide important experimental constraints on the modification of the hadronization process from low to high density partonic environments.


The $\Ds/\Dzero$ ratio at midrapidity does not change significantly from the lowest (top left panel of Fig.~\ref{fig:vsmultalice}) to the highest (top right panel) multiplicity class studied by ALICE. Within the uncertainties, it is also independent of \pt in the measured \pt interval and in line with \ee values. 
The $\rm{B^0_s}/\rm{B^0}$ ratio measured by LHCb at forward rapidity, 2 $< y <$ 4.5, may instead indicate a dependence on multiplicity of the rate of $\rm{B^0_s}$ production relative to $\rm{B^0}$ one. At high multiplicity, LHCb data favour values higher than those found in \ee collisions at LEP, which are compatible and more precise than those obtained at B factories. In collisions with relatively low charged-particle multiplicity, and for B mesons with \pt $>$ 6 GeV$/c$, the $\rm{B^0_s}/\rm{B^0}$ ratio is consistent with that measured at LEP. 
In order to properly interpret the difference of the trends observed for $\Ds/\Dzero$ and $\rm{B^0_s}/\rm{B^0}$ at low \pt, few caveats must be taken into account. First, in the $\Ds/\Dzero$ case the measurement is limited to $\pt>2$~\GeVc. As mentioned, at \pt larger than the B-meson mass, also the $\rm{B^0_s}/\rm{B^0}$ ratio does not show a significant dependence on multiplicity. Second, the multiplicity estimators used in the LHCb analysis are the total number of charged tracks reconstructed in the VELO detector ($\rm{N_{VELO}}$ tracks), and the subset of VELO tracks that point in the backward direction, away from the LHCb spectrometer ($\rm{N_{back}}$ tracks). The $\rm{B^0_s}/\rm{B^0}$ enhancement is observed when charged-particle multiplicity is measured in the full VELO detector, while it is not observed when the charged-particle multiplicity is based on $\rm{N_{back}}$ tracks. The ALICE Collaboration also performed the measurement of the $\Ds/\Dzero$ ratio, as well as that of the $\Lc/\Dzero$ ratio discussed later, with two different multiplicity estimators using  detectors located at different pseudorapidity regions; one determined by the multiplicity of charged particles at midrapidity, thus overlapping with the pseudorapidity window in which the charm hadrons are reconstructed, and one with a significant $\eta$ gap. To be noted that in the former estimator, which is the one used for the measurements reported in Fig.~\ref{fig:vsmultalice}, the contribution of the decay particles of the reconstructed charm hadron is subtracted from the estimated multiplicity. Within uncertainties, ALICE found no differences between the $\Ds/\Dzero$ ratios measured with the two estimators~\cite{ALICE:2021npz}. 
Further guidance for interpreting the two measurements is provided by the comparison with PYTHIA expectations. The $\Ds/\Dzero$ ratios at low and high multiplicity are both described by PYTHIA 8 with Monash tune as well as by tunes with CR-BLC. The differences between the predictions are relatively small and, in the measured \pt interval, all tunes expect a tiny reduction of the $\Ds/\Dzero$ ratio from low to high multiplicity.  
The $\rm{B^0_s}/\rm{B^0}$ ratio calculated by the PYTHIA 8 event generator with and without colour reconnection is compared with the experimental measurements for the \pt interval 0--6 GeV$/c$ in the right panel of Fig.~\ref{bosb0}. Both sets of PYTHIA calculations show a rise with multiplicity, which is more pronounced when colour reconnection is included. Note that this rise in PYTHIA is not due to an explicit strangeness enhancement mechanism (e.g. Rope hadronization~\cite{Bierlich:2017sxk} or close-packing~\cite{Fischer:2016zzs}) and it is already present with the standard Lund fragmentation model. However, the exact cause of this multiplicity dependence has not yet been studied. 
Both PYTHIA 8 models reproduce the data within uncertainties though both scenarios give values that are lower than the central values of the data at high multiplicity, possibly indicating a different slope of the rise with multiplicity. At higher \pt, both PYTHIA 8 simulations well reproduce the data~\cite{LHCb:2022syj}.
The enhancement of the rate of $\rm{B^0_s}$ production relative to $\rm{B^0}$ production at low \pt might be a consequence of a bias arising from the coinciding rapidity regions between the multiplicity estimator and the measurement of interest or it might signal that the effect seen is related to the \enquote{local} multiplicity of particles surrounding the B mesons. It would be interesting to further investigate these possibilities with Run 3 data and to perform similar measurements at midrapidity and to extend the measurement of the $\Ds/\Dzero$ ratio down to $\pt=0$.

Turning to baryon-to-meson ratios, the lower panels of Fig.~\ref{fig:vsmultalice} show the \pt-differential $\Lc/\Dzero$ ratio for the lowest and highest multiplicity classes studied by ALICE. The data evidence a dependence on multiplicity, with higher values of \Lc/\Dzero ratio observed at high multiplicity. 
When comparing the highest multiplicity interval to the lowest one, the effect has a significance of 5.3$\sigma$ for 1 $< \pt <$ 12 GeV$/c$. The trends and values obtained when the multiplicity classes are defined using the two multiplicity estimators mentioned above are consistent. This indicates that the enhancement from low to high multiplicity is not induced only by the possibly different correlation of \Lc and \Dzero production with the \enquote{local} multiplicity of particles produced close to them in pseudorapidity. Such a correlation could arise either from a genuine correlation between hadronization process and the multiplicity, e.g. in case \Lc production were favoured over \Dzero one when the \enquote{local} parton density is higher, or from a correlation dictated by different average local multiplicities accompanying \Lc and \Dzero production independently from the hadronization process, e.g. if \Lc and \Dzero were produced in jets with different number of constituents. Correlations induced by a bias on the multiplicity estimate related to signal reconstruction, like the contribution of a different number of decay particles to the multiplicity estimate, are instead removed by ALICE in the analysis. 

 As expected by the \Lc/\Dzero ratio results shown in Fig.~\ref{BMR}, the Monash tune does not reproduce the $\Lc/\Dzero$ ratio in either multiplicity interval, nor does it show a strong multiplicity dependence.
 By contrast, the CR-BLC tunes describe the $\Lc/\Dzero$ decreasing trend with \pt, and are closer to the overall magnitude in both multiplicity ranges. All CR-BLC tunes show a clear dependence with multiplicity, qualitatively reproducing the trend observed in data, but the evolution predicted by the three ``modes'' is different, indicating that the data can be useful for further tuning the model.
 
The predictions of a canonical-ensemble statistical hadronization (CE-SH) model~\cite{Chen:2020drg} are also compared to the $\Ds/\Dzero$ and $\Lc/\Dzero$ ratios. In the calculation, the
underlying charm-baryon spectrum is augmented to include additional excited baryon
states predicted by RQM. On the contrary, for the $\Ds/\Dzero$ ratio the prediction obtained considering only the hadron states listed in the PDG is shown since the additional RQM states do not modify significantly the D-meson relative yields with respect to the PDG set.
The CE-SH model predicts a dependence on multiplicity for both ratios, deriving from the reduced volume size of the formalism towards smaller multiplicity combined with quantum-number conservation. The resulting modification of the $\Lc/\Dzero$ ratio reproduces the multiplicity dependence observed in data and interpreted as a decrease of the ratio at low multiplicity as a consequence of the strict baryon-number conservation. However, such behaviour is also expected for charm-strange mesons relative to non-strange charm mesons because of strangeness-number conservation but the resulting prediction reproduces the $\Ds/\Dzero$ ratio only at low multiplicity, while it overestimates it in the highest multiplicity interval. 

It is worth noticing that the measured $\Lc/\Dzero$ ratio in the lowest multiplicity class is still higher (in the measured \pt range) than the average of corresponding ratios measured in \ee collisions. As it will be further discussed in Sect.~\ref{sect:expAA}, exploring lower multiplicities and extending the measurements to \pt=0 would be of fundamental importance to verify whether the ratio approaches the value measured in \ee collisions, as the \pt-integrated $\Lb/{\mathrm B^{0}}$ ratio at forward rapidity, which will be now discussed, does.

The LHCb Collaboration recently released the measurement of the $\Lb/{\mathrm B^{0}}$ cross-section ratio as a function of \pt and multiplicity~\cite{LHCb:2023wbo}. The \pt-integrated ratio is shown as a function of relative multiplicity in Fig.~\ref{Lbb0}. In the lowest multiplicity interval, the LHCb data approaches the baryon fraction measured in \ee{} collisions at the LEP. 
The distinct rise of the baryon fraction with multiplicity reaches a plateau at $\sim$ 0.5 for
collisions that produce more than twice the average number of VELO tracks.  \\
These last two measurements show that b-quark hadronization is different in \ee{} and pp collisions, thus it is not an universal process. 
This may provide indication that quark coalescence plays an important role in forming beauty hadrons at the LHC, or that one should more carefully examine string models with multiplicity dependencies. 
The trend observed for the $\pt$-integrated $\Lb/{\mathrm B^{0}}$ ratio resembles a \enquote{turn-on} curve which may mark the transition from two different regimes, with the onset of a different hadronization mechanism in pp collisions becoming very quickly effective as soon as a minimal hadronic activity is present. It would be interesting to try correlating the observed trend with the number of MPI. 
A work, in which a set of unobserved excited beauty quark states are included, seemed to be able to qualitatively explain the system size dependence of these ratios in terms of canonical suppression of baryons and strangeness toward smaller multiplicities~\cite{Dai:2024vjy}.
It must be noted that the evolution observed for the $\pt$-integrated $\Lb/{\mathrm B^{0}}$ ratio does not contrast what was previously discussed about the evolution with multiplicity of the \pt-differential $\Lc/\Dzero$ ratio measured by ALICE at midrapidity. This will be further discussed in Sect.~\ref{sect:expAA}.

\subsection{Proton--nucleus collisions}
\label{sect:exppA}

\begin{figure}[ht!]
    \begin{center}
    \includegraphics[width = 0.5\textwidth]{ 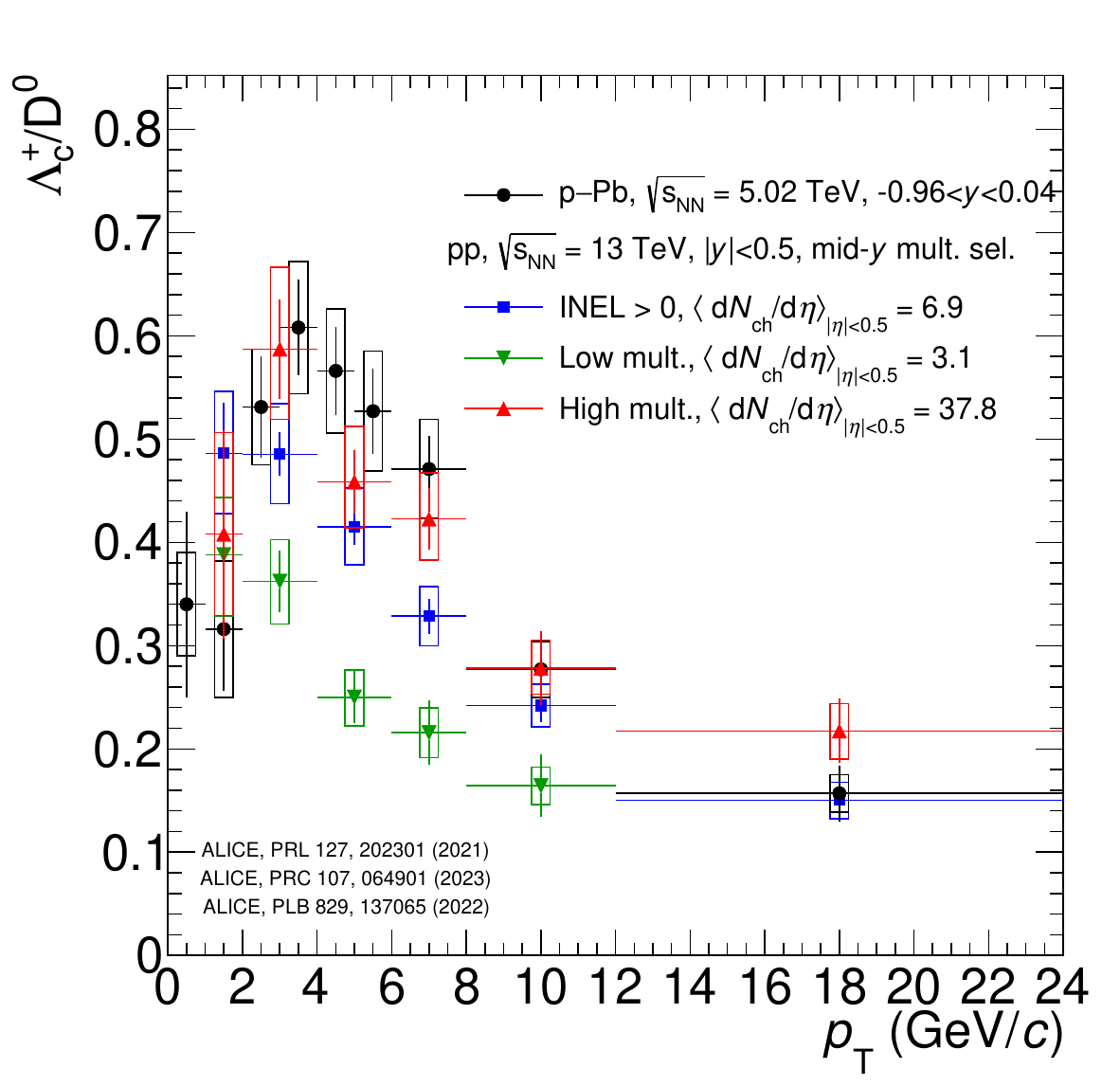}
    \includegraphics[width=0.48\textwidth]{ 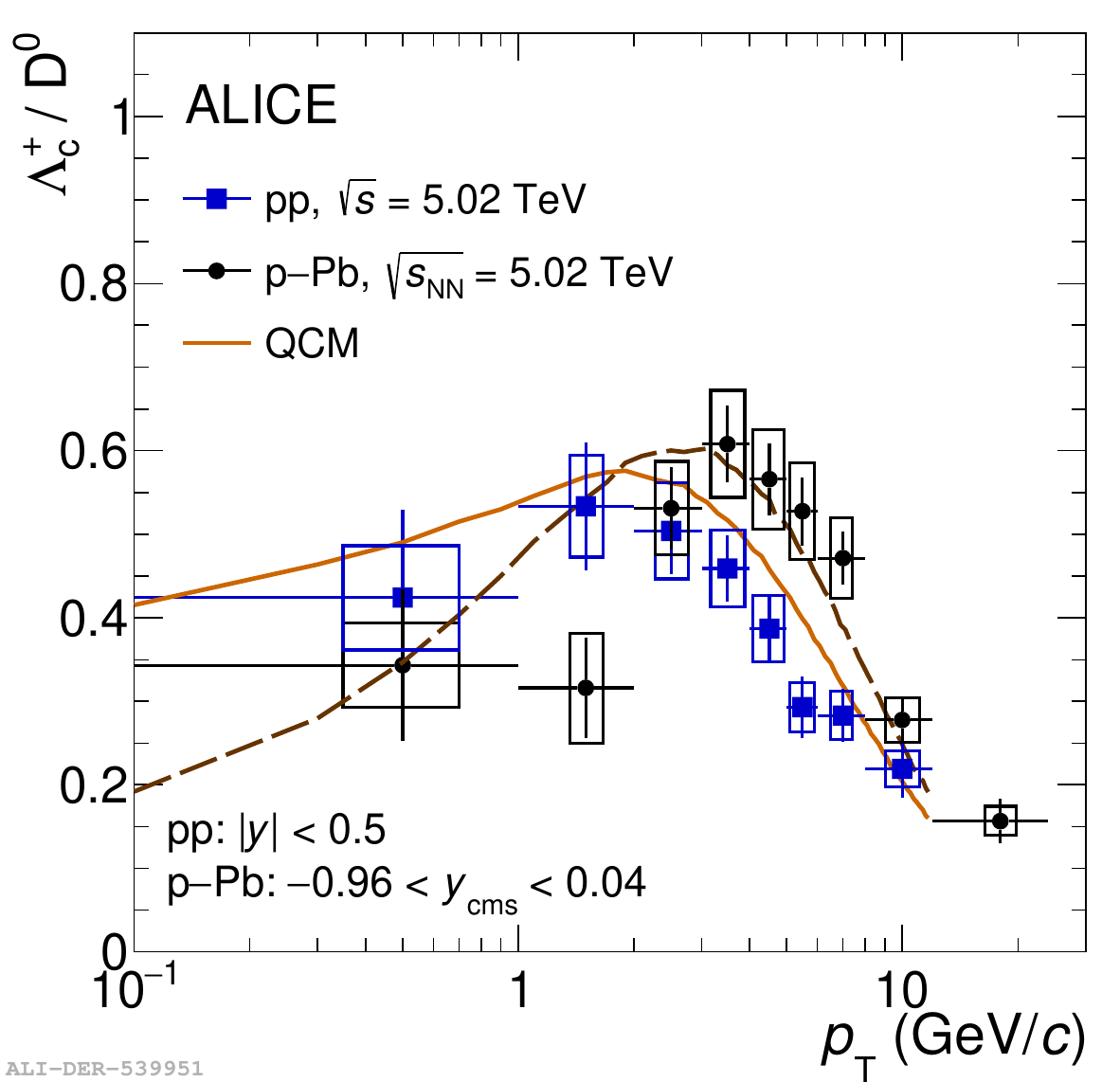}
    \end{center}
    \caption{Left: $\LcD$ ratio as a function of \pt in pp collisions (at low and high multiplicity and multiplicity integrated) and in p--Pb collisions~\cite{ALICE:2022ych,ALICE:2020wla,ALICE:2021npz}. Right: comparison of \LcD ratio in pp and p--Pb collisions with predictions from QCM model~\cite{ALICE:2022ych}.}
    \label{fig:LctoDzeroPPmultPPB}
\end{figure}

\begin{figure}[ht!]
    \begin{center}
    \includegraphics[width = 0.55\textwidth]{ 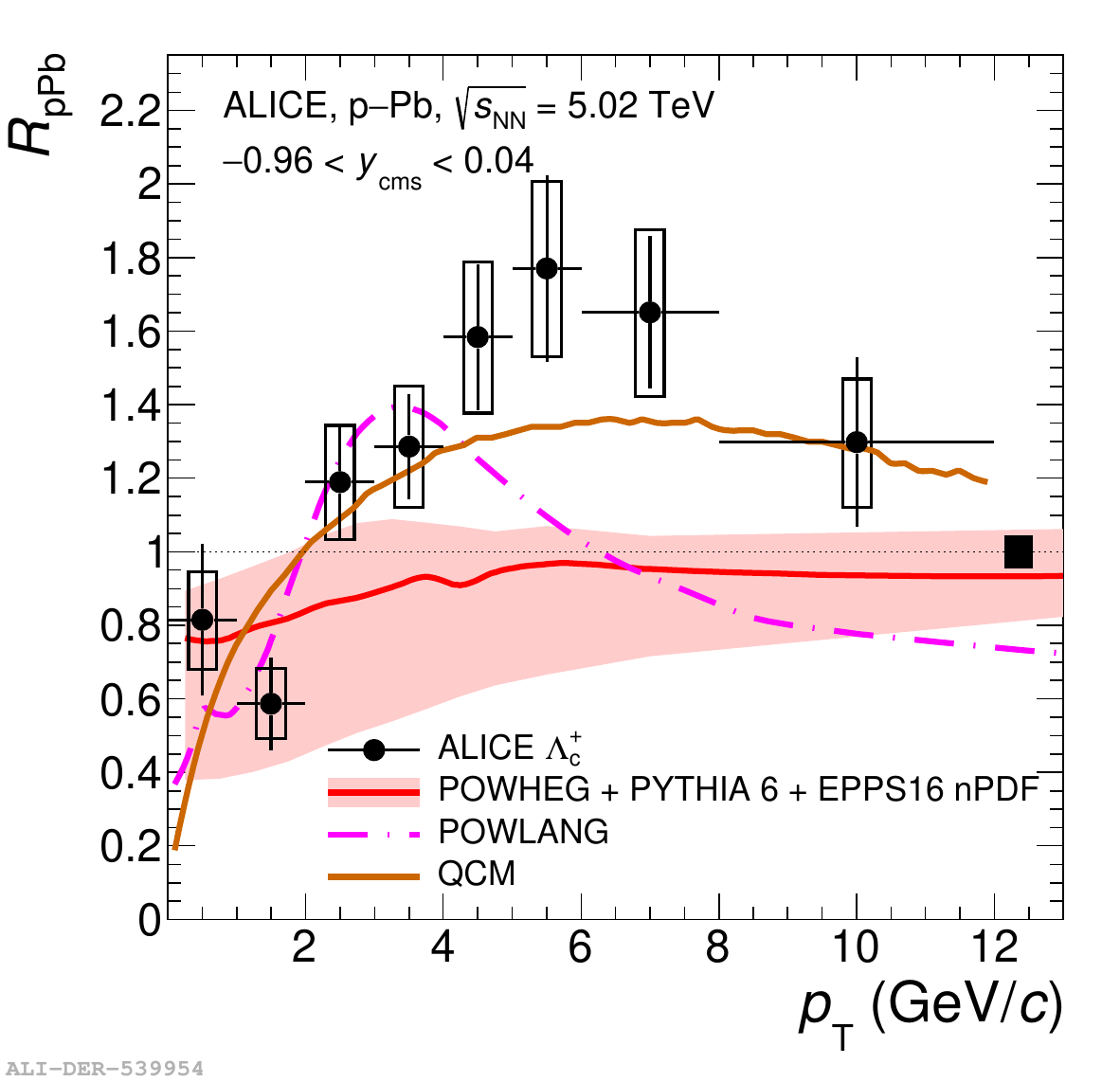}
    \end{center}
    \caption{Nuclear modification factor of $\Lc$ baryons as a function of \pt measured by ALICE~\cite{ALICE:2022ych} and compared with predictions from POWLANG~\cite{Beraudo:2015wsd}, QCM~\cite{Li:2017zuj}, and POWHEG + PYTHIA~6 + EPPS16 nPDF~\cite{Alioli:2010xd,Frixione:2007nw,Sjostrand:2006za,Dulat:2015mca,Eskola:2016oht}. See text for details.} 
    \label{fig:LcRPBcharmFlowpPb}
\end{figure}

The influence of a hadronic environment on the hadronization process is evidenced by the observed modifications of heavy-flavour baryon-to-meson ratios from \ee to pp collisions in conjunction with the observed dependence of these ratios on charged particle multiplicity in pp collisions as discussed in the previous section. From this alone we see that the study of increasing the collision system size is important, and pA collision systems provide a tantalising bridge between the small pp and large AA collision environments. 
Collisions of protons with nuclei are interesting for the study of heavy-flavour hadronization in particular for at least two reasons. 
The first is that they give access to high multiplicities without the need of selecting very rare events, unlike pp collisions whereby high-multiplicity events represent only a small fraction of the pp inelastic cross section. 
The second is the possible interplay of hadronization and collective effects, the presence of which in small collision systems emerged in the last decade, particularly from the study of particle azimuthal anisotropies and the observation of relatively large flow coefficients in high-multiplicity p--Pb collisions~\cite{Nagle:2018nvi}. These effects resemble those observed in high-energy nucleus--nucleus collisions and ascribed to the formation of a QGP, which expands and cools down following hydrodynamic laws. %

The first aspect can be better quantified by the following considerations. High-multiplicity pp events represent a small fraction of the pp inelastic cross section. For instance, the highest multiplicity class considered in ALICE measurement of $\Lc/\Dzero$ ratio reported in Fig.~\ref{fig:vsmultalice} represents a fraction certainly smaller than 0.5\%~\cite{ALICE:2019dfi,ALICE:2021npz} of all inelastic events\footnote{This can be deduced, as a conservative upper limit, from the numbers reported in Table~1 and Table~4 of Ref.~\cite{ALICE:2021npz} and ~\cite{ALICE:2019dfi}, respectively.}. Thus, it cannot be excluded that the multiplicity trends observed in these events derive from 'peculiar' features of the selected events rather than from a general dependence of the measured particle ratios from multiplicity. A quite natural next step is to measure these ratios in collision systems in which similar or higher multiplicities are reached without selecting rare events. This opportunity is offered by collisions of nuclei or of protons with nuclei, in which significantly higher multiplicities are reached with respect to pp collisions. The multiplicity of charged particles at midrapidity increases with the number of nucleons participating to the collisions (\Npart)~\cite{ALICE:2014xsp,ALICE:2018wma,ALICE:2015juo}.
In p--Pb collisions at $\snn=5.02$~\TeV, 
the average charged-particle multiplicity at midrapidity in non-single diffractive events is 
more than a factor of 3 higher than the multiplicity measured in non-single diffractive pp events~\cite{ALICE:2022kol}. The average multiplicity of the highest-multiplicity interval in which ALICE measured the $\Lc/\Dzero$ ratio in pp collisions is about a factor of two higher and is reached in 0--5\% more central p--Pb collisions. 
As shown in Fig.~\ref{fig:LctoDzeroPPmultPPB} (left panel), the $\Lc/\Dzero$ ratio measured by ALICE in p--Pb collisions is higher than that measured in pp collisions at $\s=13~$\TeV without multiplicity selections in the range $3<\pt<12$~\GeVc, and similar to that measured in high-multiplicity pp events. At lower \pt, there is a hint for a lower ratio in p--Pb collisions than in pp collisions. 
Preliminary results from CMS~\cite{CMS-PAS-HIN-21-016} suggest that the ratio does not evolve significantly with multiplicity in p--Pb collisions, contrary to what observed for the $\Lambda/\mathrm{K_{S}^{0}}$ ratio in the light-flavour sector~\cite{ALICE:2019avo}. This breaks the similarity of $\Lc/\Dzero$ and $\Lambda/\mathrm{K^{0}_{s}}$ ratios observed by ALICE in pp collisions, both multiplicity-integrated and as a function of multiplicity~\cite{ALICE:2021npz}, and, within larger uncertainties, in multiplicity-integrated p--Pb collisions~\cite{ALICE:2020wfu}.
Overall, the data suggest that the modification with multiplicity of the \pt-differential $\Lc/\Dzero$ ratio observed in pp collisions is not dictated by peculiar properties of the selected high-multiplicity events. 
If CMS preliminary data is confirmed, the \pt-differential $\Lc/\Dzero$ ratio from low-multiplicity pp collisions up to high-multiplicity p--Pb collisions, would display a stronger evolution at low multiplicities followed by saturation. Constraining better the low-multiplicity region is a major goal for the future. New measurements with reduced uncertainty and extended to even lower multiplicities are necessary to profile the onset of the modification of the baryon-to-meson ratio from \ee collisions to hadronic collisions.
In the right panel of Fig.~\ref{fig:LctoDzeroPPmultPPB}, the \LcD ratio measured in pp and p--Pb collisions are compared with QCM predictions, which are compatible with the data. 
In QCM, the ratio has a different \pt trend in p--Pb collisions and peaks at higher \pt than in pp collisions because of the larger assumed light-quark velocity, which is needed to reproduce the hardening of the light-flavour hadron spectra. 
The evolution of the \pt-differential \Lc spectrum from pp to p--Pb collisions is quantified in Fig.~\ref{fig:LcRPBcharmFlowpPb} in terms of the nuclear modification factor, $\RpPb$, which is calculated by dividing the p--Pb \pt-differential production cross section by the pp one scaled by the Pb mass number. 
A deviation of $\RpPb$ from unity is observed and corresponds to a modification of to the spectra in p--Pb with respect to pp collisions. The modification of the \Lc-baryon production spectrum in \pPb collisions has been additionally confirmed by the computation of the mean transverse momentum, \meanpt, which results larger in \pPb with respect to pp~\cite{ALICE:2022ych}.  For D mesons $\RpPb\approx 1$~\cite{ALICE:2019fhe}. Therefore, the \Lc \RpPb is almost equivalent to the double ratio $(\LcD)^{\mathrm{p--Pb}}/(\LcD)^{\mathrm{pp}}$.
The POWHEG+PYTHIA6+EPPS16~nPDF model is based on the POWHEG NLO calculation~\cite{Alioli:2010xd,Frixione:2007nw} of charm quark production cross section matched with PYTHIA 6~\cite{Sjostrand:2006za} for parton shower and fragmentation. It uses CT14NLO~\cite{Dulat:2015mca} parton distribution functions (PDF) and the EPPS16 parameterization~\cite{Eskola:2016oht} of nuclear PDF for the the Pb nucleus. The latter term causes a departure of the expected \RpPb from unity at low \pt due to the suppression (``nuclear shadowing'') at small Bjorken-{\it x} of the gluon PDF. The model, which can describe D-meson data within uncertainties, cannot describe \Lc \RpPb, evidencing that the observed modification cannot be determined by the modification of PDF of nuclei with respect to those of protons. For heavy-flavour hadrons at LHC energies, nuclear shadowing was considered the largest ``cold-nuclear matter'' effect, i.e. an effect not related to the formation of a quark-gluon plasma, that could modify the production cross sections in nuclear collisions to the simple geometrical scaling of cross sections in pp collisions. 
In the POWLANG calculation~\cite{Beraudo:2015wsd}, charm quarks are transported through a ``small-size'' expanding QGP, which affects the \pt distributions of charm hadrons. However, the calculated \RpPb value is identical for all charm-hadron species as it does not consider any modifications of the relative hadron abundances due to quark coalescence. The presence of a QGP phase has a substantial impact on the predicted \RpPb but the model cannot describe either the D-mesons nor the \Lc \RpPb in the full \pt range. 
The QCM model instead describes the data within uncertainties. 

Most recently, the ALICE Collaboration released measurements of the fragmentation fractions of charm quarks to different charm-hadron species, which are measured for the first time in p--Pb collisions at 
\snn = 5.02 TeV at midrapidity \cite{ALICE:2024ocs}.
The fragmentation fractions for all charm-hadron species are found to be consistent for pp and p–Pb collisions at the same
collision energy, indicating that there is no significant modification of the charm-quark hadronisation process due to the different hadronic collisions’ system sizes.

The LHCb collaboration measured the \Lc and \XicPlus production and their ratios to \Dzero at forward and backward rapidity in p--Pb collisions~\cite{LHCb:2018weo, LHCb:2023cwu}. When compared with the measurements at midrapidity~\cite{ALICE:2024ozd,ALICE:2022ych}, the baryon-to-meson ratios are lower for \pt $<$ 8 GeV/$c$, while at higher \pt the measurements are consistent within uncertainties. The origin of this tension is not yet fully understood, for a more detailed discussion on this open point see Sect.~\ref{sect:future}. 
Also the ratio of $\Ds$ production to that of \Dzero and \Dplus was studied in p--Pb collisions by ALICE~\cite{ALICE:2019fhe} and LHCb~\cite{LHCb:2023rpm}. Without selections on multiplicity, the ratios are observed to be consistent with those measured in pp collisions. On the other hand, while ALICE does not observe a significant dependence on centrality, LHCb reports an increase of the ratio with multiplicity, with consistent values at forward and backward rapidity. However, the data uncertainty prevent conclusions on a possible dependence on rapidity of the ratios, as well as whether LHCb values converge to the values measured by ALICE in Pb--Pb collisions~\cite{ALICE:2018lyv,ALICE:2021kfc}.

\subsection{Nucleus--nucleus collisions}
\label{sect:expAA}
 The study of heavy-quark hadronization in nucleus--nucleus collisions offers a unique window for addressing hadron formation in an extended system of quarks, which can be assumed to be in local thermodynamic equilibrium and to move collectively along with the system expansion. Indeed, since long it was suggested that in the QGP hadrons may form via recombination of existing quarks~\cite{Greco:2003vf}, with important implications for the hadron abundances and momentum distributions, expected to significantly differ from those resulting from fragmentation and hadronization in a ``vacuum-like'' system, as that of an \ee collision can be classified. Understanding the hadronization process in these collisions is also a necessary step to exploit heavy-quarks for the study of in-medium partonic energy loss and of the medium transport properties~\cite{Rapp:2018qla}. This was the initial goal motivating heavy-flavour hadron production measurements in heavy-ion collisions: even at LHC energies, the medium temperature is significantly lower than the charm-quark mass, so that charm and beauty quark thermal production is negligible compared to that in hard-scattering processes occurring in the early stage of the collision with cross sections known with relatively small uncertainty~\cite{Cacciari:1998it,ALICE:2021mgk,LHCb:2015swx,LHCb:2016qpe,ALICE:2019rmo}. Interacting with the medium, constituents heavy quarks exchange energy but preserve their flavour identity. Thus they can be regarded as ``calibrated'' probes of the medium and used as markers to trace the quark diffusion and the partonic interactions in the QGP, resolving the medium constituents and allowing to build a connection between the microscopic ``local'' partonic interactions and the medium global properties, in particular, its transport ones. However, the uncertainties in the modelling of the hadronization process, which significantly influences the momentum distribution of the final-state hadrons, can reduce the capability of inspecting the dynamics in the partonic phase. On the other hand, a QGP offers a (expanding) thermal source of light quarks and gluons whose modelling, e.g. in coalescence models, might be simpler and less sensitive to event-by-event fluctuations than that of a pp collision system. Therefore, the measurement of the production of different heavy-flavour hadron species in heavy-ion collisions can simultaneously constrain the modelling of hadronization and parton dynamics in the medium. 

Several measurements in Au--Au collisions at RHIC and Pb--Pb collisions at the LHC demonstrate that charm and beauty quarks interact strongly with the QGP constituents~\cite{ALICE:2022tji,Acharya:2018upq,Acharya:2018hre,ALICE:2012ab,Abelev:2012qh,Sirunyan:2017xss,Adam:2015jda,Adam:2015sza,ALICE:2016uid,ALICE:2021rxa,ALICE:2017pbx,ALICE:2018gif,ALICE:2021kfc,ALICE:2020pvw,ALICE:2020iug,ALICE:2020hdw,Adamczyk:2014uip,Abelev:2006db,Adare:2012px,PHENIX:2010xji,Adler:2005xv,CMS:2015sfx,CMS:2012bms,CMS:2016mah,CMS:PhysRevLett.123.022001,CMS:2017uoy,CMS:2021qqk,CMS:2020bnz,Aaboud:2018bdg,Aaboud:2018quy,ATLAS:2021xtw}.
The $\Raa$ values observed for both charm and beauty hadrons are significantly smaller than unity at intermediate $\pt$ and larger for beauty than charm, indicating a significant energy loss of both quarks in the medium and evidencing the expected dependence of energy loss on the quark mass. 
At moderate/low \pt the $v_{2}$ of prompt D mesons is large, close to that observed for charged pions~\cite{ALICE:2020iug}. This can indicate that charm quarks are strongly coupled to the system and are pushed towards kinetic equilibrium via rescattering with the medium constituents. It can also signal coalescence as a hadronization process relevant to charm quarks through which charm hadrons partly inherit the flow of light quarks, a relevant aspect that could be more deeply investigated through heavy-light correlation of the anisotropic flow $v_n$ \cite{Plumari:2019hzp,Sambataro:2022sns}. At low $\pt$, where elastic processes are the most important ones, heavy quark transport in the medium can be described as a diffusion process~\cite{bib::intro4,Dong:2019unq,Capellino:2022nvf,Altenkort:2023oms}. The main transport parameter is the spatial diffusion coefficient \Diffs, which is related to the kinetic equilibration time of a heavy-quark with mass $m_{\rm Q}$, $\tau_{\rm Q}=\Diffs m_{\rm Q}/T$~\cite{Moore:2004tg,Petreczky:2005nh}. 
The comparison of \Raa and \ellflow data with model predictions favour small values of heavy-quark \Diffs, in the range $1.5<2\pi \Diffs T_{\rm c}<4.5$ at the pseudocritical temperature $T_{\rm c}=155$~\MeV, in line with other estimates~\cite{Xu:2017obm,Dong:2019unq,Capellino:2023cxe,PHENIX:2010xji,STAR:2017kkh}. 
This interval corresponds to relaxation times for charm quarks smaller than about 10~fm$/c$, thus of the order of or shorter than the QGP lifetime, actually suggesting that charm quarks might get close to reaching kinetic equilibrium in the medium~\cite{Capellino:2022nvf,Altenkort:2023oms}.  
A recent estimate that combines lattice QCD to the heavy quark effective theory in the limit $M\gg T$~\cite{Altenkort:2023oms} implies a very small thermalization time, at least in the $p\rightarrow 0$ limit;
namely of about 1-1.5 fm/c for charm quarks and 3-5 fm/c for bottom which are significantly smaller than those extrapolated from the phenomenological analysis. This work has been developed in non-quenched QCD which is a very important advancement w.r.t. previous studies, however, it is still in the $M \rightarrow \infty$ limit.
In Ref.~\cite{Sambataro:2023tlv,Sambataro:2024mkr} it has been shown that within a quasi-particle model (QPM-Catania) one expects that $\Diffs$ has a significant mass dependence from the charm quark mass to the limit $M \rightarrow \infty$, decreasing by about a factor 1.7. This could to a large extent compensate the discrepancy between the value of $ 2\pi T \Diffs$ from the lQCD driven estimate w.r.t. the one extrapolated from the phenomenology for charm quarks.
However, an even more recent extension of the study in Ref.~\cite{Altenkort:2023oms} has done a first extrapolation to finite mass finding a very small dependence of $2\pi T \Diffs$ on $M_Q$. This, if confirmed by further studies within lQCD and heavy quark effective field theory, opens the question on how to reconcile them with the phenomenological estimates. At this stage, we only mention that a preliminary study have indicated that an initial glasma phase may lead to infer smaller values of $\Diffs$~\cite{Sun:2019fud}, and more recently it has been suggested that also memory effect (non-Markovian dynamics) may affect the phenomenological extrapolation of $\Diffs$~\cite{Pooja:2023gqt}, especially for charm quarks. Therefore, with more precise data becoming available with the ALICE upgrade (especially at low $\pt$), it will be possible to achieve a significant advancement in the understanding of the heavy quark dynamics.

However, even within the uncertainties of the mass dependence of the diffusion or drag coefficient, it remains that 
because of the larger mass, beauty quarks are expected to diffuse less than charm quarks and to have a significantly larger relaxation time that is at least a factor of two \cite{Sambataro:2023tlv,Pooja:2023gqt} if not a factor of three larger (if $\Diffs$ is nearly mass independent) as suggested by the ratio $M_b/M_c$. Energy loss is also expected to be smaller for beauty than charm quarks. In line with these expectations, experimental measurements of beauty hadrons or their decay products show higher \Raa and smaller $v_{2}$ with respect to charm or light-flavour hadrons \cite{ALICE:2023gjj,ALICE:2022tji}. 

\begin{figure}[ht!]
    \begin{center}
    \includegraphics[width = 1\textwidth]{ 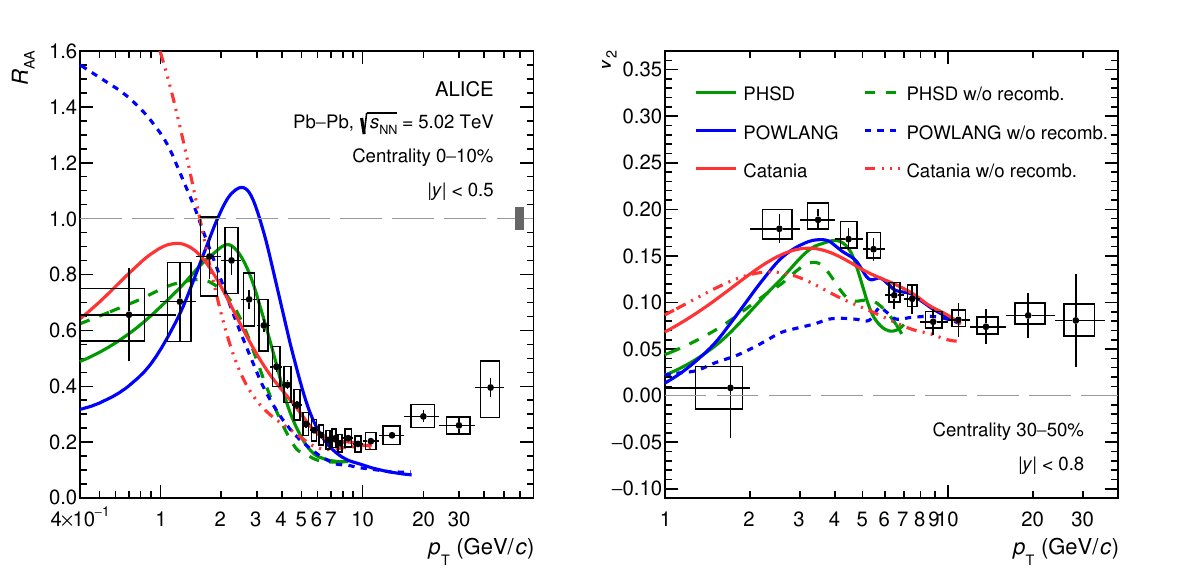}
    \end{center}
    \caption{Prompt D-meson \Raa in the 0--10\% centrality class (left panel) and $v_2$ in the 30--50\% centrality class (right panel) compared with the PHSD~\cite{Song:2015sfa}, POWLANG~\cite{Beraudo:2014boa,Beraudo:2017gxw}, and Catania~\cite{Plumari:2017ntm} predictions obtained with and without including hadronization via recombination. Figure adapted from Ref.~\cite{ALICE:2021rxa}.}
    \label{fig:Draav2_vsModels}
\end{figure}

The hadronization schemes used in transport models have a significant impact on the predicted values of \Raa and $v_{2}$ and on their dependence on \pt. 
The effect of hadronization via recombination on D-meson \Raa and $v_{2}$ is illustrated in Fig.~\ref{fig:Draav2_vsModels} where ALICE measurements in Pb--Pb collisions in the centrality intervals 0--10\% (\Raa, left panel) and 30--50\% (\ellflow, right panel) are compared with the expectations of the PHSD, POWLANG, and Catania models with and without the inclusion of recombination~\cite{ALICE:2021rxa}. In a medium that is radially expanding with a common velocity, hadronization via quark recombination tends to equilibrate particle velocities towards the medium expansion velocity. 
The combined effect of recombination, diffusion, and energy loss on the \Raa is the appearance of a maximum at momenta not much larger than the hadron mass. 
For what concerns the elliptic flow, the \ellflow of heavy-flavour hadrons formed via recombination derives from a combination of the charm and light quarks flows. From the comparison of model expectations with and without hadronization via recombination, shown in the right panel of Fig.~\ref{fig:Draav2_vsModels}, it can be appreciated the effect of recombination on charm-meson \ellflow in increasing \ellflow and pushing its maximum to higher momenta. Recombination is an essential ingredient for reproducing the data \cite{Greco:2017rro,Scardina:2017ipo,Rapp:2018qla,Dong:2019unq,Zhao:2023ucp}. 

 All models that can reproduce the large observed $v_{2}$ of D mesons~\cite{ALICE:2020iug} predict that it partly derives from hadronization via coalescence. In order to single out the contribution of hadronization from the contribution of charm-quark diffusion in the generation of charm flow, the modification in Pb--Pb collisions with respect to pp collisions of the \pt-differential yield ratios of different heavy-flavour particle species must be studied.  
In particular, models based on statistical hadronization or coalescence were initially designed to describe hadron formation in AA collisions. They were expecting an increase of the production of heavy-flavour baryons relative to heavy-flavour mesons and of hadrons with strange quarks relative to non-strange ones in heavy-ion with respect to pp collisions, the latter originally thought to deliver systems with properties closer to \ee collisions and in which \enquote{in-vacuum} fragmentation should have been the dominant hadronization mechanism. 
Such an expectation was motivated by the standard lore that only the large reservoir of partons and high energy densities of nucleus--nucleus collision systems could make it likely for quark coalescence to become a relevant process compared to the mechanism of fragmentation, which requires the creation of quarks from the vacuum. Hence, the typical signature of coalescence, that is, an enhancement of baryon-to-meson ratios, was not expected to occur in pp collisions.
However, as discussed in the previous sections, the data indicates that, particularly in the heavy-flavour sector, pp collisions exhibit significant differences compared to smaller collision systems such as \ee and \ep. These differences result in modifications to the particle ratios, providing a sensitive test of changes in the hadronization dynamics.
The coalescence plus fragmentation approach, such as the Catania model, suggests that even if pp collisions result in the formation of a QGP droplet, D meson production is likely still dominated by the fragmentation process. Meanwhile, the coalescence process becomes dominant for baryon production, significantly increasing it in pp collisions compared to \ee collisions. This outcome, though initially unexpected, naturally follows from the application of coalescence dynamics and effectively explains the large increase in the measured $\Lc/\Dzero$ ratio.
However, the enhancement in pp collisions can also be explained in the context of PYTHIA, but, also in this case, the main modification affects the baryon production that has to come from colour reconnections of quarks and gluons in topologies irrelevant in \ee collisions but emerging with an increasing number of MPI, hence particularly in high-multiplicity events, in hadronic collisions. This also shows that already in pp (in particular for high-multiplicity events) more local recombination of quarks (minimizing string energy) becomes quite relevant for baryon production. The same coalescence modelling shows that in AA collisions the large-size reservoir of high-density quarks makes a further difference making the recombination the dominant mechanism for both heavy quark mesons and baryons.

\begin{figure}[ht!]
    \begin{center}
    \includegraphics[width = 0.49\textwidth]
    { 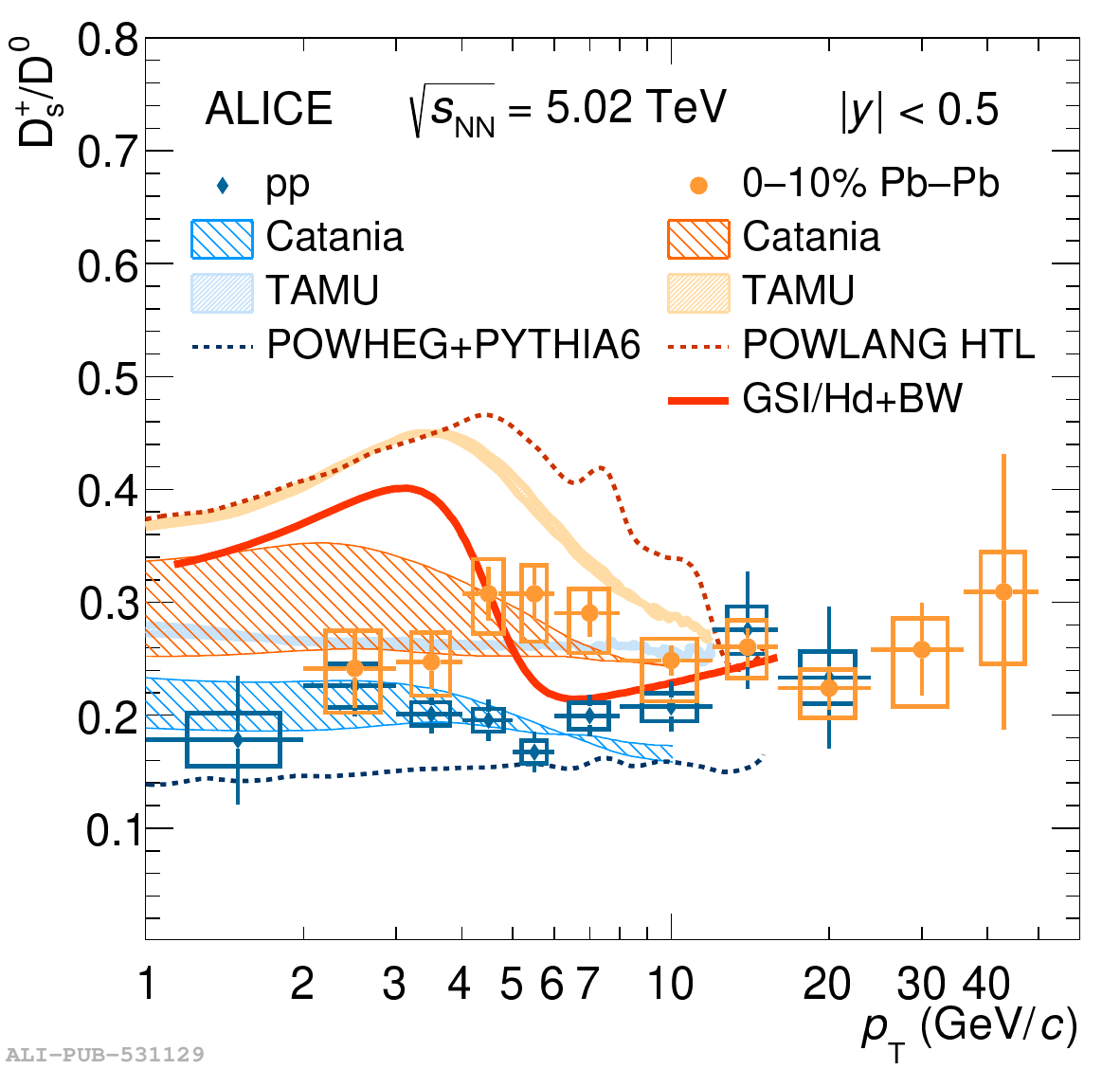}
    \includegraphics[width = 0.49\textwidth]{ 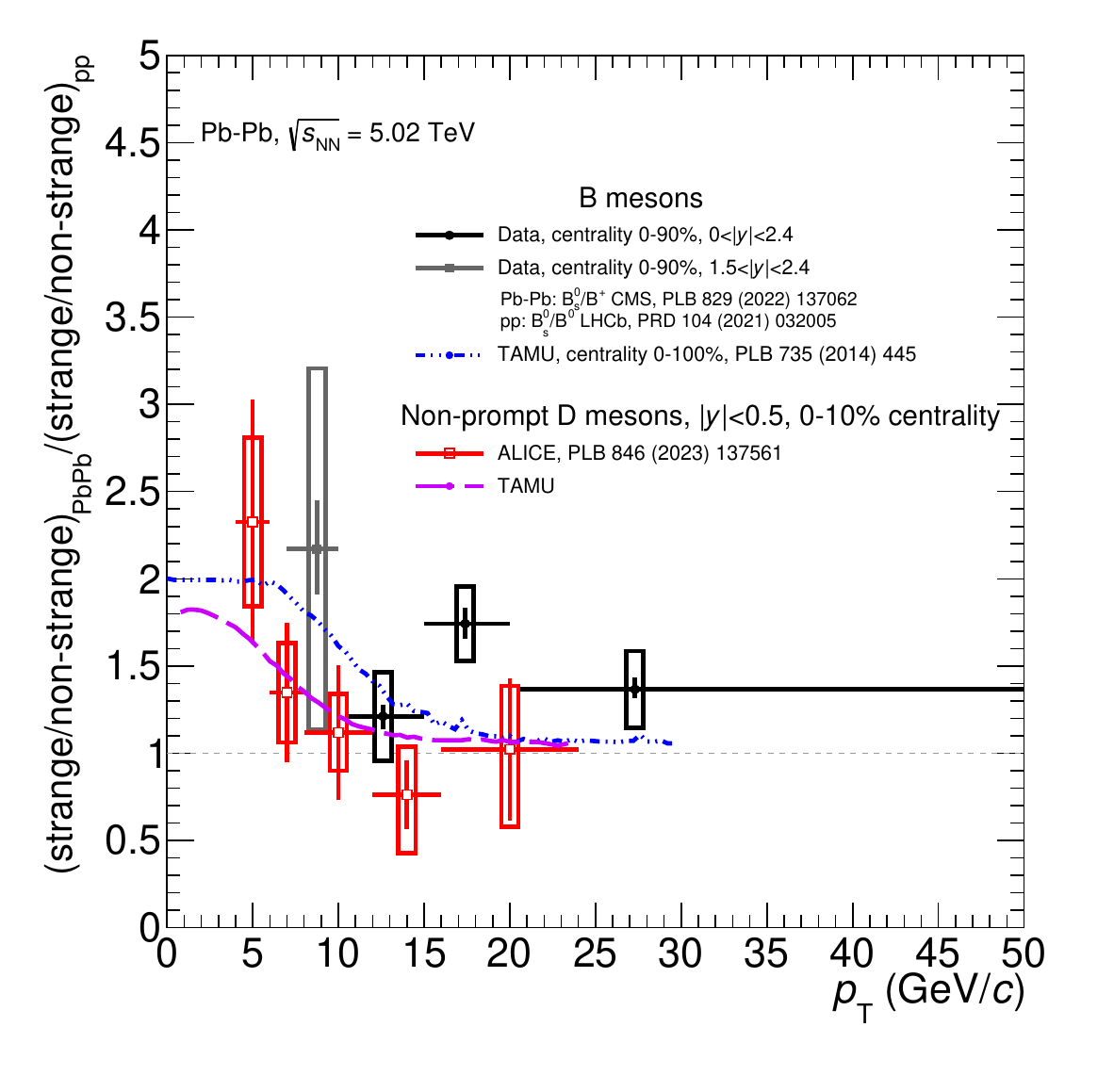}
    \end{center}
    \caption{Left (from Ref.~\cite{ALICE:2022wpn}): prompt \Ds/\Dzero ratio measured by ALICE in $|y|<0.5$ in pp and 0--10\% central Pb--Pb collisions at \snn=5.02~\TeV~\cite{ALICE:2021kfc,ALICE:2019nxm,ALICE:2021mgk} compared to model expectations~\cite{Plumari:2017ntm,Minissale:2020bif,He:2019vgs,Beraudo:2022dpz,Andronic:2021erx}. Right: red (black and grey) markers show the ratio of strange to non-strange non-prompt D mesons (B mesons) yield in 0--10\% (0--90\%) Pb--Pb collisions at \snn=5.02~\TeV measured by ALICE~\cite{ALICE:2022xrg,ALICE:2022tji} (CMS~\cite{CMS:2021mzx}) and normalised to the analogous ratio measured in pp collisions. In the B-meson case, the parametrization of the $f_{\mathrm{s}}/f_{\mathrm{d}}$ ratio measured at \s=7~\TeV by LHCb in $2<y<4.5$~\cite{LHCb:2021qbv} is used as pp reference, considering the collision-energy and rapidity dependence of the ratio, as well as the parametrization uncertainty, as negligible compared to the Pb--Pb data uncertainty. The data are compared to TAMU calculations~\cite{He:2014cla,He:2019vgs,CMS:2018eso}.} 
    \label{fig:strangenessPbPb}
\end{figure}

 The measurements of the production of prompt and non-prompt \Ds mesons~\cite{ALICE:2021kfc,ALICE:2022xrg} and of \Bs mesons~\cite{CMS:2021mzx} provide first hints supporting the presence of an enhancement of the production of heavy-strange mesons with respect to non-strange ones. Figure~\ref{fig:strangenessPbPb} shows in the left panel the comparison of the prompt $\Ds/\Dzero$ ratios measured by ALICE in central Pb--Pb collisions and in pp collisions. A hint of a higher ratio in Pb--Pb collisions is visible in $3<\pt<10$~\GeVc. The Catania model reproduces the data within uncertainties, while the TAMU and POWLANG transport models as well as the GSI/Hd+BW statistical hadronization model, for which the hadron \pt spectrum is modelled according to a Blast Wave function, tend to overestimate the data in the low \pt region. Slightly higher $\Ds/\Dzero$ values, closer to 0.4 and without a significance dependence on \pt in $1<\pt<8$~\GeVc, were reported by STAR in semicentral and peripheral Au--Au collisions at $\snn=200$~\GeV~\cite{STAR:2021tte}. In central Au--Au collisions, the ratio decreases below 0.3 for $\pt<4$~\GeVc and, within uncertainties, becomes compatible with PYTHIA expectation and pp data at LHC energies. New measurements with reduced uncertainty and lower \pt reach, as well as the measurement of \XicPlusZero production (larger than \Ds one in pp collisions), are necessary for a final assessment and understanding on the enhancement of char-strange hadron production.
 In the right panel of Fig.~\ref{fig:strangenessPbPb}, the $\Bs/\Bplus$ ratio measured as a function of \pt by CMS in 0--90\% central collisions~\cite{CMS:2021mzx} and normalized to the $f_{\mathrm{s}}/f_{\mathrm{d}}$ ratio measured at \s=7~\TeV by LHCb in $2<y<4.5$~\cite{LHCb:2021qbv} is shown together with the ratio of non-prompt $\Ds$ \Raa to $\Dzero$ \Raa measured by ALICE in the 0--10\% centrality interval~\cite{ALICE:2022xrg,ALICE:2022tji}. In pp collisions about 50\%   of non-prompt \Ds derive from the decay of a \Bs meson. Together the data provide a hint for an enhancement of strange B-meson production relative to non-strange B-meson production, in line with the expectation of the TAMU model~\cite{He:2022tod}.

\begin{figure}[ht!]
    \begin{center}
    \includegraphics[width = 0.48\textwidth]{ 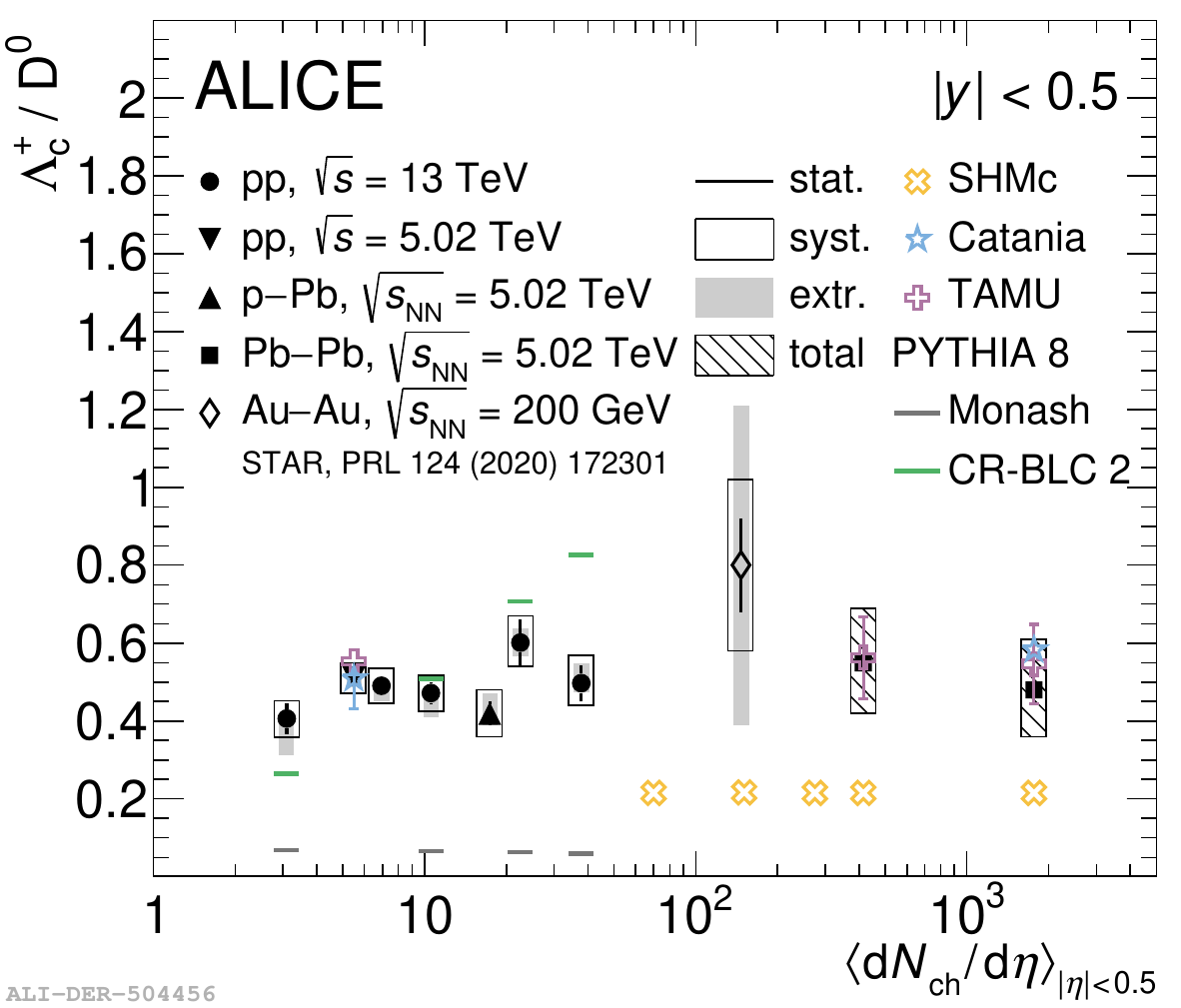}
    \includegraphics[width = 0.48\textwidth]{ 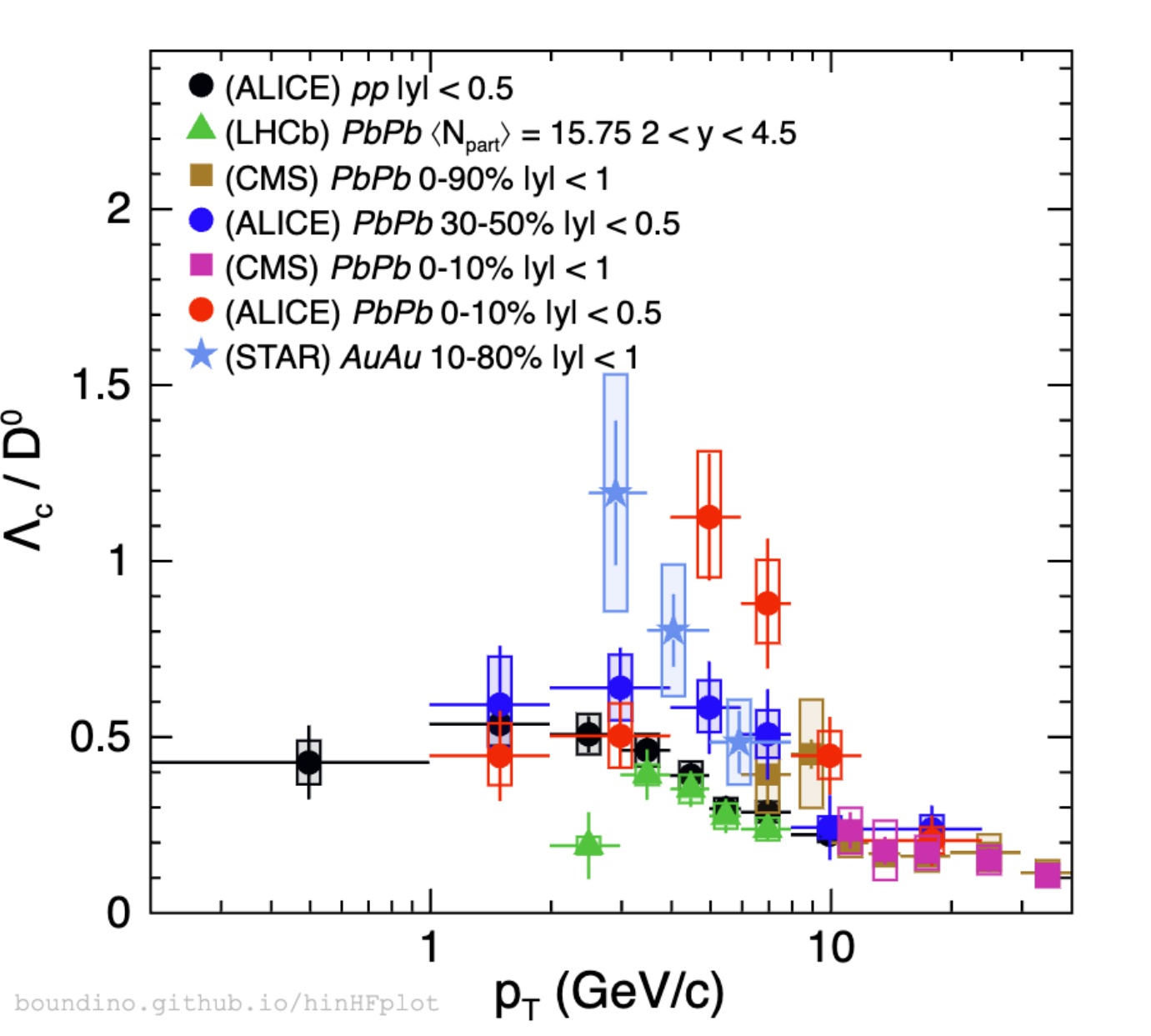}   
    \end{center}
    \caption{Left (from~\cite{ALICE:2021bib}): baryon-to-meson $\Lc/\Dzero$ cross-section ratio $\pt$-integrated for $\pt>0$ as a function of charged-particle multiplicity at midrapidity measured by ALICE in pp, p--Pb, and Pb--Pb collisions~\cite{ALICE:2021npz,ALICE:2021bib} and by STAR~\cite{STAR:2019ank} in Au--Au collisions. Right (adapted from~\cite{Rossi:2023xqa}): comparison of the \pt-differential $\Lc/\Dzero$ cross-section ratios measured in pp collisions by ALICE~\cite{ALICE:2020wfu}, in Pb--Pb collisions by ALICE~\cite{ALICE:2021bib}, CMS~\cite{CMS:2023frs}, and LHCb~\cite{LHCb:2022ddg} in different centrality intervals, and in Au--Au collisions by STAR~\cite{STAR:2019ank}.}
    \label{fig:LcToD0AllSystems}
\end{figure}

The $\Lc$ is the only heavy-flavour baryon whose production has been measured in nucleus--nucleus collisions. The left panel of Fig.~\ref{fig:LcToD0AllSystems} shows the evolution of the \pt-integrated $\Lc/\Dzero$ ratio as a function of the charged-particle multiplicity at midrapidity measured by ALICE in pp, p--Pb, and Pb--Pb collisions and by STAR in Au--Au collisions~\cite{ALICE:2021bib}. The ratio is approximately 0.5, thus much higher than the $\approx 0.11$ value measured in \ee and ep collisions, and, within uncertainties, it does not show a dependence on multiplicity. The same ratio in pp (minimum bias) and in Pb--Pb collisions is expected by the Catania and TAMU models, which reproduce also quantitatively the data. Also the SHMc model does not expect a dependence on centrality in Pb--Pb collisions. However, it predicts a ratio around 0.2 and underestimates the data. It must be noted that this version of the SHMc does not include the augmented set of high-mass baryon states considered by TAMU. In the SHMc charm is not in chemical equilibrium and the abundance of charm hadron species depends on the charm fugacity factor, which however cancels in single-charm particle ratios, and by the chemical freezeout temperature, which results in being independent of centrality. Similarly, in the TAMU, though the hadronization process is modelled differently in pp and Pb--Pb collisions, particle ratios mainly depend on statistical factors. For the Catania model, the similarity between pp and Pb--Pb values derives from the similar description of the bulk of light-flavour quarks as a QGP medium with similar temperatures in the two systems and from the applied normalization condition that coalescence probability equals unity at $p_{\rm T}^{\rm{hadron}}=0$. There is still a dependence on the volume of the hadronizing fireball and a subdominat dependence on the radial flow but the last has a weak effect on the $p_T$-integrated yield. In pp collisions, the PYTHIA 8 Monash tune is even lower than the \ee expectation and largely independent of event multiplicity.
Contrastingly, PYTHIA 8 with CR-BLC expects a significance increase of the ratio with multiplicity, deriving from the larger probability of forming junction colour connections in events with a higher number of MPI, which is not supported by the data. The region of very low multiplicities is not covered by measurements, thus leaving open the possibility that \ee values are recovered when the hadronic activity in the event is very small, in line with what the $\Lb/\mathrm{B}^{0}$ ratio at forward rapidity, discussed in Sect.~\ref{sect:expppvsMult}, suggests. It is worth recalling, that, as discussed in Sect.~\ref{sect:expppvsMult}, the average multiplicity in \ee collisions at LEP energies, where most particles are part of high-energy dijets, may not differ tremendously from the average multiplicity in minimum bias pp collisions at LHC energies. This suggests that the number of produced particles is not the only relevant parameter and rather the whole event topology and the impact of the underlying event and MPI in hadronic collisions, even at the lowest multiplicity, must be taken into account for explaining the differences observed with respect to \ee collisions. 

\begin{figure*}[tb]
\begin{center}
\includegraphics[width=0.98\textwidth]{ 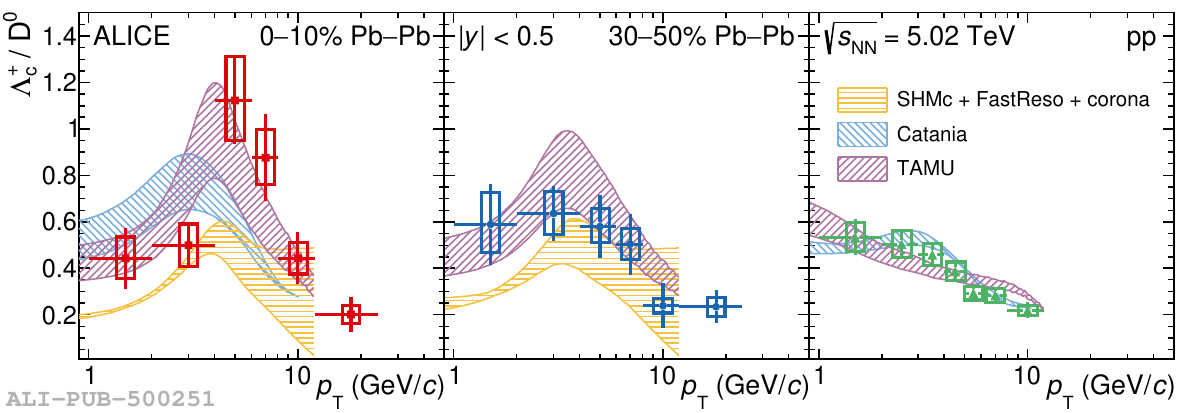}
\caption{Baryon-to-meson $\Lc/\Dzero$ cross-section ratio as a function of \pt in 0-10\% (left) and 30-50\% (middle) Pb--Pb collisions, and in pp collisions (right), compared with expectations from the Catania, SHMc, and TAMU models and from the PYTHIA~8 event generator with the Monash and CR~Mode~2 tunes. Figure from Ref.~\cite{ALICE:2023bsp,ALICE:2021bib}.} 
\label{fig:LcToD0Run2PbPbpp} 
\end{center}
\end{figure*}

 In the right panel of Fig.~\ref{fig:LcToD0AllSystems}, the evolution of the \pt-differential $\Lc/\Dzero$ ratio with centrality is shown, by comparing the measurements performed in pp collisions~\cite{ALICE:2020wfu} and in different centralities in Pb--Pb collisions at the LHC~\cite{ALICE:2021bib,LHCb:2022ddg, Sirunyan:2019fnc} and in Au--Au collisions at RHIC~\cite{STAR:2019ank}.
Contrary to the \pt-integrated case, the \pt-differential $\Lc/\Dzero$ ratio evolves substantially with centrality, as shown in the right panel of Fig.~\ref{fig:LcToD0AllSystems}, in which the Pb--Pb ALICE~\cite{ALICE:2021bib}, CMS~\cite{CMS:2023frs}, and LHCb~\cite{LHCb:2022ddg} results are reported and compared to ALICE pp data as well as to the measurement in Au--Au collisions by STAR~\cite{STAR:2019ank}. At intermediate \pt, the ratio increases with centrality. In peripheral collisions, LHCb data at forward rapidity overlap with ALICE minimum-bias pp data at midrapidity. In 30--50\% Pb--Pb collisions the ratio is higher, though the values are compatible with those measured in high-multiplicity pp and p--Pb data. In central collisions a \enquote{radial-flow}-like peak emerges, possibly shifted at lower \pt at RHIC with respect to LHC energies. The appearance of this peak can be due to hadronization via coalescence, possibly reinforced by space-momentum correlations (see Sect.~\ref{sec:theoryCoalescence}). 
In Fig.~\ref{fig:LcToD0Run2PbPbpp} the $\Lc/\Dzero$ ratio measured by ALICE in 0--10\% (left panel) and 30--50\% (middle panel) central Pb--Pb collisions is compared to model expectations. Despite the different values predicted by the models, the large experimental and theoretical uncertainties do not allow to discriminate between the $\rm SHM_c$~\cite{Andronic:2021erx}, Catania~\cite{Plumari:2017ntm}, and TAMU~\cite{He:2019tik} models. 
At $\pt>20$~\GeVc the ratio is the same at all centralities and consistent with that measured in \ee collisions, possibly signalling the transition at high-enough momentum to a regime in which \enquote{vacuum}-like fragmentation is an effective description leading to a \enquote{universal} value.

\section{Prospects and future developments}
\label{sect:future}
 The measurements of heavy-flavour baryon production carried out in the last decade in hadronic collisions have somewhat unexpectedly broken the paradigm that fragmentation modelled on \ee data could be sufficient in describing hadronization in small collision systems, and that nucleus--nucleus would be the collision system where one would search for new phenomena. 
 Rather, a quite sharp distinction between hadronization in \ee and any hadronic collision system emerged, accompanied by the need of continuity from pp to nucleus--nucleus collisions. Multiplicity, as an estimator of the event activity (hence MPI?) and system size, provides a reasonable guide for examining the evolution of the observables from small to large collision systems. While reasonable explanations and ideas about the origin of the transition from \ee to pp collisions have been put forward, together with effective modelling capable to reproduce the trends shown by the measurements, a new theoretical paradigm for hadronization is still being shaped. Certainly, as extensively discussed in the previous sections, the measurements performed so far achieved some milestones delineating features and trends that theoretical modelling of hadronization will have to consider. They can be recapped in the following list.
\begin{itemize}
    \item The first is certainly the evidence that heavy-flavour baryon production is significantly larger in pp than in \ee collisions. This implies that the assumption that hadronization can be modelled solely with universal fragmentation functions does not hold already in pp collisions. 
    \item The second concerns the boundaries out of which fragmentation seems recovered as a good description of hadronization. The observed evolution of $\Lc/\Dzero$ and $\Lb/\mathrm{B^{0,-}}$ with \pt and multiplicity suggests that this occurs at high momentum (above about 15-20~\GeVc) and at extremely low multiplicity. 
    \item The third is that the enhancement from \ee to pp of the baryon-to-meson ratio is significantly larger for charm-strange baryons (e.g. \XicPlusZero) than for \Lc, despite no significant difference is observed for the \Ds/\Dzero ratio. This highlights the problem to understand strange production. 
    \item The observation for $\Sigmac/\Dzero$ and the augmented feed-down to \Lc from \Sigmac decay. In general, one may interpret these observations as a suppression of \Sigmac and charm-strange baryon production in \ee due to ``local'' dynamical constraints (e.g. diquark production in string breaking), which are absent or mitigated by the emergence of other processes in hadronic collisions. 
    \item Finally, remarkable experimental data shows the contrast between the apparently flat evolution with multiplicity and collision centrality of the \pt-integrated \Lc/\Dzero ratio from pp to Pb--Pb collision systems, compared to the strong modification of the \pt-differential ratio. 
    \item On the theoretical side, coalescence and statistical hadronization emerged as effective models to describe some of the measured yields in pp collisions. For an extensive model like PYTHIA, which aspire at describing the full event dynamics, the central role of MPI and colour reconnection in causing and shaping the observed effects appears as an unavoidable conclusion suggested by the measurements. 
\end{itemize} 


In this section, we outline the prospects and objectives for future research, addressing additional topics and unresolved issues related to heavy-flavour hadronization that were not fully explored in earlier sections. This includes recent measurements that have yet to yield definitive conclusions, either due to their preliminary nature or because they are affected by significant uncertainties. However, the examination of various observables hints at new effects and potentials for future investigation. We also present new ideas and experimental approaches that may help resolve discrepancies between measurements conducted by different collaborations and their divergence from certain theoretical predictions. \\

\noindent \textbf{Consolidation of emerging picture} 

Many of the measurements presented in this review show striking effects, often with results so far from the original expectation of the ``baseline models'' described in Sect.~\ref{sec:MCmodels} that even large experimental uncertainties did not prevent important conclusions. However, the natural first goal for the future should be to have new measurements of the observables already inspected with a reduction of the uncertainties and with an extension of the low and high \pt and multiplicity reach. In particular, profiling the low \pt and the \pt-integrated baryon-to-meson ratios for other particles than \Lb (in particular of \Lc, \XicPlusZero, and \Sigmac) at low multiplicity in pp collisions is of uttermost importance to investigate the onset of the modification of hadronization from \ee to hadronic collisions. Measurements of charm and strange baryons in nucleus--nucleus collisions would also be particularly significant to verify whether the production rate of \XicPlusZero and \Omegac relative to \Lc increases in a system in which strange quarks are expected to be abundantly produced given that light-flavour strange hadrons yields match the expectation from chemical equilibrium without canonical suppression. In the open beauty sector, the experimental study of hadronization in nucleus--nucleus collisions has substantially only started with the first measurements of strange and non-strange B mesons and non-prompt D mesons. Though very important for first assessments, the large uncertainties, the limited and not granular \pt coverage, and the reduced sensitivity implied by the decay kinematics for non-prompt signals, and the lack of measurements of beauty-baryon production, permit only qualitative comparison with model expectations. With next LHC runs, also thanks to the LHC detector upgrades, new measurements, including first ones of \Lb or non-prompt \Lc, should allow for firmer conclusions. In pp collisions, LHCb data already provided a rather detailed picture at forward rapidity, but precise measurements down to low \pt are missing for both \Lb and \Bs at midrapidity. As described in what follows, determining the dependence on rapidity of heavy-flavour baryon-to-meson ratios must be an important objective of experimental searches. Finally, measurements of the production of strange and beauty baryons would also be very important, also given what was observed for the $\XicPlusZero/\Lc$ ratio, but they are experimentally extremely challenging and likely difficult to achieve in the near future. \\

\noindent \textbf{Rapidity puzzle}

\begin{figure}[htb!]
    \begin{center}
    \includegraphics[width = 0.45\textwidth]{ 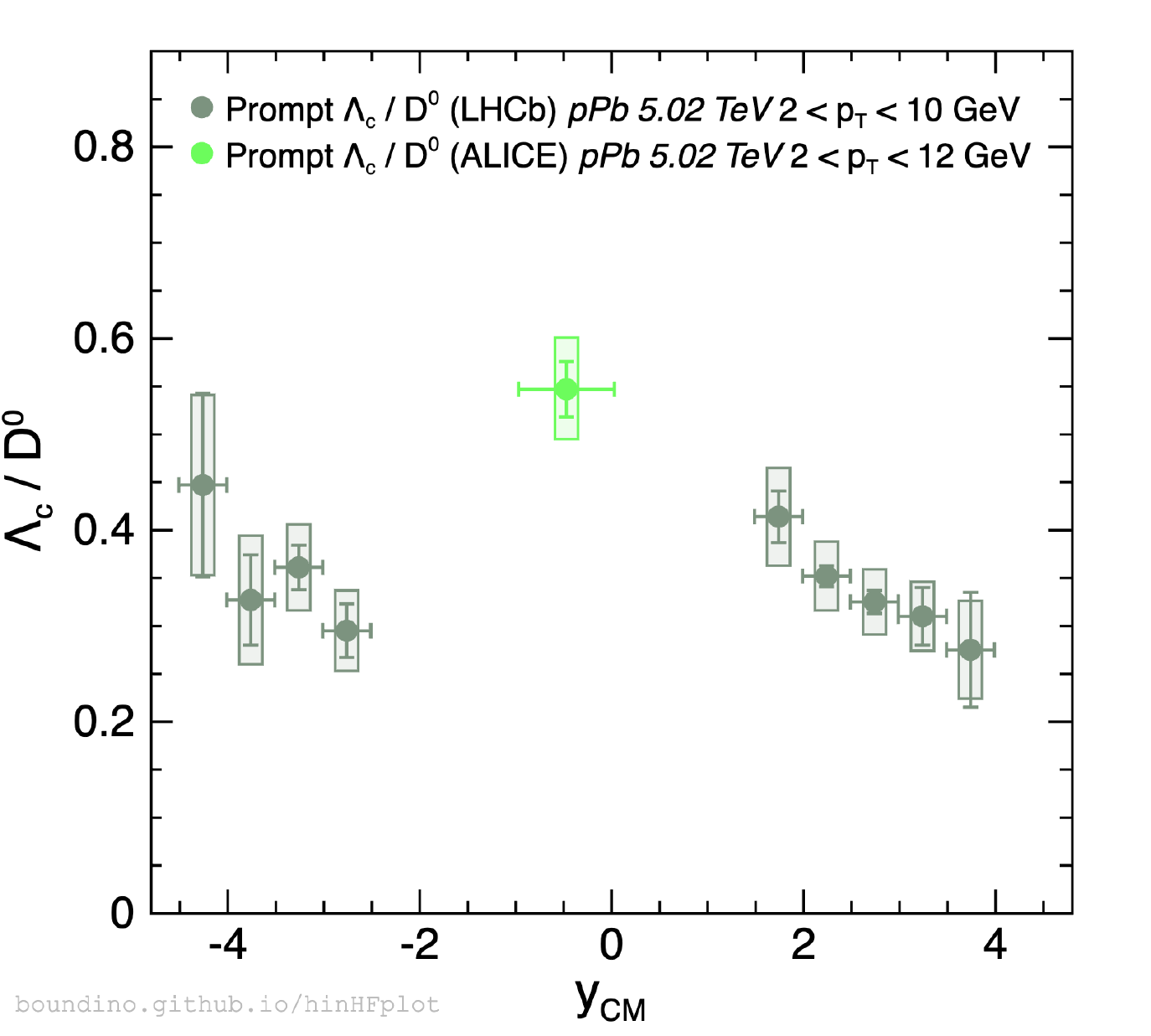}  
    \includegraphics[width = 0.42\textwidth]{ 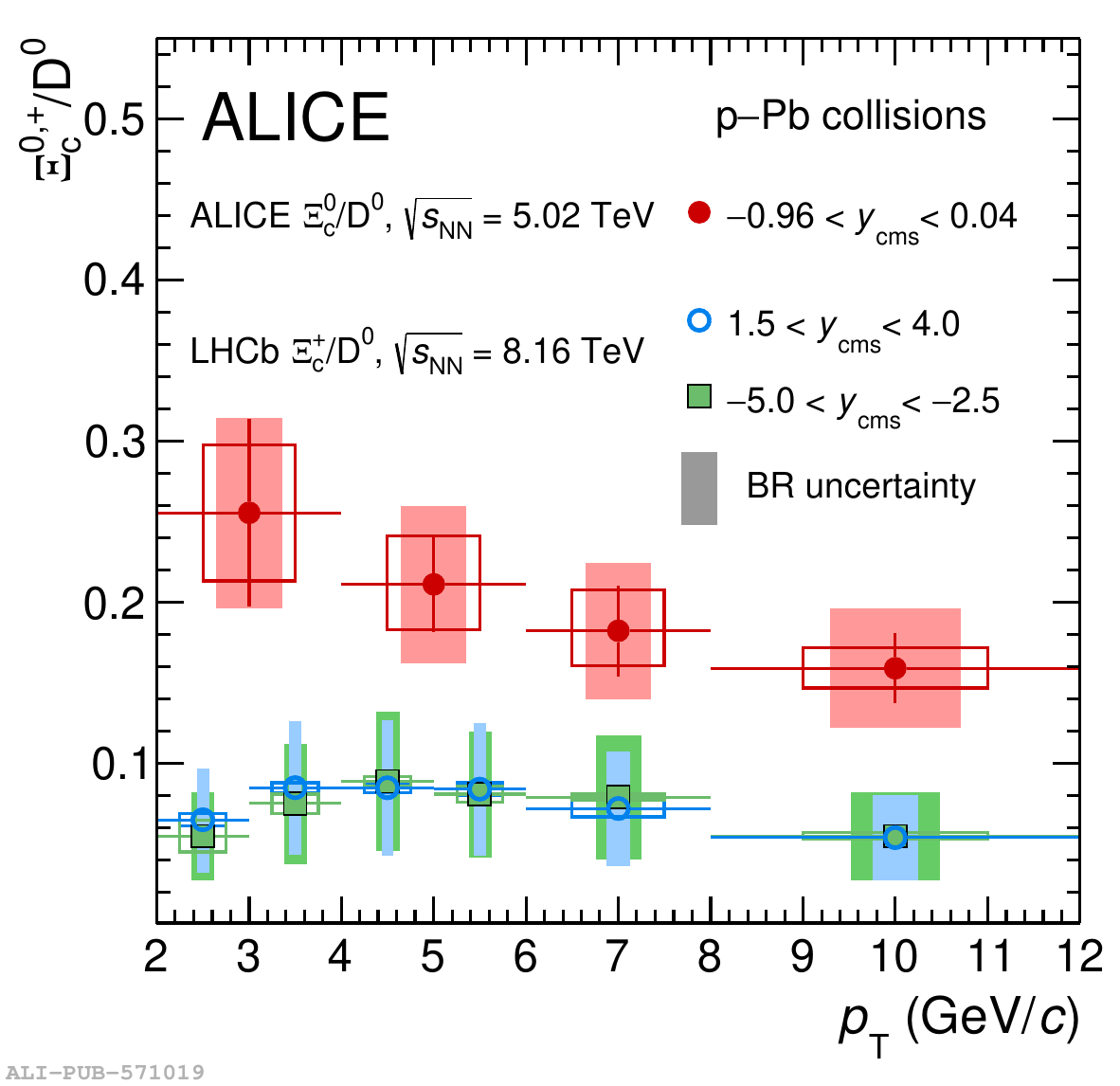}
    \includegraphics[width = 0.45\textwidth]{ 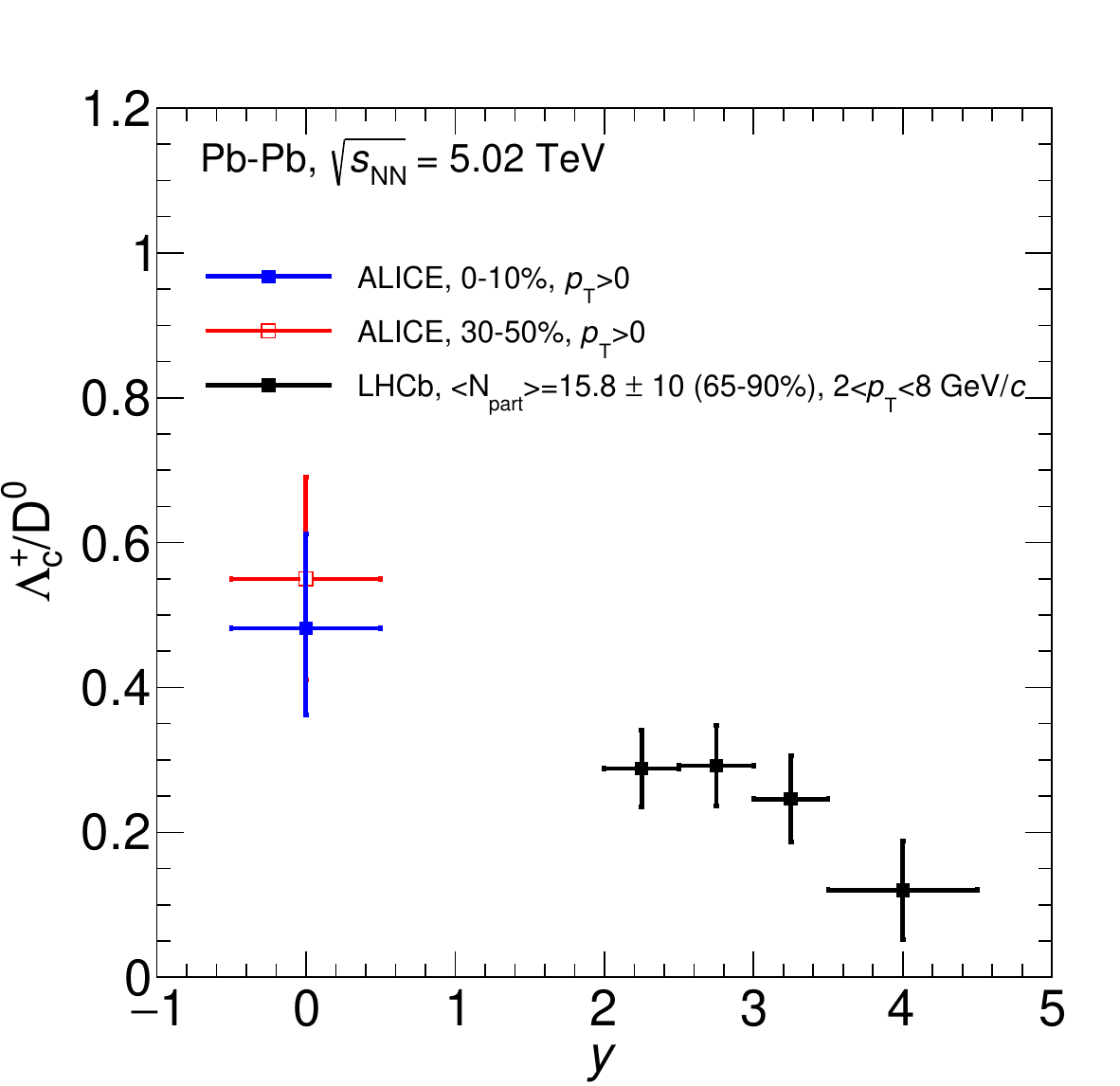}
    \end{center}
    \caption{Top left panel: the \LcD ratio measured in p--Pb collisions at \snn = 5.02~\TeV at mid, forward, and backward rapidity~\cite{ALICE:2017thy,ALICE:2020wla,LHCb:2018weo}. Top right panel (from Refs.~\cite{ALICE:2024ozd,LHCb:2023cwu}): comparison of the $\XicZero/\Dzero$ ratio measured at midrapidity in p--Pb collisions at \s = 5.02 TeV with the measurements at forward and backward rapidity at $\snn=8.16$~\TeV. Bottom panel: the \LcD ratio measured in wide \pt intervals shown as a function of rapidity in central, semicentral collisions~\cite{ALICE:2021bib}, and peripheral~\cite{LHCb:2022ddg} Pb--Pb collisions. The vertical lines represent the total uncertainties. ALICE data derive from the extrapolation to $\pt>0$ of the ratios measured in the interval $1<\pt<24$~\GeVc. In the LHCb case, the centrality interval indicated in Ref.~\cite{LHCb:2022ddg} is $\langle N_{\mathrm{part}\rangle}=15.8\pm 10$. The percentile range corresponding to this interval can be estimated to be approximately 65-90\% based on Ref.~\cite{Gu:2024xsz,LHCb:2021ysy}.
    }
    \label{rapiditypuzzle}
\end{figure}

The comparison of the measurements performed by ALICE and LHCb at mid and forward rapidity, respectively, show tantalising signs of a possible rapidity dependence of the baryon-to-meson ratio in pp, p--Pb, and Pb--Pb collisions~\cite{ALICE:2017thy,LHCb:2023cwu,xic13tev,LHCb:2018weo,ALICE:2020wla,LHCb:2022ddg,ALICE:2021bib,ALICE:2024ozd} for both the \Lc and \XicPlusZero baryons, as displayed in Fig.~\ref{rapiditypuzzle}. However, the current measurements do not allow for conclusions, either because of the large uncertainties (including those on the BR in the \XicPlusZero case) or because of inconsistencies in the accessed \pt intervals or in the definition and ranges used for multiplicity and centrality. As mentioned in Sect.~\ref{sect:exppA}, also the rapidity dependence of the evolution with multiplicity of the $\Ds/\Dzero$ ratio in p--Pb collisions is not yet experimentally established. 
It is envisioned that the LHC collaborations in the future will, as far as possible, perform measurements in the same centrality intervals and transverse momentum, facilitating easier comparisons and reducing potential ambiguities. 

In the beauty sector, the LHCb experiment measured a baryon-to-meson cross-section ratio compatible between p--Pb and pp collisions~\cite{LHCb:2019avm} at both forward and backward rapidity. However, also in this case, the data uncertainties prevent to conclude whether \pt modifications of the baryon-to-meson ratio from pp to p--Pb collisions observed in charmed sector are present also in the beauty sector. In pp collisions ALICE measured at midrapidity the ratio of non-prompt \Lb to non-prompt \Dzero, which is strictly connected to the $\Lb/\mathrm{B^{0,-}}$ ratio~\cite{ALICE:2023wbx}. However, the comparison with LHCb data at forward rapidity is quite indirect, requiring a modelling of the \pt-differential cross section of beauty production tuned on LHCb data and of the decay kinematics, which unavoidably dilutes possible effects over a broader kinematic range. The CMS collaboration measured \Lb production at 7 \TeV but only for $\pt>10$~\GeVc~\cite{CMS:2012wje}. Measurements of the baryon-to-meson ratios at low \pt in the beauty sector at midrapidity via fully reconstructed B and \Lb decays are envisioned for the next runs at the LHC: they will be essential to investigate a possible rapidity dependence of these ratios.

From a theoretical point of view, it was already briefly mentioned that PYTHIA 8 does not predict any difference with rapidity in the charm or beauty baryon-to-meson ratio in pp collisions, and expects larger baryon-to-meson ratios at forward rapidity than those measured by the LHCb Collaboration. 
It would be extremely interesting to systematically investigate models that include coalescence to see if any rapidity dependence in the baryon-to-meson emerges and to understand if a possible effect would be driven by the medium density and/or by possible differences in the heavy-quark density, which would be important for multi-charm hadron states as well as for quarkonia and the \Bc meson, between mid- and forward/backward rapidity. 
However, within a coalescence approach, one would need an appropriate description of the evolution of both the bulk medium and the charm quark momentum distribution from central rapidities to forward ones. In fact, the fireball degree of thermalization and, in general, the time evolution at forward rapidity already for the bulk matter have been less investigated in hydrodynamics or transport theory, and even much less the dynamics of charm quarks. Before making predictions of the $\Lc/\Dzero$ ratio one should check that $R_{\rm AA}(\pt)$ and $v_2(\pt)$ of D mesons at forward rapidity can be correctly predicted, as done at midrapidity. 
This would give good confidence on the level of charm-quark equilibration at forward rapidity and would provide a reasonable knowledge of the charm-quark $\pt$ distribution. 

Furthermore, the study of the directed flow of D mesons in nucleus--nucleus collisions at RHIC and LHC has revealed that the bulk has a tilt with respect to the longitudinal direction and has, therefore, a displacement with respect to the charm quark at forward rapidity~\cite{ALICE:2019sgg,STAR:2019clv,Oliva:2020doe}. This could also imply a different impact of the corona effect, or more generally of the amount of non-thermalized heavy quarks that escape from the fireball. Summarizing, a solid prediction for $\Lc/\Dzero$ within the coalescence approach should be based on a solid description of the bulk and charm distribution function in space and momentum as already done at midrapidity.\\

\noindent \textbf{The role of diquarks}

Coalescence models often calculate the formation of a baryon not as a $3\rightarrow 1$ process but rather as a two-step process, in which first two quarks form a diquark and then the diquark combines with a third quark. However, this does not necessarily correspond to requiring the formation of diquarks as real bound states nor as effective degrees of freedom of the system. Whether this occurs or not is an open question. 
The suppression of the production of \Sigmac states relative to \Lc ones in \ee collisions and its removal in hadronic collisions is striking and appears to support the picture of diquark formation in string breaking. 

While it is very hard to design what could be a \enquote{smoking gun} observation supporting the formation of diquarks is worth noticing that in~\cite{Altmann:2024icx} it was suggested that a measurement of $\Lambda_c$ elliptic flow in Pb--Pb and its difference w.r.t. the D-meson one would allow to infer the relevance of diquark degrees of freedom in the QGP. This possibility is motivated by the comparison of the \pt dependence of the difference between $v_2(\Lambda_c)$ and $v_2(D^0)$ predicted in a coalescence approach and in the POWLANG mechanism. In the coalescence model the $v_2$ of $\Lambda_c$ receives a contribution from two light quarks while the D meson only from one. In the POWLANG approach instead $\Lc$ combines to a diquark evolving hydrodynamically with the bulk medium implying a milder difference w.r.t. D, in particular at intermediate \pt.\\


\clearpage

\noindent \textbf{String-related investigations} 

There are several areas of interest in extending the current modelling of string hadronization. Some have already been mentioned above, including modelling of exotic hadrons and the heavy-ion limit of strings. Further areas of interest include considering colour reconnections in the initial state, the collective behaviour of strings, interactions between final state hadrons, studies into baryon production mechanisms, and even looking beyond Lund strings. Here we elaborate on just a few of these areas of interest.

One noticeable flaw as shown in Fig.~\ref{asymmetry} is the overprediction of the $\rm \Lb/B^0$ ratio by PYTHIA despite well predicting the $\Lc/\Dzero$ ratio due to the inclusion of junction topologies. Exploration into other ratios involving these heavy flavour hadrons such as $\Lb/\Lc$  and $\rm B^0/D^0$ would be particularly insightful into attempting to find the root of the discrepancy between model and experiment. 
Another set of observables that would be particularly interesting for studying junction modelling are baryon correlations. In the string model, baryons produced via standard diquark-antidiquark pair creation from string breaks would be highly correlated. This correlation would be reduced given the popcorn model for diquark creation, which allows meson production between diquark-antidiquark pairs. Comparatively there is generally no expectation of a correlation between junction baryons and antijunction baryons.

As presented throughout this paper, there have been several already experimentally observed behaviours correlated to the size of the system, whether it be observables as a function of charged particle multiplicities or contrasting \ee, \pp and AA results. 
Many measurements have already shown clear indications of collective-like effects in \pp collision environments, which prompts two key questions; are these effects present in \ee collisions (and/or in \ep/\eA ones), and how can we model such collective effects in the string picture? 
With regards to the former, many of these effects were not studied at LEP and hence there is little indication whether they are present in \ee or not, thus it would be intriguing to study them at future high energy \ee, \ep, and \eA experiments~\cite{AbdulKhalek:2021gbh}. 

Many collective effects in the string picture have already been studied in PYTHIA, including string tension modifications (e.g. Rope hadronization~\cite{Bierlich:2017sxk} and close-packing~\cite{Fischer:2016zzs}), string interactions (e.g. string shoving~\cite{Bierlich:2016vgw, Bierlich:2020naj}) and final state hadron interactions (e.g. hadron rescattering~\cite{Ferreres-Sole:2018vgo}). 
Such models have already shown interesting effects and compelling agreement to data. For example, the string shoving model~\cite{Bierlich:2016vgw, Bierlich:2020naj} has already been shown to produce the near-sided ridge experimentally observed in \pp collision events which is unexpected with the baseline string model. Studies of such models in AA environments will be particularly insightful to help determine whether a string model can reproduce other behaviours also typically considered QGP signatures such as flow. 
These models would also have particularly interesting effects on \pt spectra of hadrons, and further phenomenology studies would be useful to determine any signature behaviours of such models that experiments could attempt to identify. 
One can also consider interactions between final state hadrons via hadron rescattering~\cite{Ferreres-Sole:2018vgo}, which would have effects particularly for high multiplicity events. For heavy flavour hadrons in particular, one would expect effects on the shape of the \pt spectra. It is interesting to note that such rescattering effects would also be interesting for studying the $p/\pi$ ratio with respect to charged multiplicity in pp collision events, which PYTHIA currently overpredicts even without the inclusion of junctions. 

Perhaps one of the most intriguing questions for the string model is whether a string in a densely packed string system hadronizes differently from a string in vacuum. 
There are already indications that this is the case given enhancement of the light strange hadron-to-pion ratio as a function of charged multiplicity as seen by ALICE in Ref.~\cite{ALICE:2017jyt}. Higher multiplicity events either correspond the hadronization of longer strings or systems with a higher quantity of strings, and hence generally more densely packed string environments.
There are already existing models such as Rope hadronization~\cite{Bierlich:2017sxk} and close-packing~\cite{Fischer:2016zzs} which attempt to implement a strangeness enhancement mechanism by enhancing the string tension in more dense string systems. 
An increased string tension results in the reduction of the strangeness suppression via the Schwinger mechanism due to the strange quark mass. 
Both Rope hadronization and close-packing models have already been able to describe the general behaviour seen in the ALICE~\cite{ALICE:2017jyt} data for the strange hadron-to-pion ratios as a function of multiplicity.
Additionally, in \pp collisions the majority of strings are aligned along the beam axis, which in turn would result in strangeness enhancement primarily for strings at low \pt, which would lead to strangeness enhancement for low \pt hadrons. One may be hopeful that this low \pt strangeness enhancement could describe the ratios involving heavy strange hadrons such as \XicPlusZero, however as mention in Sec.~\ref{sec:pp} these strangeness enhancement mechanisms alone are unable to describe the level of strangeness enhancement needed to fit the \XicPlusZero/\Lc data whilst maintaining reasonable model parameters. 
Additional measurements involving strange b-hadrons would be vital in examining such mechanisms, so that one can make comparisons between distributions such as the $\Xi_\mathrm{c}/\Lambda_\mathrm{c}$ and $\Xi_\mathrm{b}/\Lb$ ratios alongside the $\Xi_\mathrm{c}/\Dzero $ and $\Xi_\mathrm{b}/\rm B^0$ ratios. These ratios would provide particularly insightful indications to the source of strangeness enhancement and whether the behaviour is different in the beauty and charm baryon sectors. Further data for ratios involving double strange heavy baryons such as $\Omega_\mathrm{c}$ and $\Omega_\mathrm{b}$ would also provide very useful insight. 
One could also explore the enhancement of strangeness around the junction itself which is not an unreasonable assumption as we know little about what the field around a junction itself should look like, and such a model would have significant impact as the majority of heavy flavour baryons are sourced from junctions. 
Likewise due to the sensitivity of heavy baryons on junction formation, 
further data for ratios involving $\Xi_\mathrm{b}$ and $\Omega_\mathrm{b}$ may even help identify the source of the overprediction of the \Lb/$\rm B^0$ ratio using PYTHIA CR-BLC.

To study in further detail the hadronization of a string in vacuum compared to a densely packed string system,
phenomenological studies would be insightful to examine hadron distributions for a single string compared to higher multiplets, given all the aforementioned models, in order to check if there are any identifiable signatures experiments could attempt to find. To directly compare the fragmentation of a triplet vs. octet string, it would be insightful to examine hadron spectra in the tips of gluon jets compared to quark jets, LEP hairpin configurations, and perhaps even diffractive LHC events may allow for relatively clean environments to search for such behaviours.\\

\noindent \textbf{Strings in heavy-ion collisions?} 

Through this review, models in PYTHIA useful for \ee{} up to pp collisions were discussed. PYTHIA itself does not have the ability to perform pA and AA collisions. Instead, to simulate pA and AA collisions one can use the Angantyr model~\cite{Bierlich:2018xfw}, which uses PYTHIA~8 as its base, taking inspiration from the Fritiof model~\cite{Andersson:1986gw, Pi:1992ug} and using the idea of wounded nucleons. A key feature of Angantyr is testing the assumption of continuing the usage of PYTHIA's string fragmentation mechanisms to model the hadronization process also in pA and AA collisions, rather than assuming the formation of a QGP. This includes the implementation of the QCD-CR mechanisms described in Sect.~\ref{sec:pythiaExt}, which allows CR between different nucleon-nucleon collisions~\cite{Lonnblad:2023stc, Bierlich:2023okq}. In principle one can also use Rope hadronization~\cite{Bierlich:2017sxk} and shoving~\cite{Bierlich:2020naj} in conjunction with Angantyr, however the combination of Rope hadronization and Angantyr is yet to be validated to work properly, nor have the model parameters been tuned. Similarly, the string shoving model in Angantyr does not yet perform well for full collisions due technical handling difficulties of soft gluon modelling. Nonetheless toy configurations using the model have already shown that the model itself can generate $v_2$, though further work on the string shoving implementation is required for more detailed studies.
For a review of the current state of the above models in heavy-ion collisions, see~\cite{Bierlich:2024odg}.
Though Angantyr in its current state does not fully describe the collective behaviour of high string densities for full collision events, the existing implementation already provides insight into the non-collective background of such observables and provides an important comparison between the QGP and the string model for hadronization.

This also prompts the interesting question of whether there is a some quenching limit for strings. That is, within this type of modelling paradigm, at high enough string densities, do the string states cease to exist and instead do we get a transition into some deconfined QGP state? This then begs the question when would this transition occur, what would this transitional state look like, and how would one model it. These are largely unexplored questions within the string picture, and the study of AA collisions with string hadronization using Angantyr provides the first step to exploring such questions.\\ 

\noindent \textbf{Heavy-flavour jets and correlations} 
 
 The study of beauty- and charm-tagged jets provides additional information for interpreting the modification of the baryon-to-meson ratios from \ee to hadronic collisions, giving a handle to connect it to possible modification of heavy-quark fragmentation and jet properties. Measurements of jets tagged by the presence of a fully reconstructed heavy-flavour hadron can be particularly informative. This tagging technique, with respect to others based on the recognition of a displaced secondary vertex or on lepton tagging, has the advantage of providing the momentum vector of the heavy-flavour hadron and its correlation to the jet momentum. Though used since UA1 times, in recent years it was refined and exploited to study the production and sub-structure properties of D-meson tagged jets~\cite{UA1:1990mkp,STAR:2009kkp,CDF:1989gpa,ATLAS:2015igt,ALICE:2022mur,ALICE:2022phr,ALICE:2023jgm}, as well as for the first direct observation of the ``dead-cone'' effect~\cite{ALICE:2021aqk}. These measurements allow testing the modelling of jet fragmentation, as well as of charm-production processes, in event generators like PYTHIA and HERWIG~\cite{Corcella:2000bw,Corcella:2002jc,Bellm:2015jjp}, possibly interfaced with POWHEG~\cite{Alioli:2010xd,Frixione:2007vw,Nason:2004rx}. 
Relevant variables include the fraction of jet momentum carried by the charm hadron, calculated along the jet axis direction ($\zjet$), and the transverse momenta and (pseudo)rapidities of hadrons measured with respect to the jet axis.  

\begin{figure}[ht!]
    \begin{center}
    \includegraphics[width = 0.6\textwidth]{ 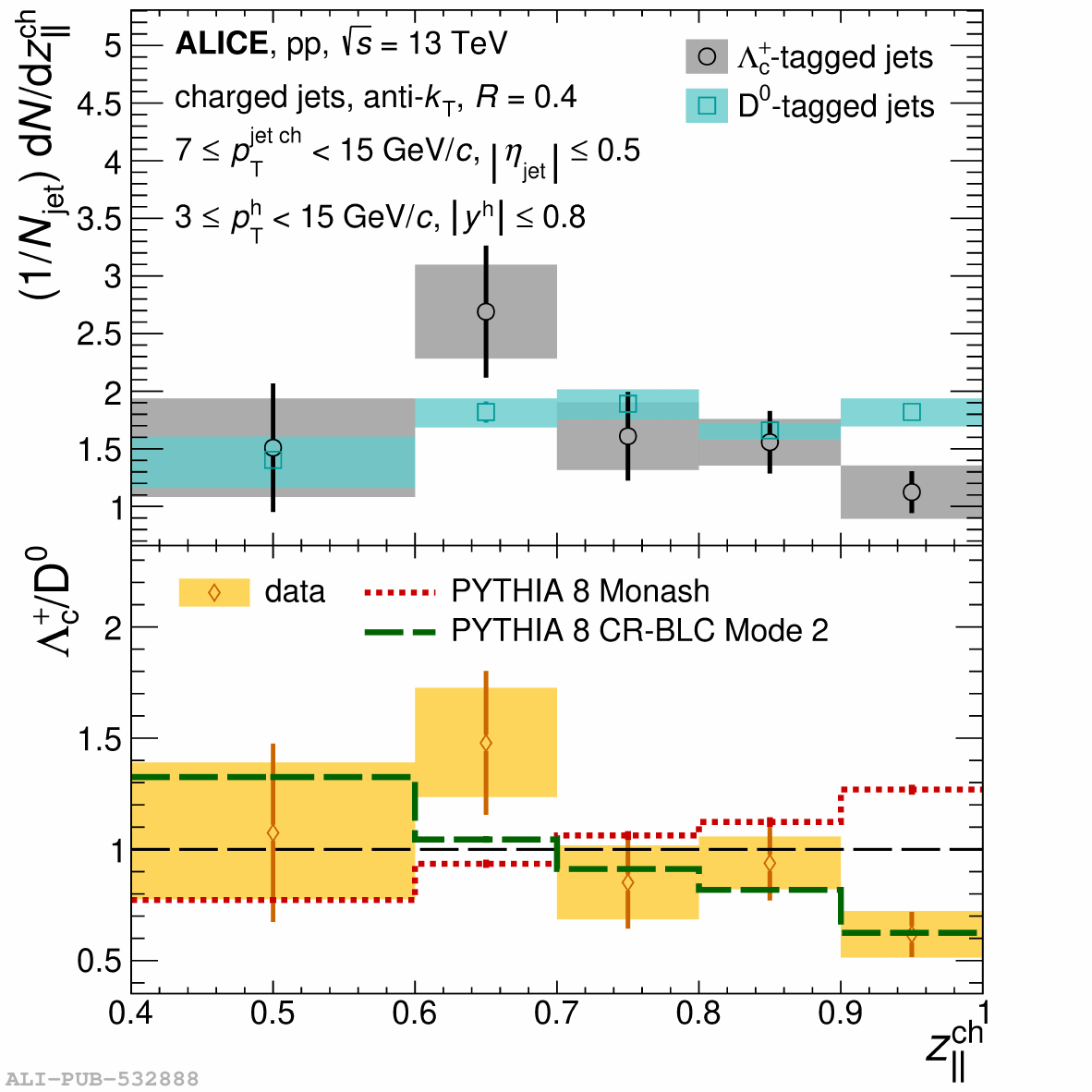}
    \end{center}
    \caption{Comparison (top panel) and ratio (bottom panel) of the distributions of the \Lc and \Dzero longitudinal track-based jet-momentum fraction ($\zjet$). Track-based jets, defined by their charged-particle constituents only, with $7<\pt<15$~\GeVc are tagged with fully reconstructed \Lc and \Dzero with $3<\pt<15$~\GeVc. In the bottom panel the ratio is compared to expectations from PYTHIA 8 default tune (Monash) and PYTHIA 8 tune with colour-reconnection implemented beyond leading colour. Figure from Ref.~\cite{ALICE:2023jgm}}
    \label{fig:LcAndDtaggedJets}
\end{figure}

As extensively discussed in Ref.~\cite{ALICE:2022mur}, the $\zjet$ distribution of D-meson tagged jets measured by the ALICE Collaboration is quite well described by Monte Carlo event generators. Recently, ALICE reported a first measurement of the same distribution for $\Lc$-tagged jets. The comparison with D-meson tagged jet data done with the same jet and hadron kinematic selection is reported in Fig.~\ref{fig:LcAndDtaggedJets}. Though the uncertainties prevent conclusions, the data may suggest that $\Lc$ baryons carry a smaller fraction of jet momentum along the jet axis with respect to D mesons. This would contrast with what expected by the default version of PYTHIA 8 (Monash tune), while it would agree with what PYTHIA~8 CR-BLC expects. The trend expected by Monash reflects the tendency in jet fragmentation that heavier particles carry a larger fraction of jet momentum, which in a collinear fragmentation picture can be related to energy conservation. The data therefore may suggest that either the $\Lc$ production is favoured in jets with higher momentum and softer fragmentation, either intrinsically or because modified by the interplay with the underlying event, or that the $\Lc$ momentum vector is less aligned with the jet one, either as an intrinsic feature of charm-jet fragmentation to $\Lc$ baryons (not expected however in Monash) or because of a angular decorrelation at hadronization time. A partly related study is that of the azimuthal correlation of \Lc baryons with charged particles produced in the event. 

A first preliminary measurement, performed by ALICE~\cite{Palasciano:2023vdw}, indicates that the yields of low-\pt associated particles are large, both in the near-side and in the opposite direction to the \Lc (away-side jet region), and resemble those observed only at much higher D-meson \pt ($\pt\gtrsim 8$~\GeVc) in correlations of D mesons with charged particles. Naively, one could expect that, except for the initial-parton energy correlation, the fragmentation (parton-shower) and hadronization of the near- and away-side jets should proceed almost independently. It is thus surprising that the larger near-side yield, which is consistent with the softer jet fragmentation suggested by the $\zjet$ analysis, is accompanied by a difference in the away-side jet. 

Taken together, the $\zjet$ and azimuthal correlation data may suggest that, with respect to D mesons, low-\pt $\Lc$ are produced on average in jets with higher \pt and softer fragmentation, with a sharing of the jet energy among several soft particles. Whether this could be due to an interplay with the underlying event is not clear, though possible, considering also the multiplicity dependence of $\Lc/\Dzero$ ratio. 

More precise measurements, possibly more differential in jet and hadron kinematics, and a detailed analysis of the jet radial profiles and substructure and of the correlation-function widths could in the future shed light on these intriguing observations. Similar measurements in the beauty sector would also be very useful. 

On top of new measurements, it would be desirable in the future to understand what hadron-formation via coalescence implies for these observables. In the coalescence process, the \Lc at a given \pt gets on average more of its momentum from the light quarks with respect to the D meson (because two light quarks are involved for forming \Lc). This can decorrelate the resulting hadron momentum vector from the jet axis. Therefore, though explicit calculations are missing, it would be in principle compatible with the observation to assume that in the jet a \Lc formed by coalescence of the charm quark with light quarks from the underlying event carries a smaller longitudinal momentum fraction (along the jet axis) relative to that carried by D mesons.\\

\noindent \textbf{High-mass excited baryon states} 
 
 Experimental search of not yet observed charm and beauty hadron resonances is of primary importance in the context of hadronization. It is needed to determine whether the additional baryon resonance states predicted by the RQM, which are largely supported by lattice-QCD computations of the vacuum spectroscopy, exist and to determine their spectral properties. As discussed extensively in this review, the existence of these states is a necessary assumption in many models to reproduce the \LcD yield ratio in pp and heavy-ion collisions, in particular for models based on statistical hadronization. 
Though it remains an open question why these high-mass resonance states have not been observed so far in \ee collisions, proving their existence and measuring the branching ratio to relevant decay channels could be pursued at \ee colliders, e.g. by the Belle II Collaboration, as well as in pp collisions by the ATLAS, CMS, and LHCb Collaborations, leveraging the large LHC luminosities. 

Actually, on top of verifying the number and spectral properties of yet unobserved states, a fundamental test for SHM would be given by a measurement of the production yields of excited mesons and baryons, both known and unknown ones. ALICE recently performed the first measurement of $\rm D^+_{s1}(2536)$ and $\rm D^{*+}_{s2}(2573)$ resonance yields in pp collisions, which provide further pieces of information to investigate the charm-quark hadronization~\cite{Faggin:2023qcf}. Cross-section measurements of excited \Lc, \Sigmac, \XicPlusZero states should be feasible with the Run 3 data at the LHC.\\

\noindent \textbf{\Bc meson} 
 
 The \Bc meson contains a b and a c quark, making it (and its excited states) the only hadron observed so far that contains heavy quarks of two different flavours. The \Bc meson has a net flavour number, which in principle secludes it from the quarkonium family. However, one could support that the proximity of its quark content to the charmonium and bottomonium states (\enquote{hidden} heavy-flavour mesons) makes it an exotic heavy quarkonium. 
 
 The \Bc meson has a particularly low production cross section in pp collisions, because two heavy quark pairs (b$\rm \overline{b}$ and c$\rm \overline{c}$) need to be produced. The obvious mechanism is for both of these heavy-quark pairs to be produced in a single hard-scattering process, but since this is quite suppressed, additional production mechanisms may play a role such as multi-parton interactions in high-multiplicity pp collisions~\cite{Egede:2022lws} and/or quark recombination in heavy-ion ones. That is, a b quark produced in a given hard scattering could recombine with a c quark from another parton-parton or nucleon-nucleon collision
(presumably with some space-time suppression since the pair needs to be sufficiently close to form a \Bc). 

If this is a significant mechanism in heavy-ion collisions, the \Bc yields might be dramatically augmented compared to the expectations from $N_\mathrm{{coll}}$-scaling, considering the small primary production expected from the pp cross section. 
The CMS Collaboration has measured for the first time the \Bc production in heavy-ion collisions and reported that the \Bc meson is shown to be less suppressed than quarkonia and most of the open heavy-flavour mesons, suggesting that effects of the hot and dense nuclear matter created in heavy-ion collisions contribute to its production~\cite{CMS:2022sxl}. 
However, the acceptance of the CMS detector only allows to reconstruct \Bc mesons of relatively high transverse momentum, typically \pt $>$ 6 GeV/$c$, while the majority of the \Bc mesons produced by recombination should have a \pt lower than the acceptance threshold of CMS. This is because most of the charm quarks present in the QGP have low \pt and because the recombining quarks need to have a small relative momentum (typically not larger than the binding energy) to be able to recombine. 

Hence, it would be extremely beneficial to measure low transverse momentum \Bc in heavy-ion collisions and in high-multiplicity pp ones.
Recently, Egede et al.~\cite{Egede:2022lws} predicted a large fraction of the produced yield in high-multiplicity pp collisions to come from MPI (albeit without accounting for possible space-time suppression effects so this should probably be taken as a maximum), and Rapp et al.~\cite{Wu:2023djn} predict a similar trend in heavy-ion collisions from regenerated \Bc mesons. In this latter study they observed that the large regeneration contributions cause a markedly rising \Raa with centrality, reaching values of up to 4-6 in central collisions. In this model, the inclusive \pt-dependent \Raa for the \Bc is dominated by regeneration contributions for \pt $<$ 10-15 GeV/$c$ in semi-/central collisions, reaching values of around 10 at low \pt. It is pointed out in the publication that the results for the \Raa are rather sensitive to the \Bc production cross section in pp collisions. A more precise measurement of this quantity, and a significant reduction of the statistical and systematic uncertainties of \Bc in heavy-ion collisions will provide unprecedented insights into heavy-flavour in-medium and is a promising new probe to study the interplay between the suppression and enhancement mechanisms in the production of heavy-flavour mesons in the QGP.\\

\noindent \textbf{Multi-charm hadron states and charm nuclei} 
 
 Analogously to the \Bc case, another window to investigate colour-reconnection and coalescence 
 are the measurements of multi-charm baryons and of charm nuclei. Measurements of multi-charm hadrons, such as the $\Xi^+_{\rm cc}$, $\Xi^{++}_{\rm cc}$, $\Omega^+_{\rm cc}$, $\Omega^{++}_{\rm ccc}$ would provide a direct window on hadron formation from a deconfined quark--gluon plasma. In fact, the yields of multicharm baryons relative to the number of produced charm quarks are predicted to be significantly enhanced in AA relative to pp collisions~\cite{Andronic:2021erx,Minissale:2023bno,Cho:2019syk,He:2014tga,Becattini:2005hb,Zhao:2016ccp,Yao:2018zze}. Enhancements are expected by as much as a factor $10^2$ for $\rm \Xi_{cc}$ baryon and even by as much as a factor $10^3$ for the as yet undiscovered $\rm \Omega_{ccc}$ baryon. Most recently, the SHMc and the Catania models~\cite{Andronic:2021erx,Minissale:2023bno} provided predictions for the production of multi-charmed hadrons in Pb--Pb collisions and they explored the system size dependence through Kr--Kr, to Ar--Ar and O--O collisions. The yields are predicted to be quite similar in the case of full thermalization of the charm quark, a condition that might not be achieved in light systems like O-O. 
 The coalescence picture coupled to a realistic dynamical simulation for charm quarks reveals that $\rm \Omega_{ccc}^{++}$ yield, and even more its \pt spectrum, would provide a sensitive probe of charm thermalization and its evolution with the system size of the colliding system \cite{Minissale:2023bno}. Such studies are currently far beyond the experimental reach and a comprehensive program of measurements in the multi-charm sector is discussed to be carried out with the proposed future ALICE 3 detector~\cite{ALICE:2022wwr}. 
 
 The unique combination of high rates, wide acceptance, strong particle identification capabilities, and excellent secondary vertex determination of the proposed ALICE 3 detector, also provide a toolbox for the search for light nuclei with charm. At LHC energies, the most promising candidates are the c-deuteron, c-triton and c-$^{3}$He. The existence of such weakly decaying bound states with lifetimes similar to those of other unbound charmed baryons is debated in the literature. In case these states are bound and under the assumption that their abundance in collisions of nuclei is described by the SHM \cite{Andronic:2021erx,ExHIC:2017smd,Hosaka:2016ypm,Krein:2017usp,Krein:2019gcm}, the yields are large enough to bring their experimental discovery within reach in the next decades and to additionally test different hadronization models.\\

\noindent \textbf{Exotic hadrons} 

The \chiThreeEight is an exotic particle that was first observed by the
Belle Collaboration~\cite{Belle:2003nnu}, and then confirmed and studied by other experiments at electron-positron~\cite{BESIII:2013fnz} and hadron colliders~\cite{CDF:2003cab,D0:2004zmu,LHCb:2011zzp,CMS:2013fpt,ATLAS:2016kwu}. The nature of this particle is still not fully understood and interpretations
in terms of conventional charmonium (a bound state of charm-anticharm quarks), D-meson molecule, tetraquark, or an admixture these states have been proposed~\cite{Chen:2022asf}. Heavy-ion collisions, as well as high multiplicity pp collisions, offer a unique window into the properties of this poorly understood hadron. In these systems, promptly produced \chiThreeEight hadrons can interact with other particles in the nucleus and/or those produced in the collision. The influence of these interactions on the observed \chiThreeEight yields provides information that can help discriminate between the various models of its structure, as well as give insight into the dynamics of the bulk of the particles produced in these collisions. The LHCb Collaboration has measured a significant decrease in the ratio of prompt \chiThreeEight to $\psi$(2S) cross sections in pp collisions with increasing charged-particle multiplicity~\cite{LHCb:2020sey}. These data were interpreted in terms of breakup of the \chiThreeEight hadron due to interactions with comoving particles produced in the event, for both compact and molecular models of \chiThreeEight structure~\cite{Esposito:2020ywk,Braaten:2020iqw}. A relatively-compact tetraquark is preferred in this case, but it is currently discussed how conclusive this comparison is~\cite{Wu:2020zbx,Lee:2023ysk,Esposito:2021vhu,Grinstein:2024rcu}.
The LHCb Collaboration has also recently performed the same measurement in p--Pb collisions~\cite{LHCb:2024bpb}. In this collision system the ratio is measured to be higher than the one measured in pp collisions. It must be noted that the evolution of $\psi$(2S) production with particle multiplicity is not fully understood yet. This complicates the interpretation of \chiThreeEight data, when $\psi$(2S) is used as a reference. 
The CMS collaboration has measured the $\rm \sigma_{\chiThreeEight}/\sigma_{\psi(2S)}$ ratio in Pb--Pb collisions at high \pt, and found that the ratio is further enhanced relative to pp and p--Pb collisions, although the uncertainties and the significantly different \pt intervals of the measurements preclude
drawing firm conclusions~\cite{CMS:2021znk}. 
Calculations based on quark coalescence show that production rates of \chiThreeEight hadrons in AA collisions are sensitive to its structure. In these models, the production of compact tetraquarks is expected to be enhanced over hadronic molecules~\cite{ExHIC:2010gcb,ExHIC:2011say}. However, a recent transport calculation reaches the opposite conclusion~\cite{Wu:2020zbx}. Late-stage interactions in the hadron gas phase of a heavy-ion collision can also modify the yields~\cite{Abreu:2016qci}. It is currently unknown where the crossover between the suppressing effect of breakup and the enhancing effect of coalescence may occur. The increase of the ratio of cross sections $\rm \sigma_{\chiThreeEight}/\sigma_{\psi(2S)}$ from pp to p--Pb to Pb--Pb collisions may indicate that the exotic \chiThreeEight hadron experiences different dynamics in the medium than those relevant for a \enquote{conventional} charmonium state like $\psi$(2S). If the experiments could in the future cover similar \pt intervals in the different collision systems and multiplicities/centralities, possibly better assessing the lower \pt region, these measurements will provide new constraints to determine the internal \chiThreeEight structure as well as to model parton transport and hadronization in nuclear collisions. 
In this context, it will also be important to precisely measure the production cross section of other exotic states, like the $\rm T^+_{cc}$.

\begin{figure}[ht!]
    \begin{center}
    \includegraphics[width = 0.8\textwidth]{ 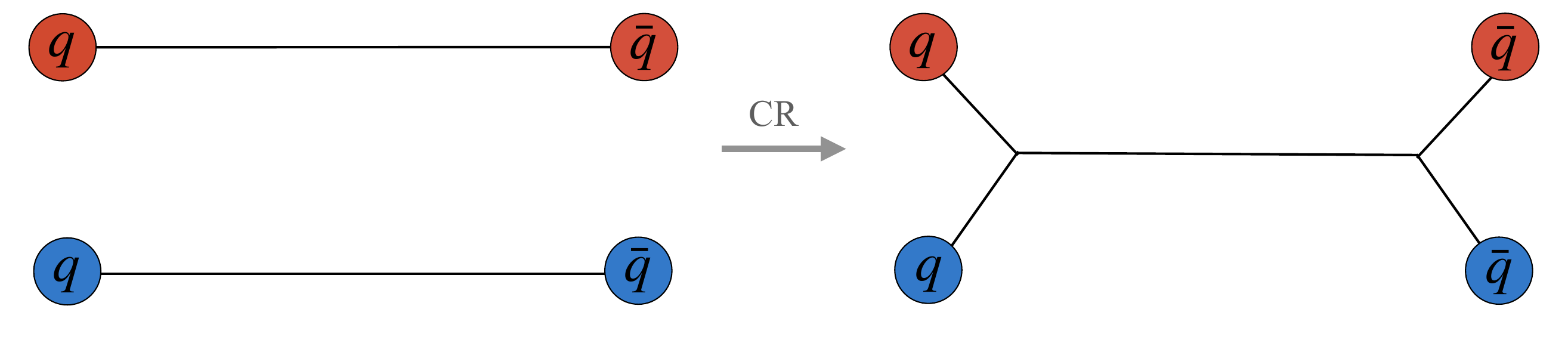}
    \caption{Junction-antijunction formation in the string model. The left images show the initial string configuration in the LC limit, and the right images depict the possible alternative configuration given the respective colour reconnection models.}
    \label{fig:JA}
    \end{center}
\end{figure}

 A related possible future implementation in PYTHIA is the inclusion of exotic hadrons. Not only does QCD-CR allow beyond LC topologies such as junctions, but it also allows for more complicated structures called junction networks. These junction networks are string configurations that consist of multiple junctions within the structure and provide a natural way within the string picture to represent tetraquarks, pentaquarks, etc. For example, the simplest of these structures are junction-antijunction configurations which consist of two colour-charged endpoints and two anticolour-charged endpoints as shown in Fig.~\ref{fig:JA}. When these junction-antijunction configurations are encountered as high mass systems, PYTHIA currently handles these structures by making breaks on the string piece connecting the junction to the antijunction, and then fragmenting each junction system separately. However, should a junction-antijunction system have a low invariant mass, one could propose that the system does not fragment and instead can be treated as a tetraquark. The same argument holds for larger junction networks. The inclusion of exotic hadrons is not yet implemented in PYTHIA and it would not be straightforward to implement the required significant modifications to the existing code, however junction networks are already present and could provide a sensible starting point for modelling such structures.\\ 

\begin{figure}[t]
 \centering
    \includegraphics[width=0.61\textwidth]{ 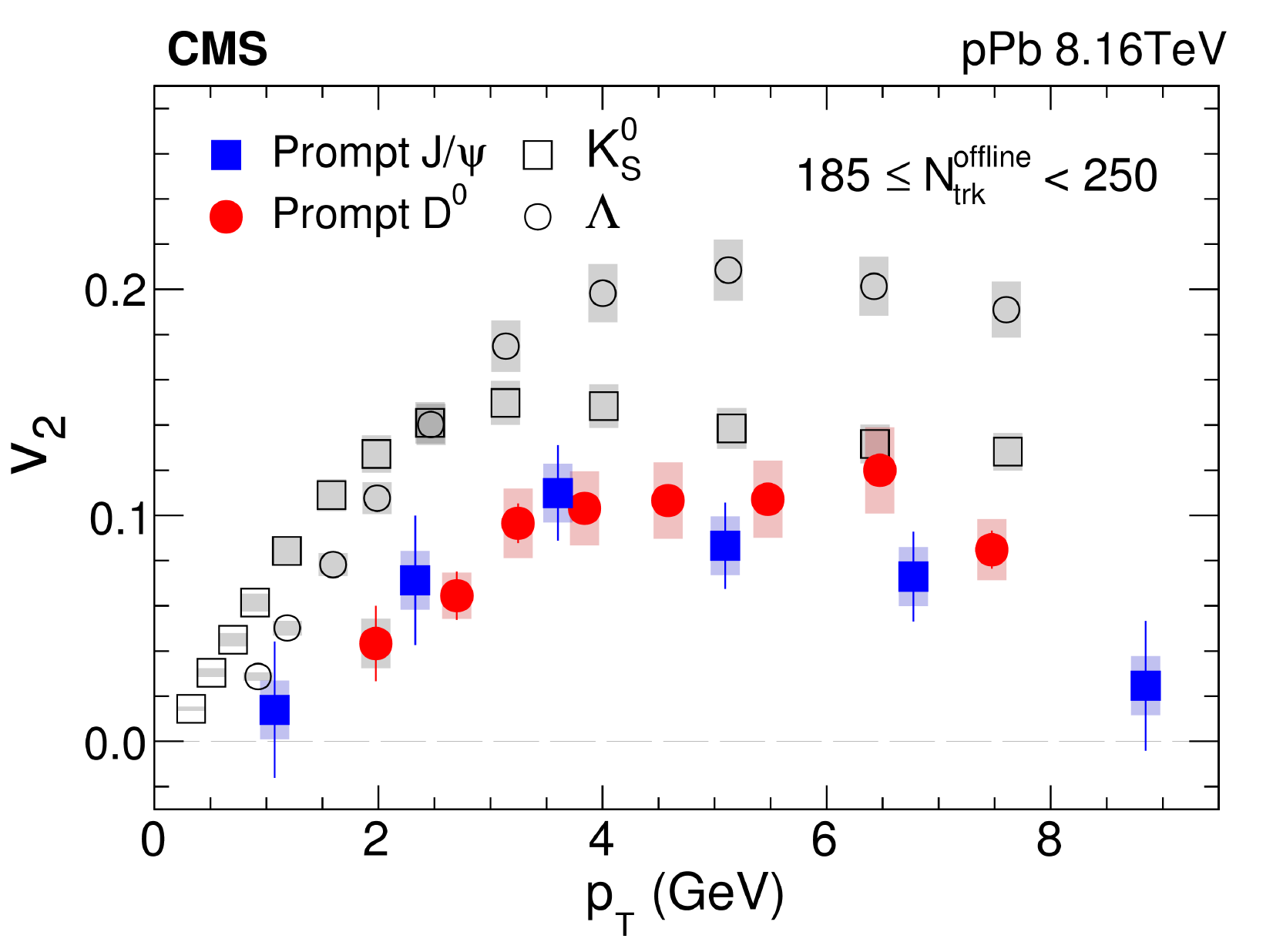}
    \caption{
    Elliptic flow of prompt \Jpsi at forward+backward rapidity ($-2.86 < y_{\mathrm{cm}} < -1.86$ or $0.94 < y_{\mathrm{cm}} < 1.94$), and of \Dzero, $\Lambda$ and $\mathrm{K^{0}_{s}}$ at midrapidity shown as a function of \pt in high-multiplicity p--Pb collisions at 
    \s=8.16\TeV~\cite{CMS:2018duw,CMS:2018loe}.}
    \label{fig:flowsmall}
\end{figure}
\noindent \textbf{Collectivity in small systems} 
 
 Positive \ellflow values have been observed for both open charm~\cite{CMS:2018loe,ALICE:2018gyx,ALICE:2022ruh} and \Jpsi~\cite{ALICE:2017smo,CMS:2018duw} also in high-multiplicity p--Pb collisions, as well as in high-multiplicity pp collisions~\cite{ATLAS:2019xqc}. As shown in Fig.~\ref{fig:flowsmall}, in high-multiplicity p--Pb collisions \Dzero and \Jpsi \ellflow reach similar maximum values, which are about a factor of two smaller than the maximum \ellflow of light hadrons~\cite{ALICE:2013snk} and significantly smaller than the maximum observed in semi-central Pb--Pb collisions for D mesons. Though the maximum \Jpsi \ellflow values are similar in high-multiplicity p--Pb collisions and semi-central Pb--Pb collisions, in the latter case positive values, close to open-charm and light-flavour ones, are observed up to higher \pt. The origin of flow in small collisions systems are still debated, even for light-flavour particles~\cite{Noronha:2024dtq}. Hadronization via recombination could induce a positive \ellflow for open-charm hadrons even in the case of a small/negligible \ellflow of charm quarks \cite{Greco:2003vf}. This is of course not the case for charmonia. The absence of clear experimental evidences of charm energy loss, makes the observation of open charm and charmonia flow particularly interesting to understand the interplay of initial-state and final-state effects for the generation of azimuthal anisotropy in momentum space. A possible dynamic not yet firmly identified could lead to an $R_{\rm AA}(\pt)$ close to unity together with a sizeable $v_2(\pt)$ is the super-diffusion in the initial stage glasma phase that would lead to an increase of the $R_{\rm AA}(\pt)$ \cite{Sun:2019fud,Liu:2019lac,Avramescu:2023qvv} followed by a moderate energy loss that generate a sizeable elliptic flow.

The points above outline an exciting program of future measurements and theoretical developments that will require significant effort from both the experimental and theoretical communities involved in the topic. Understanding hadronization may also shed light on the properties and degrees of freedom of the system in the partonic phase before and during the formation of hadrons. From the experimental side, some of the mentioned measurements may not be within reach for another 10-15 years, i.e., until the completion of the ALICE and LHCb detector upgrades proposed for Run 5 at the LHC. Nonetheless, achieving a satisfactory understanding and modelling of such a fundamental process of nature is certainly worth the effort.

\section*{Acknowledgements}
J.A.~and P.S.~acknowledge support from The Royal Society and the Wolfson Foundation (UK), from the Australian Research Council Discovery Project DP230103014, and from the Monash-Warwick Alliance for Particle Physics. V.G. acknowledge support PRIN2022 under fundings EU Next Generation (project code 2022SM5YAS).


\bibliography{bibliography}





\end{document}